\documentclass[11pt]{article}
\usepackage{amssymb,lscape,cite,multirow}
\usepackage{epstopdf}
\usepackage{amssymb}
\usepackage{amscd}
\usepackage{amsmath,graphicx}
\usepackage{xypic}
\usepackage[matrix,arrow,curve]{xy}
\def\1ad{\mbox{\normalsize $^1$}}
\def\2ad{\mbox{\normalsize $^2$}}
\def\3ad{\mbox{\normalsize $^3$}}
\def\4ad{\mbox{\normalsize $^4$}}
\def\5ad{\mbox{\normalsize $^5$}}
\def\6ad{\mbox{\normalsize $^6$}}
\def\7ad{\mbox{\normalsize $^7$}}
\def\8ad{\mbox{\normalsize $^8$}}

\makeatletter
\@addtoreset{equation}{section}
\makeatother

\def\beq{\begin{equation}}                     %
\def\eeq{\end{equation}}                       %
\def\bea{\begin{eqnarray}}                     
\def\eea{\end{eqnarray}}                       
\def\nn{\nonumber}

\def\stt {SU(3)$\times$SU(3)}
 
\def\0 {\nonumber} 
\def\del{\partial}

\def\minicent#1#2{
  \begin{minipage}{#1 cm}
    \begin{center}
     #2 
    \end{center}
  \end{minipage}
}
\newcommand{\sla}{\slash\!\!\!\!}

\def\N{\mathcal N}
\def\J{\mathcal J}
\def\G{\mathcal G}
\def\O{\Omega}

\def\Im{{\rm{Im}}}
\def\tts{$T \oplus T^*$ }
\def\l{\|}
\def\gs{g_s}

\newcommand{\I}{{\rm{Im}}}
\newcommand{\R}{{\rm{Re}}}

\def\stt {SU(3)$\times$SU(3)}

\def\slas#1{\rlap{\begin{picture}(10,10)(-5,0)
\put(0,0){\line(2,1){15}}
\end{picture}} #1 }
\def\slashh#1{\rlap{\begin{picture}(10,10)
\put(0,0){\line(5,1){40}}
\end{picture}}#1}

\newcommand{\yes}{{\rm yes}}
\begin{document}
\setcounter{page}{0}
\begin{titlepage}
\titlepage
\rightline{hep-th/0609124}
\rightline{SPhT-T06/094}
\rightline{SU-ITP-06/23}
\vskip 3cm
\centerline{{ \bf \LARGE A scan for new  $\N=1$ vacua on twisted tori
}}
\vskip 1.5cm
\centerline{Mariana Gra{\~n}a$^{a}$, Ruben
Minasian$^{a}$, Michela Petrini$^{b}$
and Alessandro Tomasiello$^c$}
\begin{center}
\em 
$^a$ Service de Physique Th\'eorique,                   
CEA/Saclay \\
91191 Gif-sur-Yvette Cedex, France  
\vskip .4cm
$^b$ LPTHE, Universit\'es Paris VI et VII, Jussieu \\
75252 Paris, France
\vskip .4cm
$^c$ITP, Stanford University, Stanford CA 94305-4060
\end{center}
\vskip 1.5cm  
\begin{abstract}

We perform a systematic search for $\N=1$ Minkowski vacua   of type II string theories on  compact 
six--dimensional parallelizable  nil-- and solvmanifolds (quotients of six--dimensional nilpotent and 
solvable groups,  respectively). 
Some of these manifolds have appeared in the construction  of string backgrounds and are typically called 
twisted tori. We look for vacua directly in ten 
dimensions, using the a reformulation of the supersymmetry condition in the framework of generalized complex geometry.
Certain algebraic criteria to establish compactness of the 
manifolds involved are also needed. 
Although the conditions for preserved $\N=1$ supersymmetry fit 
nicely in the framework of generalized complex geometry, they are notoriously
hard to solve when coupled to the Bianchi identities.
We find solutions in a 
large--volume, constant--dilaton limit.  Among these, we identify those 
that are T--dual  to backgrounds of IIB on a conformal $T^6$ with self--dual  
three--form flux, and hence conceptually not new. 
For all  backgrounds of this type fully localized solutions can be obtained. The 
other new solutions need  multiple intersecting 
sources (either orientifold planes or  combinations of O--planes and D--branes) 
to satisfy the Bianchi identities; the full list of such new solution is given.  These 
are so  far only 
smeared solutions, and their localization is yet unknown. Although valid  in 
a large--volume limit, they are the first examples of 
Minkowski vacua  in supergravity 
which are not connected by any duality
to a Calabi--Yau. Finally,  we discuss 
a class of flat solvmanifolds that may lead to $AdS_4$ vacua of type 
IIA strings.

\end{abstract}

\vfill
\begin{flushleft}
{\today}\\
\vspace{.5cm}
\end{flushleft}
\end{titlepage}

\newpage

\tableofcontents

\section{Introduction}

The interest in flux compactifications was originally driven primarily by 
practical considerations --- they allow for a stable supersymmetry 
breaking, have mechanisms for moduli stabilization and may even hold a key 
to the solution of the hierarchy problem (see \cite{MG} for a review and references).  A new motivation is now emerging:  understanding the phase space of string theory, in which an important
role is played by flux vacua.  In spite of much progress registered in 
the last few years, our understanding of these backgrounds is based essentially on a 
rather limited collection of classes of examples.

Supersymmetric solutions are most commonly looked for in a four--dimensional 
effective gauged supergravity framework (some of the references directly related 
to the present context are \cite{Kachru,VZ, Dere, DF, Hull1, DF2, VZ2, K-taylor, Camara, Dalpre,ferrtri, Dere2, trigiante,
Hull2,Camara2}) . This has succeeded in producing many solutions. 
It is not always obvious however that the models produced this  way can be 
embedded consistently into string theory. For example, the effective 
field theory
is always derived in a large--volume limit, in particular not taking into 
account warp factors. On the other hand, it is not  clear whether a solution can be
made compact at all. We will see in particular that there are certain 
obstructions to this that are invisible from a four--dimensional perspective.

In this paper, we will look for vacua directly in ten dimensions, using 
geometrical methods that have recently introduced 
 some systematics in  
the classification of supersymmetric string backgrounds.  
This systematics has been proven helpful for finding and classifying solutions
where the ``internal'' six--dimensional part is 
non--compact. Their large number is in contrast to the scarcity of compact backgrounds. (An obvious 
explanation for this difference is found in the tadpole condition, which is 
notoriously hard for the compact internal spaces.) The conditions for  
$\N=1$ supersymmetry can be reformulated 
 in terms of certain 
combinations of even or odd differential forms called pure spinors. 
The differential equations that supersymmetry imposes on 
them are best analyzed in the framework of generalized complex geometry (GCG) \cite{hitchin, gualtieri}, 
which was developed concurrently with the progress on the physics side of 
the problem (an applied review of GCG will be given in Section \ref{Gentwist}).

The search for $\N=1$ vacua proceeds in three steps. The first one, 
consists in solving one of the conditions for a supersymmetric compactification
to four--dimensional Minkowski space, namely   
the existence of a closed pure spinor on the internal space (see eq.~(\ref{int})). 
This is equivalent to a
reduction of the structure group on $T \oplus T^* (M)$ to SU(3,3) and the 
integrability of the associated generalized complex structure \cite{gmpt, gmpt2}. Spaces 
with this property  have what is called a generalized Calabi-Yau (GCY) structure. 
This, however, 
is only half of the story. The full match with supersymmetry conditions 
requires the existence of a second compatible pure spinor (or in other 
words a further reduction of the structure group on $T \oplus T^* (M)$ to 
SU(3)$\times$SU(3)) whose real part is closed, and whose imaginary part is 
the RR field (see eq.~(\ref{Im})). This is the second step 
of the procedure. The NS flux $H$ enters the equations 
through the differential $(d-H\wedge)$.
 The metric and the B--field in the internal space are determined by
the two pure spinors.

Notice that the RR fields are completely determined by 
the geometry. In a related way, the RR equations of motion 
automatically follow from the supersymmetry conditions. 
Up to this point, finding a supersymmetric string 
background is a perfectly algorithmic procedure. Indeed, starting from a 
generalized CY structure, i.e.~a twisted closed pure spinor, one has to 
find a compatible partner for it, and calculate the RR flux by acting on 
the latter with $(d-H\wedge)$. 
However in order 
to promote a configuration satisfying the supersymmetry conditions to a 
full solution, one has to check  the NS three--form equation of motion 
and the Bianchi Identities (BI) for all the fluxes. 
This is the final step
in the search for $\N=1$ vacua.

The first step in this program is where ``twisted tori'' come in. It has 
been shown \cite{CG} that all six--dimensional {\it nilmanifolds} are generalized
Calabi--Yau. Nilmanifolds are iterated torus fibrations over tori, or 
alternatively, as the name suggests, quotients of nilpotent groups; we will 
review them in Section \ref{sec:solvm}. Nilmanifolds are sometimes used as toy models in 
mathematics (because of their tangent bundle being trivial, as we will see) 
to answer general questions; for example they provided the first
example of a symplectic non--K\"ahler manifold. In fact, they are fully classified
in dimension six, and some of them are neither complex nor symplectic:  
they are however all  generalized Calabi--Yau (see Figure \ref{fig:nil} in Appendix \ref{app:Nil}).
This class of geometries will be the 
principal target of our investigation. We also considered a certain class of 
solvmanifolds, 
i.e.~quotients of six--dimensional (algebraic) 
solvable groups, which have also 
been fully classified. While there are many more 
six--dimensional solvable algebras than  nilpotent ones (actually, nilpotent
algebras are a subset of the solvable ones), only few 
of them yield compact six-manifolds. Some of these manifolds have already 
appeared in the flux compactifications literature \cite{Dalpre, sugra1, sugra2, sugra3, sugra4}. The connection between
compactifications on group manifolds
with discrete identifications and 
Scherk-Schwarz reductions of supergravity is explained in \cite{Hull1}.
 Here we present a systematic study of this class of geometries.

It should be stressed at this point that most of these arguments deal with 
left--invariant (or {\it invariant}, for short) structures. Namely, all the forms are 
taken to have constant
coefficients in the basis of forms left--invariant under the group actions (all the
 manifolds we are considering are homogeneous). This is what the counting 
in Figure \ref{fig:nil} refers to.  This is also what we do in most of the paper: 
taking the coefficients to be constant corresponds physically to working in 
a large--volume limit  in which all the space variations (in particular, that
of the warp factor) become negligible.
This is the same limit in which four--dimensional analysis works, and
hence the ten--dimensional perspective might seem to have little advantage over 
it. But in ten dimensions one can at least hope to find later the ``true'' 
solutions with non--constant warping. This can actually be done in the T--dual
cases, as we will see (see also \cite{Schulz}); a recent attempt in an AdS$_4$ case
can be found in \cite{bobby}.

The second and third steps in the search are more challenging. As we mentioned, the total RR field $F$ 
is determined by the geometrical input of a second pure spinor compatible 
with the one defining the generalized CY structure;  
and nothing guarantees a priori that $F$ constructed 
this way will satisfy the Bianchi identities, $(d-H\wedge)F=0$. 
The best we can do is to work out the most general second pure spinor and try
out all the resulting RR fields. 

Let us outline some general features of 
the analysis of the BI. This reads in general 
$$(d-H\wedge)F=\mathrm{source}\ ;$$
 taking the singlet under \stt\ 
(see (\ref{tadpolesinglet})), one reproduces a no--go theorem normally
obtained by using the  four--dimensional Einstein equation 
(but see \cite{GMPW}). 
This no--go theorem says, for the case of Minkowski vacua,  
that orientifold planes are necessary for supergravity 
compactifications with fluxes. There is however
obviously more to the BI than just the singlet, in general. The 
most popular flux background so far is a conformally CY manifold 
with O3--planes and an imaginary self--dual three-form flux \cite{GP,GKP,BB} (usually called type B). 
This one is lucky enough to 
have just the singlet component in the tadpole, as the left hand side is 
naturally a top-form: $ dF_5 - H_ 3 \wedge F_3$. 

Consider now the BI in the general case. 
In the large--volume limit, we can compute the left hand side as a sum
$\sum_i c_i \eta^i_n$
of invariant forms  $\eta^i_n $ on the internal manifold with constant 
coefficient $c_i$. (See for example (\ref{dfmodel1}).)
Projecting to the singlet, as mentioned, we get a positive 
net contribution which says that we need at least one source of net negative 
charge to cancel it. 
The singlet projections of the individual terms in the sum can be both positive 
(and thus O--planes have to be placed there) or negative 
(need D--branes sources for cancellation). 

A true D--brane or O--plane source would actually give a $\delta$--function on the right 
hand side, which does not look like the left hand side $\sum_i c_i \eta^i_n$ we 
computed.  Hence, the BI can be satisfied only after the introduction of a
 non--constant warp factor, which gives rise to extra terms in the left hand side. 
In the large--volume limit, though, we can still at least check that the delta's on the right hand side are in the same directions
as the constants on the left hand side: that is, they multiply the
same $\eta^i_n$, and that the numbers also match. We will call these 
limits ``global'' solutions, because of this ``integration". We will 
{\it assume} in this paper that this is a necessary 
condition for there to be a honest, ``local'' solution to the BI identities {\it
connected} to the large volume. This assumption is inspired by 
the T--dual cases, in which the condition turns out to be also 
sufficient. However, for example, we cannot rule out that there are solutions 
completely disconnected to the ones in the large volume we are considering here.
\vskip .4cm 

We find the following three types of compact solutions:
\vskip .2cm 
\noindent $\bullet \,\,\,\,$ {\sl T--duals} $\,\,\,\,$ These are the configurations 
T--dual to IIB solutions on a conformal
$T^6$ with O3--planes and 
 self-dual three-form flux. One can take the T--dual by first 
going to the large--volume limit, then by using the isometries thus gained to 
perform some T--dualities. This changes the topology of the torus, giving rise to a nilmanifold, in fact not the most general but one with a so--called nilpotency 
degree up to 2, as we will see. One can then go back to a solution with a 
non--constant warping. The T--dual solutions still enjoy the property (that we 
saw for the $T^6$ solution) that the 
BI for the RR flux, even if no longer a top--form, has only the singlet 
component.  In each T--dual case there is  a single direction for the O--planes. 
The full solutions are found both in IIA and IIB 
(with different types of pure spinors and orientifold planes). There are 
nine different algebras that give rise to these solutions.
\vskip .2cm
\noindent $\bullet \,\,\,\,$ {\sl Multi-source solutions} $\,\,\,\,$  These 
are Minkowski vacua that are {\it not} T--dual to a IIB  solution on $T^6$ with  
self--dual three--form flux; hence they are conceptually new. They can be 
found in 
both IIA and IIB, have multiple (intersecting) source terms. There is only one
solution for internal 
nilmanifolds  and three on a single solvmanifold that admits a flat 
metric. Their lifting to a 
full solution (with non-trivial warp factor) is not known. There are special points 
or lines in moduli space where the solutions on 
the solvmanifold have no flux. There are also other few solutions on solvmanifolds
admitting flat metrics which have no flux. 
\vskip .2cm
\noindent $\bullet \,\,\,\,$ {\sl $AdS_4$ solutions}  $\,\,\,\,$  Demanding that the internal space is GCY, we find AdS$_4$ solutions in IIA. The internal space
is flat (a very special case of GCY indeed; other than $T^6$ \cite{K-taylor}, the examples 
involve solvmanifolds).
The only non-vanishing components of the fluxes  ($H$, $F_0$ and $F_4$) are singlets, 
and they conspire to generate the cosmological constant. The warp 
factor and dilaton are constant and O6--planes are required.

\vskip .4cm 
\noindent
The labels we just introduced will be used as a shorthand throughout the text 
hopefully without confusion. We had discussed so far everything in a regime of constant warp 
factor and dilaton, or in other words  global solutions.  Completing the 
solutions 
by turning on the warp factor and the dilaton (of 
course, without violating the closure of the pure spinor defining the GCY 
condition), is the final problematic issue. We succeed in finding the 
lifting for the T--dual configurations to full localized solutions. We have 
not managed to do this for the second type, where the nil(solv) manifold 
in question is not related by T--duality to $T^6$. Thus the heuristic 
summary of the current state of affairs is that a local solution is 
found in all and only cases where the RR flux determined by the geometry 
happens to generate only one component in the BI.

The structure of the paper is as follows. We start by reviewing the nil- 
and solvmanifolds. For the later the necessary condition for the 
compactness is explained in detail. We then turn to a rather detailed 
``practical guide'' into GCG and pure spinors.  Conditions for supersymmetry preservation  and the Bianchi Identities for RR fluxes are considered in Section 4. We present the new global multi-source solutions in Section 5.  Section 6 discusses the  T--dual models. It contains some material about T--duality transformations of pure spinors that goes beyond the direct applications to twisted tori.   
Sections 7 are devoted to AdS$_4$ vacua and a special case of flat solvmanifolds. We conclude by a brief discussion of some of the possibilities for 
further constructions that stay outside the scope of this paper. In Appendix  \ref{ap:N=1calc}, we give the proofs of equivalence of the supersymmetry conditions with the pure spinor equations that we use for our analysis. The basic 
data of the six--dimensional nilmanifolds and compact solvmanifolds are 
collected in Appendix \ref{app:Nil}. Appendix \ref{app:eqs} collects the specific supersymmetry
equations and form of pure spinors for all possible cases catalogued by the types of the pure spinors and the orientifold planes.

\section{Nilmanifolds and solvmanifolds}
\label{sec:solvm}
Since the nilmanifolds (and, more generally, solvmanifolds) are going to be the main geometrical ingredients of our solutions, let us start by briefly reviewing these. We will in particular try to explain why we restricted ourselves to this class of manifolds. 

\subsection{From algebra to geometry}

Let us start with a Lie group $G$ of dimension $d$ 
viewed as a  manifold (also called a group manifold). 
From the Maurer--Cartan equations, $G$ has  
a set of $d$ globally defined one forms $e^a$ that satisfy
\beq \label{strdef}
de^a=\frac{1}{2} f^a{}_{bc} \, e^b \wedge e^c
\eeq
with  $f^a{}_{bc}$ the structure constants of the group $G$. Such a basis is obviously
very useful: it can reduce many differential problems to algebraic ones. One can 
also define a basis of vectors $E_a$ dual to the $e^a$ (i.e.~$\langle E_a, e^b\rangle=
\delta_a{}^b$). This basis obeys 
\beq
[E_b,E_c]=-f^a{}_{bc}\, E_a\ .
\eeq  

Conversely, if we are looking for a manifold $M$ with a basis $e^a$ of one--forms
defined everywhere, we are providing a global section to the frame bundle, hence
trivializing it. Hence the cotangent and tangent bundle will be topologically trivial.  
Such manifolds are called {\it parallelizable}. One can of course always expand
$d e^a$ in the basis of two--forms $e^b\wedge e^c$, which would give
us  (\ref{strdef}) but with $f^a{}_{bc}$ not necessarily constant.
If  they are constant, the manifold is homogeneous.
Imposing $d^2 e^a=0$ results in
\beq
f^a{}_{[bc} f^e{}_{d]a}=0
\eeq
i.e.~the $f$'s satisfy Jacobi identities, and are therefore structure constants of a real Lie 
algebra ${\cal G}$. The vectors $E_a$, when exponentiated, give then an action
of $G$ over $M$. One can see that this action is transitive (it sends any point into
any other) and hence $M=G/\Gamma$, where $\Gamma$ is a discrete subgroup. 
Manifolds of this type are a small subset of all possible parallelizable 
manifolds: for example, all orientable three--manifolds are parallelizable, but only a few are
quotients of Lie groups. 

So far we have simply reviewed the fact that discrete quotients of Lie groups are
manifolds particularly simple to deal with.  There are many $G$ one could consider. 
Actually, Levi's theorem tells us that any Lie algebra is a semi--direct sum of a semisimple algebra and of a {\it solvable} one (to be defined shortly). While 
semisimple algebras are used
in many physical applications, in this paper we are going to focus on the solvable ones.
This is mainly because we know
some of the mathematics better -- in particular, the criteria for compactness,
one of which we will review in the next subsection.
Solvable algebras
have previously played a role as gauge groups of four--dimensional supergravities
\cite{sugra1, sugra2,sugra3,sugra4}. 

A solvable algebra can be defined as follows. Consider the series defined recursively
by  $\mathcal{G}^0=\mathcal{G}$ and $\mathcal{G}^s \equiv [\mathcal{G}^{s-1},\mathcal{G}^{s-1}]$. If this series becomes zero after a finite number of steps ($\exists k |
\mathcal{G}^k=\{0\}$), then the Lie algebra is said to be solvable. 
There is a special class of solvable algebras that we will find particularly useful: 
{\it nilpotent} algebras. In this case, the condition is that the series defined recursively as $\mathcal{G}_0=\mathcal{G}$ and $\mathcal{G}_s \equiv [\mathcal{G}_{s-1},\mathcal{G}]$ converges to zero in a finite number of steps  ($\exists k | \mathcal{G}_k=\{0\}$). The integer $k$ is called the nilpotency degree
of the manifold. Since we are taking commutators
of $\mathcal{G}_{s-1}$ with the whole of $\mathcal{G}$ rather than with itself, this
series is clearly decreasing more slowly, and hence it might not reach zero even if
the $\mathcal{G}^s$ did. Nilpotent algebras have nicer properties, for instance they are  all generalized 
complex \cite{CG}, as we will see later. This was one of the main 
motivations for the present work. 

The most often mentioned example of a nilpotent algebra is  the so--called
Heisenberg algebra. 
The structure constants in  the form language of (\ref{strdef}) are:
\beq \label{ex}
de^1=0 \ ; \qquad  de^2=0 \ ; \qquad  de^3= N e^1 \wedge e^2 \     .
\eeq
We will also use the compact notation $(0,0, N\times12)$ to refer to (\ref{ex}).
One can see already that the third direction is fibred over the second two ($N e^1\wedge
e^2$ being the curvature of the fibration). To see this more clearly, let us choose a  gauge where
\beq \label{heis}
e^1=dx^1\ ; \qquad  e^2=dx^2 \ ; \qquad e^3=dx^3+ N x^1 e^2  \ .
\eeq
We can compactify $G$ by making the identifications  $(x^1,x^2,x^3)\simeq (x^1,x^2+a,x^3)\simeq (x^1,x^2, x^3+b)$, with $a,b$ integer, but cannot do the same 
for $x^1$, because the one form $e^3$ would not be single--valued. For that, we need
to ``twist" the identification by $ (x^1,x^2,x^3)\simeq (x^1+c,x^2,x^3-N\,c\, x^2)$. 
In this way, the resulting $G/\Gamma$ turns out to be an $S^1$ fibration over
$T^2$ whose $c_1=N$, and hence topologically distinct from $T^3$. Such 
a quotient of a nilpotent group by a discrete subgroup is called a  {\it nilmanifold} or
sometimes more loosely a ``twisted torus".
The structure constants are often referred to in the literature of flux compactifications
as ``metric fluxes". 
Some twisted tori are T--dual to regular tori with  NS 3-form fluxes, as we will see. For instance,
the manifold given in this example T--dual to a $T^3$ with $N$ units of $H$ flux. 
A general nilmanifold is always an iteration of torus fibrations: a torus fibration, over which another torus is fibred, 
and so on. In the example, only one step was required.

Nilmanifolds are also non--Ricci--flat, and therefore suitable for compactifications in the presence of fluxes 
\cite{Kachru,VZ, Dere, DF, Hull1, DF2, VZ2, K-taylor, Camara, ferrtri, Dere2, trigiante,
Hull2,Camara2}.  The Ricci tensor is given in terms of 
the structure constants by
\beq
\label{ricci}
R_{ad}=\frac12 \left(\frac12 f_{a}{}^{bc} f_{dbc} - f^b{}_{ac} f_{bd}{}^c - f^c{}_{ab} f^b{}_{dc}\right)
\eeq
where indices are lowered and raised with $g_{ab}$ and its inverse: for example, $f_{abc} = g_{ad} f^d{}_{bc}$. Notice that we do not require that the one-forms $e^a$ are vielbeine
 and therefore the metric  $g_{ad}$ is not necessarily $\delta_{ad}$ (see Section \ref{Gentwist}).
For nilmanifolds the last term in (\ref{ricci}) (which is the Killing metric) is zero 
and it is not difficult to check that the Ricci tensor is 
never vanishing. Moreover, the Ricci scalar is given by  $R= -1/2 f_{abc} f_{abc}$ and it is always negative.
On the contrary solvmanifolds have no definite sign for the curvature and can be Ricci-flat. This difference will be very important 
later, in particular in  Sections \ref{sec:fluxless} and \ref{flat}.

Let us now come back to the general discussion. 
A key fact in the systematic search of $\N=1$ vacua performed in this paper
 is that solvable algebras of dimension up to six are classified.
This is done by using their  {\it nilradical} (the largest nilpotent ideal). 
For example, the nilradical of solvable algebras of dimension six (those that concern us in this paper) can be of dimension 3,4,5 or 6
(in the latter case, the algebras being nilpotent). 
Those with three-dimensional nilradical are decomposable as sums of two solvable algebras, of which there are nine.
There are 40 equivalence classes of six--dimensional
solvable algebras with four--dimensional nilradical \cite{Tarko} and 99 with  
five--dimensional nilradicals \cite{Mubara5}.  
Finally, for nilpotent Lie groups, those up to  dimension 7 have been classified, and 6 is the highest dimension
where there are finitely many. There are 34 isomorphism classes of simply--connected 
six--dimensional nilpotent Lie groups \cite{MOMA1, MOMA2}. These are the nicest playgrounds for generalized complex geometry and flux compactifications, as  some nilmanifolds admit an integrable complex or symplectic structure, and some do not,  but they all admit generalized complex structures \cite{CG} -- something that is necessary for
admitting a supersymmetric Minkowski vacuum, as we will see. 

These classifications, however, do not take into account whether the group $G$ can 
be quotiented in such a way as to produce a compact manifold. This is the issue
we turn to next.

\subsection{Compactness}
\label{sec:compact}
Not any set of structure constants gives rise to a compact manifold. 
A necessary condition ($f^a{}_{ab}=0$) \cite{ss} is commonly used in the physics literature.
In this subsection we review what is known about this problem in the mathematical literature. In 
particular, in the case of nilmanifolds and more generally solvmanifolds, the problem has
been solved. 

A set of structure constants $f^a{}_{bc}$ corresponds to a certain Lie algebra $\G$; by exponentiating
it, one can always produce a (simply connected) Lie group $G$. In some special cases, $G$ might happen to be compact
already:  for example, if $G$ is semisimple, this will be the case if and only if the Killing form $f^c{}_{ad}f^d{}_{bc}$
is negative definite. If $G$ is noncompact (which is far more often the case) it is still possible to 
produce a compact manifold by modding it by a discrete compact subgroup $\Gamma$. This manifold
$M\equiv G/\Gamma$ has still obviously the property that the tangent space at every point is isomorphic to  
the Lie algebra $\G$. The dual formulation (on the cotangent bundle) of this statement  is that there 
is a basis of one--forms 
$e^a$  that obey $d e^a= f^a{}_{bc} e^b \wedge e^c$.  

In this paper we restrict ourselves to solvable algebras (which make up
the bulk of all algebras anyway), essentially because we know how to determine 
whether a solvable $G$ can be compactified or not. If it can, $M=G/\Gamma$ is a  compact solvmanifold. 
An important subtlety must be noted here. 
A compact solvmanifold is something more general. It can also be obtained by quotienting by subgroups which 
are {\it not} discrete but closed in the topological
sense, so that the quotient be a manifold. 
One could consider e.g.~a seven--dimensional $G$ quotiented by a $\Gamma$ that, in addition to a discrete part, has a one--dimensional continuous part, so that the quotient  $G/\Gamma$ stays six--dimensional. An example of such a type can already be found in two dimensions: 
the Klein bottle is a solvmanifold which is  a quotient of a three-dimensional solvable group, $G=
{\Bbb R} \ltimes {\Bbb C}$, by a non--discrete subgroup $\Gamma$. (For details see \cite{auslander}, Ex.~3.) 
It is clear that this phenomenon makes
the number of compact six--dimensional solvmanifolds very large, possibly  infinite.\footnote{We do not know how to estimate the number of such manifolds.  The only place where we will deal with this more general class is in the discussion of flat compact solvmanifolds (see Section \ref{flat}), and we will see that the number of these is surprisingly high. } We will restrict
our attention to solvmanifolds with ``discrete isotropy group", which is a way to say that 
$G$ is already six--dimensional and $\Gamma$ is discrete. 

Hence we will now discuss  when a six--dimensional $G$ admits a discrete subgroup $\Gamma$ so that
$M= G/\Gamma$ is compact. Such a $\Gamma$ is called a cocompact subgroup. 
Let us  remark  that, whereas we are going to determine whether  such a $\Gamma$ exists, we will not  try to find out how many  such $\Gamma$'s there are. Already in three dimensions there would be infinitely many nilmanifolds, namely the ones we saw in 
the previous subsection, see eq.~(\ref{ex}). However, the algebras are all
isomorphic, via a rescaling of the generator $e^3$. The information lost in 
this rescaling is which subgroup $\Gamma$ is being modded out. The reason 
this choice does not matter much to us is that we work for most of the paper
with left--invariant forms, which hence have constant coefficients in the
basis given by the $e^i$. Were one to work with a non--trivial dependence, 
the non--constant coefficients would be well--defined for a certain choice
of $\Gamma$ and not for another. From now on, we will loosely refer to the
various (finitely many) classes of nilpotent algebras as of classes of
nilmanifolds, and similarly for solvmanifolds.  

We now come back to the existence of at least one $\Gamma$. 
As a warm--up, it is easy to guess a necessary condition.
 Suppose $f^b{}_{ab}\neq 0$. Then one can see that
the top form $\mathrm{vol}\equiv e^1\wedge\ldots\wedge e^6$ is exact. Indeed, if $\alpha\equiv
\epsilon_{a_1\ldots a_6} \alpha^{a_1} e^{a_2}\wedge \ldots \wedge e^{a_6}$ with $\alpha^{a_1}$ constant, 
one has $d\alpha=
(f^b{}_{ab} \alpha^a) \mathrm{vol}$. That means of course that $\mathrm{vol}$ cannot be 
a volume
form for $G/\Gamma$, since a compact manifold  needs to have a top--form 
non--trivial in cohomology. 

This argument is not complete in that it leaves open the possibility 
that $f e^1\wedge\ldots\wedge e^6$, with $f$ some function, might be non--trivial 
in cohomology. 
For this reason \cite{ss} relates the condition $f^b{}_{ab}=0$
to the presence of a left--invariant volume form. In general, it is not clear
that computing the cohomology using left--invariant forms gives the same as using
all forms. On group manifolds (before any quotient) this is true due to an 
argument based on the Haar measure. On nilmanifolds (after the quotient by $\Gamma$) 
this is actually also true \cite{nomizu}, but less trivial. At any rate,  it is not hard to see that for nilmanifolds
$f^b{}_{ab} = 0$ is automatically satisfied.
Nevertheless, one can still prove that $f^b{}_{ab}=0$ (which is referred to as $G$ 
being {\it unimodular})
is indeed a necessary condition.

One might wonder if this condition is also sufficient. This would imply that all nilpotent groups can be 
compactified by quotienting. This is almost true: it turns out that for nilpotent groups
it is enough to require the structure
constants to be rational in some basis \cite{malcev}. 

What about solvmanifolds? Criteria for this more general case exist as well, but they are considerably 
more complicated \cite{auslander, saito}. We will now describe, with the help of 
some examples, the criterion obtained by Saito \cite{saito}, which seems to be the
easiest. The price to pay for this simplicity is that it can only be applied to a 
certain class of solvable groups --- those that are {\it algebraic} subgroups of
Gl$(n,\Bbb R)$ for some $n$. This means that they have a representation on ${\Bbb R}^n$
which is faithful (that is, one to one), and that once so realized as a subgroup of
Gl$(n, \Bbb R)$ they can be characterized by polynomial equations. For example, the 
orthogonal group O$(n)$ is algebraic, because it is described by equations 
$O_{ik} O_{kj}=\delta_{ij}$ which are quadratic in the entries $O_{ij}$.

We need first a few definitions. 
The {\it nilradical} $N(\G)$ of a Lie algebra is its largest nilpotent ideal. It is often
used to classify solvable algebras, following \cite{Mubara3}. In our case it can have dimension from three to six (in which case the algebra is nilpotent itself).
Now consider the usual adjoint representation of $G$ over $\G$, but restrict it to $N(\G)$.
This is a group of matrices of dimension $n\times n $, where $n=\dim(N(\G))$,   
that we call $P(G)$. This has a natural discrete subgroup: 
$P(G)^{\Bbb Z} =P(G)\cap Gl(n,{\Bbb Z})$. Here $Gl(n,{\Bbb Z})$ is defined as the
group of matrices with coefficients in ${\Bbb Z}$ and with determinant $\pm 1$. (The latter
requirement is necessary to have a group.) As we will see clearly later in our examples, the group $P(G)^{\Bbb Z}$ depends on our choice of basis in the Lie algebra $\G$. It is convenient to restrict one's attention to bases that are {\it adapted} to the series of subalgebras $\{\G_k\}$ (in the sense that, for any $k$, there is a subset of the elements of the basis of $\G$ that is a basis of $\G_k$) and {\it rational} (in the sense that the structure constants are rational in such a basis). 
Now the result of \cite{saito} is that
the problem can be reduced to one in the adjoint representation: a cocompact 
$\Gamma$ exists if and only if there exists an adapted rational basis of $\G$ such that the quotient $P(G)/[P(G)^{\Bbb Z}]$ is compact. 

It is time to give some examples. For simplicity, we will look here at two 
three--dimensional solvable algebras which are relevant as examples of string compactifications.
The first algebra is known as $E_{2}$ and is defined by 
$[e^1,e^3]=-e^2$ and $[e^1,e^2]=e^3$. A convenient short notation is $(0,-13,12)$. 
First of all, the group obtained by exponentiating this algebra is algebraic: it can be
described as the group of $3\times 3$ matrices ${{O\ v}\choose{1\ 0}}$, where $O$ is
an element of O$(2)$. Hence we can apply the criterion in \cite{saito}.
The nilradical $N(\G)$ 
is spanned by $e^2, e^3$: it certainly is an ideal (it is left invariant by the action of all 
elements, since commutators can only give back $e^2$ and $e^3$);
it is nilpotent (it is even abelian); and it is maximal (nothing can be added to it without
getting the whole algebra). To compute the adjoint representation of $G$ on 
$N(\G)$, let us first compute the adjoint representation of its Lie algebra $\G$. A general element 
has the form $a\equiv a_1 e^1+a_2 e^2+a_3 e^3$, and its action reads
$[e^2,a]=-a_1 e^3$, $[e^3,a]=a_1 e^2$. In other words, the adjoint representation of
$\G$ on $N(\G)$ is the group of matrices of the form $a_1{{\ 0\ \  -1}\choose {1\ \ 0}}$.
Notice that two of the three dimensions of the algebra have been trivially represented.
To get the adjoint representation of the group $G$ on $N(\G)$, we just have to exponentiate: 
\begin{equation}\label{e2}
\G=E_{2}: \qquad P(G)=\left\{\left(\begin{array}{cc}
\cos(a_1) & - \sin(a_1) \\ \sin(a_1)&\cos(a_1) 
\end{array}\right)\right\}\ \, .
\end{equation}
This rather simple group of matrices is obviously isomorphic to 
$S^1$, which is already compact. Just for illustration $P(G)^{\Bbb Z}$ is also non--trivial, being generated
by $a_1= k \pi/2$, with $k$ an integer. The quotient by this subgroup merely reduces the radius.  
So $P(G)/[P(G)^{\Bbb Z}]=S^1/{\Bbb Z}_4=S^1$ is compact, 
and this time  $G$ does have a cocompact discrete subgroup, although the theorem does not tell us explicitly what it is.

Next we look at the algebra known as $E_{1,1}$, which is defined by
 $[e^1,e^2]=-e^2$, $[e^1,e^3]=e^3$, 
or $(0,-12,13)$. The group obtained by exponentiating this algebra is again algebraic: it can be
described as the group of $3\times 3$ matrices ${{O\ v}\choose{1\ 0}}$, where $O$ is
an element of O$(1,1)$. The nilradical $N(\G)$ 
is again spanned by $e^2, e^3$, for the same reasons as in the previous example. 
Now we compute the adjoint representation of $G$ on 
$N(\G)$. A general element 
of the form $a\equiv a_1 e^1+a_2 e^2+a_3 e^3$ acts as
$[e^2,a]= a_1 e^2$, $[e^3,a]=-a_1 e^3$. In other words, the adjoint representation of $\G$ on $N(\G)$ 
is the group of matrices of the form $a_1{{1\ \  0}\choose {0\ -1}}$.
The exponential of this representation now gives
\begin{equation}\label{e11}
\G=E_{1,1}:\qquad P(G)=\left\{\left(\begin{array}{cc}
e^{a_1} & 0 \\ 0 & e^{-a_1}
\end{array}\right) \right\}\  \, .
\end{equation}
This time $P(G)$ is isomorphic to
 ${\Bbb R}$. According to Saito's criterion we reviewed earlier, in order to establish compactness, we then have to compute $P(G)^{\Bbb Z}$. Requiring that a matrix 
of the form (\ref{e11}) has integer
coefficients implies that 
both $e^{a_1}$ and $e^{-a_1}$ are integer, which is only true for $e^{a_1}=1$, the identity. 
This is the trivial group: hence $P(G)/[P(G)^{\Bbb Z}]={\Bbb R}/\{e\}$ is noncompact. However, this might be an artifact of our choice of basis for the Lie algebra; as we remarked earlier, $P(G)^{\Bbb Z}$ depends on our choice of basis. There are several possible choices of such a basis. One can be extracted
 from \cite[Eq.~(4)]{bock}: defining $\tilde e^1 = \sqrt{2}e^1$, 
$\tilde e^2 = (1/2)((1-3/\sqrt{5})e^2 + (1+3/\sqrt{5})e^3)$, 
$\tilde e^3= (e^2 - e^3)/\sqrt{5}$, we obtain a basis in which the structure constants are rational: the group $N(\G)$ is now the group of matrices of the form $\tilde a_1 {{-3\  -2}\choose {2\ \ \ 3}}$. The  exponential of such a matrix is equal to ${{0\  -1}\choose {1\ \ 3}}$ 
for $\tilde a_1 = (1/\sqrt{5})\log(\frac{3+\sqrt{5}}2)$. (Another construction of such a basis can be found in \cite[version 1]{haque-shiu-underwood-vanriet}.) In this basis, $P(G)^{\Bbb Z}$ is actually ${\Bbb Z}$, and $P(G)/[P(G)^{\Bbb Z}]= {\Bbb R}/{\Bbb Z}$ is compact. Hence, the group $E_{1,1}$ can be compactified too, although realizing it requires some ingenuity.\footnote{\label{foot:mistake}In previous versions of this paper, 
{we did not appreciate the fact that it is sufficient to find just {\it one} basis in which $P(G)/[P(G)^{\Bbb Z}]$ is compact},  and  erroneously claimed  that $E_{1,1}$ cannot be compactified. For this reason, our search for compact solvmanifolds is not as systematic as previously claimed.}  
 
These two examples are particularly easy to analyze: $P(G)$ is a one parameter group, 
and deciding whether a discrete quotient of a one--dimensional group is 
compact cannot be too complicated. In higher dimensions, it is often just as easy as here; but for a few algebras, 
$P(G)$ has many parameters, and determining compactness
can be challenging. This is most commonly the case for those algebras whose
nilradical is not abelian. 

In Table \ref{ta:nil} of Appendix \ref{app:Nil} we give the list of the 34 isomorphism 
classes of six--dimensional nilmanifolds, taken from \cite{CG, MOMA1, MOMA2}. In Table
\ref{ta:solv} we give thirteen six--dimensional solvmanifolds; eight of them 
can be made compact by quotienting by a discrete subgroup, and possibly the six
remaining ones too (more on this in Appendix \ref{app:Nil}). These are all 
the solvmanifolds we found with a simple search: namely, we applied Saito's criterion directly to the basis given to us by the classifications \cite{Tarko,Mubara5}.\footnote{A recent work \cite{bock} gives a complete list of compact solvmanifolds up to dimension five.} Notice that some of 
the algebras are lower dimensional, and are brought to
six--dimensions by adding trivial directions. The associated manifolds are 
therefore products
of a lower dimensional nil or solvmanifold with $T^n$. This is the case for example 
for the
nilpotent algebras denoted 3.9 and 3.10 in Table \ref{ta:nil}, where the direction 4 is trivial.\footnote{Throughout the paper we will use  double-number labels for the compact manifolds collected in Tables   \ref{ta:nil} and  \ref{ta:solv}. This nomenclature is explained in  Appendix \ref{app:Nil}. Wherever it is not clear from context as  to what Table we are referring a letter $n$ or $s$ will indicate that the corresponding algebra is nilpotent or solvable.}
Regarding the solvable algebras, 3.2 of Table \ref{ta:solv} is constructed out of 
the only compact (not nilpotent) four-dimensional algebra adding two trivial directions, 
while 2.5, 2.6, 3.3 and 3.4 are 
built using five-dimensional solvable algebras. Finally, 2.4, 3.1 and 4.1 are 
constructed as a direct sum of 
the  three-dimensional algebra $E_2$ and 
 respectively another  copy of it, the Heisenberg algebra that we have seen earlier 
(the only three--dimensional nilpotent algebra),
 and 3 trivial generators. It is not hard to check using (\ref{ricci}) that the 
three--dimensional solvable algebra (23, -13, 0) is a Ricci--flat manifold if the $e^a$ are 
taken to be vielbeine (i.e. for $f_{abc}=f^a\,_{bc}$). In fact, not only the Ricci tensor vanishes, but so does the full Riemann tensor. 
The manifold is therefore flat. Similarly, the manifolds corresponding to the algebra 2.4 and 4.1 admit a flat metric. Their flatness should then not come as a surprise, since \cite{flat} showed that homogeneous parallelizable Ricci--flat manifolds  are flat.

We should also notice that, while all nilmanifolds are parallelizable, not all solvmanifolds are \cite{auslander-szczarba}. Even though the Lie group $G$ before the quotient is parallelizable (as we remarked at the beginning of this section), some of the one--forms may be non well--defined on the quotient $G/\Gamma$. However this does not happen  in the class of solvmanifolds  with  discrete isotropy group $\Gamma$, and all manifolds considered in this paper are  parallelizable (see \cite{bock} and references therein). 
For the example discussed in  section \ref{sec:scope} we will describe explicitly how the basis of globally defined vielbeins is related to $\Gamma$.

\section{Generalized complex structures and pure spinors}
\label{Gentwist}

Four--dimensional $\N=1$ supersymmetric vacua of type II theories in 
presence of NS and RR fluxes involve generalized Calabi-Yau manifolds, as we will 
see in Section \ref{N1vacua}. 
In this Section we review some aspects of generalized complex geometry we will need 
in the rest of the paper. Much of what is contained in this Section is known \cite{hitchin,gualtieri,witt}; we have tried to write down some arguments implicitly present in the literature and that we need, and in some cases we have simplified more usual arguments or proofs.

\subsection{Generalized complex structures}

Generalized complex geometry is the generalization of complex geometry 
to \tts,  the sum of the tangent and cotangent bundle of a manifold. 
A manifold $M$ of real, even dimension $d$
is generalized complex if it has an 
integrable {\it generalized almost complex structure}.
A generalized almost complex structure is a map $\cal  J$: \tts $\to$ \tts 
that squares to $-\mathbb I_{2d}$ and satisfies
the 
hermiticity condition ${\mathcal J}^{t} {\mathcal I} {\mathcal J} = {\mathcal I}$, 
with respect to the natural metric on $T \oplus  T^*$,
\beq \label{natmetr}
{\mathcal I}={{0 \ \ \mathbb I_d} \choose {\mathbb I_d \ \ 0}} \ .
\eeq 
This metric is just the pairing $(\ ,\ )$ 
between vectors and one--forms, and it has signature
$(d,d)$. It reduces the structure group of \tts to O$(d,d)$. 
The hermiticity condition implies that a generalized almost complex structure
should have the form
\beq
\label{formJ}
 {\mathcal J} = 
\left(
\begin{array}{cc}
I& P\\ L & -I^t 
\end{array}
\right) \, ,
\eeq
with $P$ and $L$ antisymmetric matrices. The condition 
${\mathcal J}^2=-\Bbb I_{2d}$ imposes further constraints on $I,P$ and $L$; for example, $I^2+PL=-\Bbb I_d$.  ${\cal J}$ reduces the structure group of \tts
further, to U$(\frac d2,\frac d2)$.

Just as for almost complex structures, it is possible to give an integrability
condition for a generalized almost complex structure. Let us recall how the definition
of integrability works for an ordinary complex structure $I$. Within the complexified
tangent bundle $T\otimes\Bbb C$ one defines the holomorphic 
$T^{1,0}$ even without integrability; a $(1,0)$ vector satisfies 
$I^m{}_n v^n= i v^m$. Integrability can be formulated as the requirement
that $T^{1,0}$ be integrable under the Lie bracket: $[T^{1,0},T^{1,0}]_\mathrm{L}\subset T^{1,0}$, 
or, in other words, 
\begin{equation}\label{eq:nij}
\bar P [P(v), P(w)]=0\ , 
\end{equation}
where $P=\frac12(\mathbb{I}_d-iI)$ is the projector on $T^{1,0}$. One can see that both real and imaginary part of the left hand side of (\ref{eq:nij}) are actually proportional 
to $\mathrm{Nij}(v,w)$, the Nijenhuis tensor. We will see later
an alternative (slightly stronger) definition of integrability for $I$, which is known to geometers and somehow more natural in the context of generalized complex geometry.

Turning now to integrability for a generalized almost complex structure ${\cal J}$, 
we can again consider the ``$(1,0)$" part of the complexified \tts in a similar way defining the projector  on ${\cal L}_\J$ 
\beq
\label{proj}
\Pi = \frac{1}{2} (\mathbb{I} _{2 d} - i \J) 
\eeq
and imposing  ${\Pi} A=  A$, where $A =v + \zeta$ is a section of \tts. We will call this $i$--eigenbundle $L_{\cal J}$. It is null with respect to 
the metric ${\cal I}$ in (\ref{natmetr}), since for $A,B\in L_\J$, 
\beq
( A , B) = A{\cal I} B= 
A\J^t {\cal I} \J B= (i A) {\cal I} (iB)=-A{\cal I} B = - (A,B) \, .
\eeq 
Also it has the maximal dimension that a null space
can have in signature $(d,d)$, namely $d$  since $\Pi A \in L_\J$ for any real $A$.  This is often also phrased by saying that
$L_{\cal J}$ is 
a {\it maximally isotropic} subbundle of $T\oplus T^*$.

We now need a bracket on 
\tts to impose integrability, similarly to what we did for almost complex structures. 
There is no bracket satisfying the Jacobi identity on \tts, but fortunately there is one
that satisfies it when restricted on isotropic subbundles of \tts
This is the {\it Courant bracket}. A discussion of the Courant bracket 
as an extension
of the Lie bracket can be found in  \cite{gualtieri}, here
we will introduce it in an alternative way, as a particular case of so--called {\it derived brackets} (see for example \cite{ks}).
This alternative definition will be more useful later when considering pure spinors. 
Let us start by defining the  Lie bracket $[\ , \ ]_\mathrm{Lie}$ as a derived 
bracket
\begin{equation}\label{eq:Lie}
[\{\iota_v,d\} , \iota_w]=\iota_{[v,w]_\mathrm{Lie}}\ . 
\end{equation}
Here and in the following $\iota_v$ denotes the contraction by the vector $v$. 
All the variables in this equation are to be understood as {\it operators} acting on 
differential forms.  
 The brackets on the left hand side are meant to be commutators and anticommutators; the one on the right hand side is the Lie bracket. 
 We can now define the  Courant bracket analogously: 
\begin{equation}\label{eq:Cou}
\frac12 \Big([\{A\cdot ,d\}, B\cdot]-[\{B\cdot ,d\}, A\cdot] \Big)\equiv[A,B]_\mathrm{Courant}\cdot  \ , 
\end{equation}
 where $A$ and $B$ are sections of \tts, and again all variables are considered as operators on differential forms: $A\cdot= \iota_v +\zeta\wedge$, namely
 vectors act by contraction, and one--forms act by wedging. From now on 
 $[\ ,\ ]_\mathrm{Courant}=[\ ,\ ]_\mathrm{C}$. 
  One can compute explicitly 
 \beq
\label{courant}
[v+ \zeta , w + \eta]_\mathrm{C}=[v,w] + {\mathcal L}_v \eta 
- {\mathcal L}_w \zeta - \frac12 d(\iota_v \eta -\iota_w \zeta) \, .
\eeq 
In this formalism the definitions of the Lie
(\ref{eq:Lie}) and Courant (\ref{eq:Cou}) brackets are very similar (indeed, Courant
contains Lie as a particular case, when $A=v$ and $B=w$).  
The main feature of a derived bracket is that it contains
a differential. For both Lie and Courant the differential is $d$, but 
one can generalize it to other differentials. A generalization that will appear naturally in 
generalized complex geometry is the inclusion a closed 
three--form $H$, to form the differential $d-H\wedge$. The bracket will of course be 
modified as a consequence (see eq. (\ref{intsp}) below).
It is also possible to extend the 
definition of derived bracket  in an obvious way, to different spaces of operators.

A generalized almost complex structure $\cal J$ is integrable if its $i$ eigenbundle
$L_{\cal J}$ is closed under the Courant bracket
\beq
\label{Jintegrability}
\bar\Pi \,  [\Pi (v+\zeta), \Pi (w+\eta)]_\mathrm{C}=0 \, ,
\eeq
where $\Pi$ is the projector 
on $L_{\cal J} \subset$\tts. In this case, ${\cal J}$ is called a generalized complex structure, dropping the ``almost". A manifold on which such a tensor exists 
is called a generalized complex manifold.

The simplest examples of generalized complex structures are provided by the embedding  in $T \oplus  T^*$
of the standard complex and symplectic structures. Take the matrices 
\beq \label{gcscom}
{\mathcal J}_I \equiv 
\left(
\begin{array}{cc}
I & 0 \\ 0 & - I^t 
\end{array}
\right) \, \ , \qquad \qquad
{\mathcal J}_J\equiv
\left(
\begin{array}{cc}
0 & J \\ -J^{-1} & 0 
\end{array}
\right) \, 
\eeq
where $I=I_m\,^{n}$ obeys $I^2=-\Bbb I_d$, i.e. it is a regular almost complex structure for the tangent 
bundle, and $J=J_{mn}$ is a non degenerate 
two--form $J_{mn}$, i.e. an almost symplectic structure for the tangent bundle. 
In these two examples, the integrability of $\J$ 
turns into a condition on the building blocks, $I_m\,^n$ and $J_{mn}$. 
Integrability of ${\mathcal J}_I$
forces $I$ to be an integrable almost complex structure
on $T$ and hence a complex structure. In other words
the manifold is complex.
For ${\mathcal J}_J$, integrability
imposes $dJ=0$, thus making $J$ into a symplectic form, and the manifold
a symplectic one. 

We can construct explicitly ${\cal J}_{I,J}$ and their $\pm i$ eigenbundles for the very simple case of 
a two-torus.
If we call $e^1$ and $e^2$ the vielbein on the two-torus, we can define the fundamental 
and the 
holomorphic forms 
as $J=e^1 \wedge e^2$ and  $\Omega_1=e^1+ie^2$, respectively. Then 
\beq \label{trex1}
\J_I=\left( \begin{array}{cccc} 0 & 1 & 0 & 0 \\ - 1 & 0 & 0 & 0 \\ 0 & 0 & 0 & 1 \\
0 & 0 & - 1 & 0 \end{array} \right) \ , \qquad 
\J_J=\left(  \begin{array}{cccc} 0 & 0 & 0 & 1 \\ 0 & 0 & -1 & 0 \\ 0 & 1 & 0 & 0 \\
-1 & 0 & 0 & 0 \end{array} \right) \ .
\eeq
The holomorphic eigenbundles  are
\bea
\label{trex}
&& L_{\J_I}=\left< \left(\begin{array}{c} 1\\ i \\ 0 \\ 0 \end{array} \right) , \left(\begin{array}{c} 0\\ 0 \\ 1 \\ i \end{array} \right) \right>= T^{1,0} \oplus (T^*)^{0,1} \ , \nn\\
&& L_{\J_J}=\left< \left(\begin{array}{c} 1\\ 0 \\ 0 \\ i \end{array} \right) , \left(\begin{array}{c} 0\\ 1 \\ -i \\ 0 \end{array} \right) \right>= \{ v^m + i v^m J_{mn} \} \ .
\eea

\subsection{Pure spinors}
\label{purespinorsx}
On $T$ there is a one-to-one correspondence between almost complex structures and 
Weyl spinors. An analogous property holds on \tts between generalized almost complex structures and pure spinors.
 
The metric (\ref{natmetr}) reduces the structure group of \tts to O$(d,d)$ and the corresponding Clifford 
algebra is Cliff$(d,d)$.  The spinor bundle is isomorphic to the bundle of differential forms $\Lambda^{\bullet} T^*$. Indeed, an 
element of \tts\, $(v+\zeta)$, acts on a spinor $\Phi$ by 
the Clifford action $\cdot$
\beq \label{Cliff1}
(v+\zeta) \cdot  \Phi = v^m \iota_{\del_m} \, \Phi + \zeta_m dx^m \wedge \, \Phi \ ,
\eeq 
and it is easy to check that 
\beq
\label{Cliff2}
\left((v+\zeta) \cdot (v+\zeta)  \Phi\right) = -(v+\zeta, v+\zeta) \Phi\ , 
\eeq
where the inner product $( , )$ is defined with respect to
 the metric (\ref{natmetr}). Hence, differential forms can be thought of as spinors
 for Cliff$(d,d)$, a fact which can be at first confusing.  On forms the gamma matrices of the
 Cliff$(d,d)$  algebra are vectors $v$
(acting by contraction, $\iota_v$) and one--forms $\zeta$ (acting by 
$\zeta\wedge$). As a basis, we can consider $\iota_{\del_m}$ and $dx^m\wedge$.
As operators on differential forms, they satisfy
\beq
\label{cliff}
\{dx^m\wedge \, ,\, dx^n\wedge\} = 0\ , 
 \quad \{dx^m\wedge\, , \,\iota_{\del_n}\} = \delta^m{}_n\ ,  \quad  
 \{\iota_{\del_m}\, , \,\iota_{\del_n}\} = 0 \ .
\eeq
This is exactly the Clifford algebra with metric ({\ref{natmetr}). 

One can further decompose the exterior algebra into the 
spaces of even and odd forms $\Lambda^\pm T^*$: a positive (negative) 
chirality spinor is an even (odd) form. We will denote them by $\Phi_{\pm}$.

The inner product between two forms $A$ and $B$ can be obtained from the pairing
\beq \label{defMukai}
\langle A , B\rangle\equiv (A\wedge 
\lambda(B))_d \ , 
\qquad \lambda(A_n)=(-1)^{\mathrm{Int}[n/2]} A_n
\eeq
where the subindices $d$ and $n$ denote the degree of the form.
A top form is proportional 
to the volume form "${\rm vol}$", which means that using the volume form one can extract
a number from the Mukai pairing (the constant of proportionality). 
In $d=6$ this pairing is antisymmetric; it 
is then convenient to define
the norm of $\Phi$ as  
\begin{equation}
    \label{eq:norm}
\langle \Phi,  \bar\Phi \rangle=
- i ||\Phi||^2 {\rm vol} \ .    
\end{equation}

From the action of the Clifford algebra  (\ref{Cliff1}), one defines the {\it annihilator} of a spinor
as 
\beq
\label{annih}
L_{\Phi} = \{ v + \zeta \in T \oplus T^* \, |  \, ( v + \zeta) \cdot \Phi =0 \} \ .
\eeq
From  (\ref{Cliff2}), it follows that the annihilator space $L_\Phi$ 
of any spinor $\Phi$ is isotropic. 
It can have at most dimension $d$, in which case it is maximally isotropic. If $L_\Phi$ is maximally isotropic
$\Phi$ is called a {\it  pure spinor} \footnote{An alternative definition is
much used in the world--sheet literature \cite{berkovits}: it says that all bilinears
$\eta^t \gamma_{m_1\ldots m_k} \eta$ vanish for $k<d/2$. The equivalence among the
two is proven in \cite{Chevalley}.}.

This observation can be used to state a correspondence between 
pure spinors and generalized almost complex structures on \tts:
\begin{equation}\label{eq:JPhi}
{\cal J}\ \leftrightarrow \Phi \qquad {\rm if} \ \ L_{\cal J}=L_\Phi\ \ ,
\end{equation}
which means, we recall, that the $i$--eigenbundle of ${\cal J}$ is equal to the 
annihilator of $\Phi$. An alternative definition of the generalized almost
complex structure ${\cal J}$ associated to $\Phi$, maybe more suitable for computations, can be given by
\begin{equation}\label{eq:cJbil}
{\cal J}_{\pm\,\Lambda\Sigma}
=\langle\mathrm{Re}(\Phi_\pm)), \Gamma_{\Lambda\Sigma} \mathrm{Re}(\Phi_\pm)\rangle \ ,
\end{equation}
where $\Lambda$, $\Sigma$ are indices on \tts, and $\Gamma_\Lambda$ denote collectively the gamma matrices of Cliff$(d,d)$.
This definition is essentially the application to \tts of the usual definition of an almost
complex structure $I$ from a (pure) spinor\footnote{An ordinary Cliff$(d)$ spinor   
$\eta$ is always pure for $d\le 6$.}
in the usual Cliff$(d)$, $I_m{}^n=\eta^\dagger \gamma_m \gamma^n \eta$. 

The correspondence is not exactly one to one because rescaling $\Phi$ does not 
change its annihilator $L_\Phi$. Hence, it is more convenient to think about a correspondence between a $\J$ and a {\it line bundle} of pure spinors. This 
line bundle need not have a global section, in general; when it does, the structure group
on \tts is further reduced from U$(d/2)\times$U$(d/2)$ (which was already accomplished by $\J$) to SU$(d/2)\times$SU$(d/2)$.

What is remarkable about this correspondence is that integrability of ${\cal J}$ can be
reexpressed in terms of $\Phi$. To see this, let us consider an integrable ${\cal J}$.
Then, if $A, B\in L_{\cal J}$, we should have $[ A, B]_\mathrm{C} \in L_{\cal J}$. But, by 
(\ref{eq:JPhi}), we have $[A,B]_\mathrm{C} \Phi=0$; then by (\ref{eq:Cou})
\[ 
0=[A,B]_\mathrm{C} \Phi=(AB-BA)\cdot d\Phi\ .
\] 
So, for example, imposing $ d\Phi=0$  
implies that $[A,B]_\mathrm{C}\in L_\Phi=L_{\cal J}$, and hence by definition $\cal J$ is integrable. The condition $d\Phi=0$ can be relaxed:
if we think of $A$, $B$ as gamma matrices, then we are imposing that $d
\Phi$ be annihilated by two gamma matrices, so that it be at most at level one
starting from the Clifford vacuum $\Phi$. In other words
\beq
\label{intsp}
d\Phi= (\iota_v + \zeta \wedge) \Phi \qquad \Leftrightarrow \qquad {\cal J} \ {\rm integrable}
\eeq
for some $v$ and $\zeta$. This proof is from \cite{gualtieri}, except that here we
used the derived bracket definition (\ref{eq:Cou}) of Courant to streamline it. 
In particular, the conclusion (\ref{intsp}) is now not too surprising: $d$
is present from the very beginning in the definition of the Courant bracket. In this perspective, 
the Courant bracket is almost defined {\it ad hoc} so that integrability corresponds
to something very similar to closure. In this paper, $\Phi$ will have a more 
prominent role than $\J$. 

A  {\it  generalized Calabi-Yau} (GCY) is a manifold that has a pure spinor closed under
$d$ whose norm
does not vanish.\footnote{1 is a differential form of degree zero which also has an annihilator of dimension 6, 
$L_1=T$; so it is a pure spinor, but it has zero norm, since $1\wedge 1$ has no top--form part, compare with (\ref{defMukai}). So the norm requirement is essential, or else
any manifold would be a generalized Calabi--Yau.}
In general, a GCY  does not admit a unique
closed $\Phi$. Given a closed two--form $B$, one can transform $\Phi$ into  
\beq\label{eq:eb}
\Phi_B= e^B\wedge \Phi \  \ \Leftrightarrow \  \J_B = {\cal B} \J {\cal B}^{-1} \, , \quad e^B=1+ B \wedge + \frac{1}{2} B\wedge B \wedge +... \ , \quad {\cal B}=\left( \begin{array}{cc} 1 & 0 \\ B & 1 \end{array}
\right) \ ;
\eeq
$\Phi_B$ is clearly still closed. In fact, by a few simple modifications of the formalism seen so far, we can accommodate more general 
$B$s:   non--closed ones, 
but even ones which are not well--defined as two--forms, but rather as connections 
for a gerbe. In other words, we can extend $B$ in (\ref{eq:eb}) to a B--field. 

Obviously $e^B \Phi$ is not closed if $B$ is a B--field. But it is closed under $d-H\wedge$, 
where $H$ is the curvature of $B$\footnote{If the B--field is defined in each open
set $U_\alpha$ of a covering 
by $B_\alpha$, with gluing $B_\alpha= 
B_\beta+d\omega_{\alpha\beta}$, then $H_\alpha=d B_\alpha$. For a $B$ which is actually a globally defined two--form, we can simply write $H=dB$.}. In type II theories without NS--fivebranes (so that $dH=0$)  $d-H\wedge$ is a differential (it squares to zero). 
We can use it to define a modified Courant bracket, the {\it twisted Courant} ,  $[\ , \ ]_H$ 
\beq
\frac12 \Big([\{A\cdot ,(d-H\wedge)\}, B\cdot]-[\{B\cdot ,(d-H\wedge)\}, A\cdot] \Big)\equiv[A,B]_H \cdot  \, .
\eeq
In components it reads
 \beq
[v+ \zeta ,w + \eta]_H=[v,w] + {\mathcal L}_v \eta 
- {\mathcal L}_w \zeta - \frac12 d(\iota_v \eta -\iota_w \zeta) +\iota_v
\iota_w H  \, .
\eeq
This new bracket has all the good properties to be a derived bracket.
Moreover, we can now retrace all the steps of the correspondence between generalized complex structures and pure spinors, to get
 \beq\label{eq:twistedint}
(d-H\wedge)\Phi= (\iota_v + \zeta \wedge) \Phi \qquad \Leftrightarrow \qquad {\cal J} \ {\rm twisted\ integrable}\ .
\eeq
In generalized complex geometry the word ``twisted" is usually associated with the insertion of the three--form 
$H$; it has nothing to do with the occurrence of that word in 
the expression ``twisted torus" except that, as we will see, the two get exchanged by 
T--duality.  So  a manifold on which there exists 
a pure spinor $\Phi$ which is closed under $d-H\wedge$ is called {\it  
twisted generalized Calabi--Yau}. 

We can use the two-dimensional example discussed above as a toy model to construct
pure spinors and their annihilators.
Let us consider first the generalized complex structure $\J_I$. 
$\Phi_I$ is determined (up to a factor) 
by  having $L_{\Phi_I}=L_{\J_I}$ as annihilator  
\beq \label{puremi}
(\iota_{\del_1}+ i \, \iota_{\del_2} ) \Phi_I=0 \ , \qquad 
(e^1 + i \, e^2) \wedge  \Phi_I =0 \ , 
\eeq
which gives
\beq \label{puremi2}
\Phi_I = c_- \, (e^1 + i \, e^2) = c_- \, \O_1 \ ,
\eeq
where $c_-$ is a complex number that gives the normalization of $\Phi_I$.
Similarly for $L_{\J_J}$ one has 
\beq \label{purepl}
\left.
\begin{array}{c}
(\iota_{\del_1} + i \, e^2 \wedge \, ) \Phi_J=0 \\
(\iota_{\del_2} - i \, e^1 \wedge\, )  \Phi_J=0 
\end{array} \right\}
\, \Rightarrow 
\Phi_J = c_+ \, (1-i e^1 \wedge e^2)=c_+ \, e^{-i J}  \, .
\eeq
The generalized Calabi-Yau condition $d\Phi=0$  implies in these examples that either 
$J$ or 
$\Omega_1$ are closed\footnote{In two dimensions, $J$ is trivially closed, and closure
of $\Omega_1$ implies that all 1-forms are closed, but the situation is different in 
higher dimensions, and a generalized Calabi-Yau is generically not Calabi-Yau.}.

Remarkably, one can actually prove \cite{gualtieri} that in any dimension, a pure spinor must have the form
\beq \label{type}
\Phi= \Omega_k \wedge e^{B+ij}
\eeq
where $\Omega_k$ is a complex $k$--form and $B$, $j$ are two real two--forms. 
Hence the most general pure spinor is a hybrid of the two examples we just constructed 
on $T^2$. $k$ is also called {\it  type} of $\Phi$. It can be any integer from 0 to 
 $d/2$, where $d$ is the dimension of the manifold. (If it were bigger than $d/2$, the
 norm of $\Phi$ would be zero.) It can also be defined by looking at $\J$: it is then 
 the dimension of the intersection of the annihilator $L_{\J}$ with the tangent bundle
 $T$.

The two extreme cases, type 0 and $d/2$ are the most popular ones, for a number of reasons. 
They correspond to the generalizations 
to arbitrary dimensions of the two-dimensional examples given above.
A generic type 0 spinor, $\Phi_+=e^{B+iJ}$, is the  $B$--transform 
(see eq.~(\ref{eq:eb})) of $e^{iJ}$, and it has non--zero norm if 
 $J$ is non--degenerate ($J\wedge J\wedge J\neq 0$ everywhere).  
For it to be closed, we must have $dJ=0$. These are precisely
the conditions for a manifold to be symplectic. In fact, this was to be expected: the 
generalized complex structure associated to $e^{iJ}$ is $\J_J$, as we have seen in 
the example (\ref{purepl}). But integrability of $\J_J$ required indeed that 
the manifold
be symplectic, as seen in (\ref{gcscom}). 

The correspondence (\ref{intsp}) works in a more interesting way in the case of a 
$\Phi$ of type $d/2$. Let us consider for example the case $\Phi=\Omega_{d/2}$. 
First of all, non--zero norm requires again non--degeneracy, namely $\Omega_{d/2}\wedge\bar\Omega_{d/2} \neq 0$ everywhere. More importantly, 
one can see that purity \cite{hitchin} requires that the form be decomposable, namely
that it can be written in every point as $\omega_1\wedge \ldots \wedge \omega_{d/2}$, with $\omega_i$ one--forms. $\Omega_{d/2}$ then describes an Sl$(d/2,{\Bbb C})$ 
structure\footnote{An 
example of a complex three--form with non--zero
norm and not pure is, in ${\Bbb R}^6$, $dx^1\wedge dx^2 \wedge dx^3+i dx^4
\wedge dx^5\wedge dx^6$. This one defines an Sl$(3,{\Bbb R})\times$Sl$(3,{\Bbb R})$
structure.}.  and defines an almost complex structure $I$: 
$T^*_{1,0}$ is the span of the $\omega_i$. This almost complex
structure has by construction $c_1=0$, since the bundle $K$ of $(3,0)$--forms 
(which is defined even if $I$ is not integrable)
is trivialized by  $\Omega_{d/2}$. 
Integrability is easy to impose for $\Omega_{d/2}$. We take for simplicity $d=6$, and call
$\Omega_3\equiv \Omega$. In this case integrability of the complex structure 
( $\mathrm{Nij}(I)=0$) is equivalent to $(d\Omega)_{2,2}=0$.
Clearly if $I$ is integrable, $d$ acting on a $(3,0)$--form cannot 
give a $(2,2)$--form so that the condition $(d\Omega)_{2,2}=0$ is satisfied.
The other direction of the implication is less obvious (see for example 
\cite{cs}). But in the present context, this is implied by 
(\ref{intsp}): indeed the condition on $\Phi$  in this case just says that $d\Omega$
is a $(3,1)$--form. 
A final comment: if one imposes $d\Omega=0$, one also gets that
$K=0$  ---  namely, that the canonical bundle is trivial holomorphically, and not only
topologically (which is already guaranteed by the existence of $\Omega$).

These two examples are the main motivation behind the introduction of generalized 
complex geometry and the definition of generalized Calabi--Yau manifolds.
The crucial observations are that symplectic geometry can be defined by $d e^{iJ}=0$,
that complex geometry (with trivial $K$) can be defined by $d\Omega=0$, and
that both $e^{iJ}$ and $\Omega$ are pure spinors. 

The existence of an integrable pure spinor also allows to know the local geometry of the 
manifold. If the integrable pure spinor has type $k$, the generalized 
complex manifold is locally equivalent to a product
$\Bbb C^k \times (\Bbb R^{d-2k},J)$, 
where $J=dx^{2k+1} \wedge dx^{2k+2} + ... + dx^{d-1} \wedge dx^d$ is
the standard symplectic structure and $k$ is again the type. This is a 
complex--symplectic ``hybrid".

The type needs not remain constant over the manifold $M$.  
Generically it is as low as it is allowed by parity, and will jump up in steps of 
two at special loci. 
Thus a generic even pure spinor $\Phi_+$ will be of type 0, and jump to type 2 at loci 
(and possibly even higher on subloci, if the dimension $d$ of $M$ is $\geq 8$.) 
An odd one, $\Phi_-$, will generically have type 1, and similarly jump to 3 at loci (and, again, even higher when possible). 
The case of maximal type, $d$, is therefore highly non generic.

\subsection{Metric from pure spinor pairs}
\label{sec:metric}

We have seen how the existence of a generalized almost complex structure reduces the structure group of \tts from O$(d,d)$ to U$(d/2,d/2)$. 
The structure group can be further reduced to its maximal compact subgroup, 
U$(d/2)\times$U$(d/2)$
if it is possible to define two generalized almost complex structures, 
$\J_a$ and $\J_b$ that
commute and such that $M\equiv{\cal I} \J_a \J_b$ is a positive definite metric on \tts.
 Two such structures are said to be {\it  compatible}. 
 The fact that two such structures give rise to a metric can be seen from  the product 
 \[
 G=- \J_a\J_b \ .
 \]
$G$  squares to 1 (because $\J_{a,b}$ square
 to -1 and commute) and hence is a projector.  It divides  \tts
 in two subbundles $C_\pm$.  Since $\J_a $ and $\J_b$ commute, one 
can divide the complexified 
 \tts in four sub--bundles $L_{\pm,\pm}$ which are the intersections of $\pm i$ eigenspaces for the $\J_{a,b}$.  The subbundles $C_\pm$ are then given by $C_\pm=L_{+\pm}\oplus L_{-\mp}$; 
since 
 $L_{+\mp}=\bar L_{-\pm}$, $C_\pm$ have both rank $d$, and all four 
 $L_{\pm\pm}$ have rank $d/2$. This implies that the dimension of $M$ be even.\footnote{Moreover two compatible generalized almost complex structures 
must have opposite parity for $d=4n+2$, while for $d=4n$, they have the same parity.}
 The condition on positivity of  ${\cal I} G $ 
 is to ensure that one gets U$(d/2)$ and not U$(k,d/2-k)$; we will see shortly
 why this is important for us.  

If two compatible $\J_{a,b}$ are also integrable, they define a generalized K\"ahler structure, 
as defined in \cite{gualtieri}. This condition is equivalent to the existence
of $(2,2)$ $\sigma$--models; but we will not review that here, as the focus of this paper
is on $\N=1$ vacua with RR fields $\neq 0$. Nevertheless, some of the results in 
generalized K\"ahler geometry in \cite{gualtieri} do not depend on this integrability assumption,  and thus apply to manifolds with U$(d/2)\times$U$(d/2)$ structure too.  

From the fact that $G^2=1$ and from its hermiticity (${\cal I} G= G^t{\cal I}$) one can see that the most general form of $G$ is 
\bea
G = - \J_a \J_b&=& \left( \begin{array}{cc} -g^{-1} B & g^{-1} \\ g-B g^{-1} B & 
B g^{-1} 
\end{array}\right)= 
\left( \begin{array}{cc} 1 & 0\\B &1 
\end{array}\right)
\left( \begin{array}{cc}  0 & g^{-1} \\ g & 0  
\end{array}\right)
\left( \begin{array}{cc} 1 & 0\\-B &1 
\end{array}\right) \nonumber\\
&=&{\cal E}\left( \begin{array}{cc} -1 & 0\\ 0&1 
\end{array}\right){\cal E}^{-1}\ ; \qquad\qquad 
{\cal E}= \left( \begin{array}{cc} 1 & 1\\ g+B& -g+B  \ .
\end{array}\right)\label{eq:E}
\eea
So a U$(d/2)\times$U$(d/2)$ structure provides automatically a metric $g$ 
and B--field $B$. Notice that $B$ appears indeed as in a 
$B$--transform, eq. (\ref{eq:eb})\footnote{
It is easy to see that the transformation $\exp{{-\omega^t\ \beta}\choose{\ B\ \ \omega}}$ on \tts is
induced by conjugation by $\exp[(B\wedge+ \frac12\omega_m{}^n [dx^m\wedge, \iota_{\del_n}]+
\iota_\beta]$. Hence the matrix ${{1\ 0}\choose{B\ 1}}$ is induced by
$\exp[ B\wedge]$.}.
From the positivity condition on ${\cal I} G$, we know that 
the metric $g$ on $T$ is positive. 
Another remark on $G$
is that the metric $M={\cal I} G$ appeared in T--duality 
(see for example \cite{gpr}) as a combination that transforms
by conjugation under Sl$(2,\Bbb R)$.

Going back to the examples (\ref{trex1}) it is easy to prove that the two complex structures
are compatible and define a metric $G$ 
\beq \label{G} 
G=-\J_I \J_J = 
\left(
\begin{array}{cccc}
0 & 0 & 1 & 0 \\
0 & 0 & 0 & 1\\
 1 & 0 & 0 & 0 \\
 0&1&0&0
\end{array}
\right)  \ .
\eeq
This gives $g$ to be just the  $2\times 2$ identity matrix, and $B=0$. 
Note that
if we change the sign of one of the generalized complex structures, the pair  would still 
commute and $L_{\pm,\pm}$ would have dimension 1, but the metric $g$ would not be positive definite. 

Coming back to the general compatible pair, both $\J_a$ and $\J_b$ are now
acting on the bundles $C_\pm$ defined by the projector $G$.  
Hence they will define almost complex structures $I_1$ on $C_+$ and $I_2$ 
on $C_-$ (the choice of subscripts ${}_1$, ${}_2$ is for later 
convenience). It follows that
\beq\label{eq:JI}
\J_{a,b}=
{\cal E}
\left( \begin{array}{cc} I_1 & 0\\ 0&\pm I_2 
\end{array}\right){\cal E}^{-1}\ ;
\eeq
hence $\J_{a,b}$ are determined by $g$, $B$ and two almost complex structures
$I_{1,2}$. 


\subsection{Compatible  pairs from  spinor tensor products}
\label{sec:comp}

Pure Cliff$(d,d)$ spinors (which are sums of forms of different degrees, as we have
 seen) can be obtained from tensor products of Cliff$(d)$ spinors.  
The idea is that bispinors are isomorphic to differential forms, via the familiar
Clifford (``$\slash$") map: 
\beq
\label{clifmap}
C\equiv\sum_k \frac{1}{k!}C^{(k)}_{i_1\ldots i_k} dx^{i_i}\wedge\ldots\wedge dx^{i_k}\qquad
\longleftrightarrow\qquad
\sla C \equiv
\sum_k \frac{1}{k!}C^{(k)}_{i_1\ldots i_k} \gamma^{i_i\ldots i_k}_{\alpha\beta} \ .
\end{equation}

Under this isomorphism, 
the Cliff$(d,d)$ action  on forms (\ref{Cliff1}) translates into the action of two copies 
of Cliff$(d)$, one from the left and the other from the right of the bispinor.
A $\gamma^m$ matrix acts on the left and on the right of 
the $\gamma^{m_1...m_k}$ as 
\bea \label{gg}
\gamma^m \gamma^{m_1 ... m_k}&=& \gamma^{mm_1...m_k} + k g^{m[m_1} \gamma^{m_2...m_k]} \ \nn \\
 \gamma^{m_1 ... m_k} \gamma^m&=& 
(-1)^k \left( \gamma^{mm_1...m_k} - k g^{m[m_1} \gamma^{m_2...m_k]} \right)  \ .
 \eea
For $v^m=\pm g^{mn} \xi_m$ (with 
a plus or minus sign for the left and the right action, respectively) this is precisely the Cliff$(d,d)$ action $\cdot$ on forms\footnote{The factor $k$ 
in (\ref{gg}) comes from the definition of the contraction, 
namely $\iota_{\del_m} dx^{m_1} \wedge ... \wedge dx^{m_k} = 
k \delta^{[m_1}_m dx^{m_2} \wedge ... \wedge dx^{m_k]}$.}. 
In other words, the gamma matrix action on bispinors
is mapped to the following action on forms $C_k$ of degree $k$:  
\beq
\label{eq:gammamap}
\gamma^m \ \sla C_k = 
\begin{picture}(10,10)(-10,5)
\put(0,0){\line(6,1){100}}
\end{picture}
[(dx^m\wedge + g^{mn}\iota_{\del_n})C_k]
\ , \qquad 
 \sla C_k \ \gamma^m = (-)^k
\begin{picture}(10,10)(-10,5)
\put(0,0){\line(6,1){100}}
\end{picture}
[(dx^m\wedge - g^{mn}\iota_{\del_n})C_k] \ .
\eeq 
The use of the $\slash$ can clutter formulas
considerably, and is best avoided when it does not give rise to 
confusion. In the main text, most of the time we will not distinguish, with an abuse of
notation, a form by the corresponding bispinor. More precision is needed
in Appendix \ref{ap:N=1calc}, and there the $\slash$ is restored.

Consider now a bispinor (say, an even one)
which is a tensor product of two spinors: 
\[
\Phi_+=\eta^1_+\otimes \eta^{2\,\dagger}_+\ ;
\]
in components, $\Phi_{+\,\alpha\beta}= \eta^1_{+\,\alpha}\eta^{2*}_{+\,\beta}$.  
(This is of course not always the case  ---  bispinors live in the tensor of two spinor bundles, which means that they can be written as 
$C= \sum_{\alpha\beta} C_{\alpha\beta} \eta^{(\alpha)}\otimes \eta^{(\beta)}$, with $\eta^{(\alpha)}$ a basis of Cliff$(d)$ spinors; in the case we are considering, there is only one summand in this sum.)
Suppose now the spinors $\eta^{1,2}_+$ are also pure as ordinary Cliff$(d)$ spinors. 
We have so far talked about purity for Cliff$(d,d)$ spinors only, but in fact the
same definition can be applied to spinors in any signature and dimension: once again
a Cliff$(d)$ spinor $\eta$ is called pure if there are $d/2$ linear combinations 
of gamma matrices that annihilate it. 
In other words, a pure spinor is one that can be
taken as Clifford vacuum; the $d/2$ annihilators
 are then by definition the holomorphic gamma matrices
$\frac12(1-iI)^m{}_n\gamma^n\eta_+=0$ (or, in 
case $I$ is integrable, $\gamma^i\eta_+=0$). So every pure Cliff$(d)$ spinor 
defines an almost complex structure $I$. If $\eta^a_+$ are both pure, $\eta^1_+\otimes \eta^{2\,\dagger}_+$ is annihilated by 
$d/2$ gamma matrices acting from the left and by $d/2$ acting from the right. 
Thanks to (\ref{eq:gammamap}), we can translate these $d/2+d/2$ annihilators into $d$ annihilators in Cliff$(d,d)$. This means $\Phi_+$ is pure. 

For $d\leq 6$, Cliff$(d)$ spinors are always pure.\footnote{The space of pure spinors is the complex 
cone over the space of almost complex 
structures  compatible with a certain metric (also known as space of {\it twistors}), 
which is $\frac{\mathrm{SO}(d)}{\mathrm{U}(d/2)}$; for $d=2,4,6$ this has
dimension $0,2,6$ respectively, and summing two real dimensions because of the
complex rescaling, the total matches the dimension of the space of all spinors, 
$2,4,8$.

In eight dimensions and on, this coincidence stops, and spinors which are not pure exist. 
For example, 
a spinor that defines a Spin$(7)$ structure is not pure: such a structure can be
reformulated by starting from a four--form, and there is no natural 
almost complex structure associated to it. More explicitly, in flat space, 
the Weyl spinor  $|\uparrow\uparrow\uparrow\uparrow\ \rangle+
|\downarrow\downarrow\downarrow\downarrow\ \rangle$ is not pure: it is not annihilated
by four gamma matrices. On the contrary, the analogous spinor in four-dimensions is pure:
 $|\uparrow\uparrow\ \rangle + |\downarrow\downarrow\ \rangle$
is annihilated by 
$\gamma^1-\gamma^{\bar 2}$ and $\gamma^2-\gamma^{\bar 1}$.} 
From now on we will restrict to the case
$d=6$, which is the case of relevance to four--dimensional compactifications anyway. The $\J_{a,b}$ defined in the previous subsection will now be referred
to as $\J_\pm$, since in six dimensions one of them has even type and one
of them odd type. 

Once one has the pure spinor $\Phi_+$, 
it is useful to expand it in the basis of bispinors
$\gamma^{m_1\ldots m_k}$. The coefficients of this expansion are simply
$\frac1{8k!}\mathrm{Tr}(\Phi_+\gamma^{m_k\ldots m_1})= \eta^{2,\dagger}_+\gamma^{m_k\ldots m_1}\eta^1_+$. Hence we have
\begin{equation}
\label{fierz}
   \eta^1_{+}\otimes \eta^{2\,\dagger}_\pm =
      \frac{1}{8} \sum_{k=0}^6 \frac{1}{k!}
      \left({\eta}^{2\,\dagger}_\pm\gamma_{m_k\dots m_1}\eta^1_+\right)
         \gamma^{m_1\dots m_k} 
\end{equation} 
which is known as Fierz identity. At this point, one can apply the Clifford map (\ref{clifmap})
back and obtain an expression for $\Phi$ as a differential form.
One can also relate the trace over the bispinors with the pairing (\ref{defMukai}), by 
using\footnote{\label{fo:star}We have inherited from \cite{democratic} the somehow awkward
sign convention $* e^{a_1}\wedge\ldots \wedge e^{a_k} = \frac1{(d-k)!} 
\epsilon_{a_{k+1}\ldots
a_d}{}^{a_1\ldots a_k} e^{a_{k+1}\ldots a_d}$, where $e^a$ is the vielbein.}  
\begin{equation}
    \label{trace}
\langle A_k, B_{6-k}\rangle=\frac18(-)^{k+1}\mathrm{Tr}(\sla A_k \slas *B_{6-k}) \,  {\rm vol} \ .
\end{equation}

So far we have talked about one pure spinor only. It is also easy to obtain a 
compatible pair. Consider
\beq \label{pureforms}
\Phi_+= \eta^1_+ \otimes \eta^{2\, \dagger}_+ \ , \qquad  
\Phi_-= \eta^1_+ \otimes \eta^{2\, \dagger}_- \, .
\eeq
In our conventions, $\eta_-=\eta_+^*$. Let us write down the corresponding 
annihilators, based on what we have just learned about a single
pure spinor. They are given in terms of 
the two copies 
of Cliff$(6)$ acting on the left and on the right, (\ref{eq:gammamap}),
 \beq
 \label{eq:genann}
(\delta +i I_1)_m{}^n \gamma_n \,  \eta^1_+\otimes \eta_{\pm}^{2 \, \dagger}=0 \ ,
\qquad  
   \eta^1_+\otimes \eta_{\pm}^{2 \, \dagger}\,\gamma_n 
(\delta \mp iI_2)_m{}^n =0 \ ,
\end{equation}
where $I_1$ and $I_2$ are the almost complex structures defined 
by $\eta^1$ and $\eta^2$. (We are going to see that they are also the same
as the $I_{1,2}$ in (\ref{eq:JI}).)
The three annihilators acting on the left are the annihilators of $\eta^1_+$, the other three
on the right annihilate $\eta^{2\, \dagger}_-$.

Hence $\Phi_+$ and $\Phi_-$  share three annihilators:  the three gamma matrices 
$(\delta +i I_1)_m{}^n \buildrel\to\over{\gamma^n}$. Call $L_{++}$ 
this subbundle of $(T \oplus T^*)\otimes\Bbb C$. Similarly, $\Phi_+$ and $\overline{\Phi}_-=
 \eta^1_- \otimes \eta^{2\, \dagger}_+$
are both annihilated by the three gamma matrices $(\delta -i I_2)_m{}^n \buildrel\leftarrow\over{\gamma^n}$. Call this bundle $L_{+-}$. In this way we construct
four bundles $L_{\pm\pm}$ of dimension 3 each. Now define $\J_+$ to be $i$ on 
$L_{+\pm}$ and $-i$ on $L_{-\pm}$; and $\J_-$ to be $i$ on 
$L_{\pm+}$ and $-i$ on $L_{\pm-}$. These two commute by definition; the bundles 
$L_{\pm\pm}$ are just the ones we encountered in the previous subsection, while
describing U$(3)\times$U$(3)$ structures.
We conclude that (\ref{pureforms}) is a compatible pure spinor pair: the 
$\J_\pm$ defined
by $\Phi_\pm$ are compatible. In this situation, the structure group on \tts is actually
reduced a bit further, to SU$(3)\times$SU$(3)$. 

In fact, the most general pure spinor pair is a $B$--transform of a 
pair as in  (\ref{pureforms}). To see this, remember that two compatible 
$\J_\pm$ are determined
as in (\ref{eq:JI}). 
Note also that ${\cal E}$ in (\ref{eq:E}) can be decomposed as
\begin{equation}
{\cal E}= \left(
\begin{array}{cc}
 1 & 0     \\
 B &  1  
\end{array}
\right)\left(
\begin{array}{cc}
 1 & 1     \\
 g & -g  
\end{array}
\right)\ .
\label{eq:EE}
\end{equation}
Consider first the case $B=0$. Let us compute for example $L_{++}$. One can interpret ${\cal E}$ as a change of basis. Then, an $i$ eigenvector of both ${{I_1\ \ 0}\choose 
{0\ \pm I_2}}$ would be given by $(\delta+iI_1)^m{}_n v^n$. The change of basis ${\cal E}$ (with $B=0$) makes this $(\delta+iI_1)^m{}_n (dx^n+ g^{np} \del_p)$. Remembering
(\ref{eq:gammamap}), this coincides with (\ref{eq:genann}). (Morally, the change of basis
${\cal E}$ reproduces the Clifford products in (\ref{eq:gammamap}).) Hence we
have that for $B=0$ the pair can be written as in (\ref{pureforms}). 
Finally, when $B\neq 0$, the action of ${{1\ 0}\choose{B\ 1}}$ can be reproduced by
$e^{B}$ on the pure spinors in (\ref{pureforms}).

Besides being the most general pure spinor pair up to B--transform, 
(\ref{pureforms}) has the property that the two norms are the same. 
Indeed, 
\begin{equation}
    \label{eq:eqnorms}
   \langle \Phi_-, \bar \Phi_- \rangle=  \langle \Phi_+, \bar \Phi_+\rangle=-\frac i8 \mathrm{Tr}(\sla \Phi_\pm 
\sla \Phi_\pm^\dagger) \,  {\rm vol} =-\frac i8 ||\eta^1_+||^2||\eta^2_\pm||^2 \,  {\rm vol} \   ,
\end{equation}
where we have used that $\sla C^\dagger= \slas\lambda(\bar C)$ for any 
complex form $C$, and (\ref{eq:star}), (\ref{trace}).

We now give some examples, again in the $T^2$ setting. 
We will first look at how (\ref{pureforms}) works. Let us denote the two spinors by 
$|\uparrow\ \rangle={1\choose 0}$ and
$|\downarrow\ \rangle={0\choose 1}$, in the usual spin--up spin--down picture. The gamma matrices in two dimensions can be taken to be Pauli matrices: $\gamma^{1,2}=
\sigma^{1,2}$. 
Then, in a complex basis, a creator $\gamma^z= \gamma^1+i \gamma^2 = 
{{0\ 1}\choose {0\ 0}}$  and an annihilator $\gamma^{\bar z}= \gamma^1-i \gamma^2 = 
{{0\ 0}\choose {1\ 0}}$. Then $\gamma^{z\bar z}=
[\gamma^z,\gamma^{\bar z}]={{1\ \ \ 0}\choose{0\ -1}}$.
Let us take, in (\ref{pureforms}), $\eta^1_+=\eta^2_+=|\uparrow\ \rangle$, 
and $\eta^1_-=\eta^2_-=|\downarrow\ \rangle$. We now can compute the pure spinors by taking a trivial
tensor product,
\[ 
|\uparrow\ \rangle \langle\ \uparrow |= \left(\begin{array}{cc}1 & 0 \\ 0 &0\end{array}\right)
=\frac12(1+\gamma^{z\bar z})= \frac12(1+dz\wedge d\bar z
\begin{picture}(10,10)(30,0)
\put(0,0){\line(4,1){30}}
\end{picture})
\]
and 
\[
|\uparrow\ \rangle \langle\ \downarrow |= \left(\begin{array}{cc}0& 1 \\ 0 &0\end{array}\right)
=\gamma^z= \sla dz\ .
\]
 
Obviously one can choose $\eta^1=\eta^2$ in six dimensions too. In that case, 
the structure group on $T$ is SU(3). More generally, that is the case if
the two spinors are parallel, namely
\beq
\label{SU3st}
\eta^1_+=a \eta_+ \ , \qquad \eta^2_+=b\eta_+
\eeq
 where $a$ and $b$ are complex numbers; the redundancy is for later 
convenience. 
$\eta_\pm$ are normalized $\eta_\pm^\dagger \eta_\pm=1$. 
 Inserting (\ref{SU3st})  in (\ref{pureforms}), 
we get the pure spinors 
\beq
  \label{puresu3}
{\rm SU(3):} \qquad 
\Phi_-= 
-\frac{ia  b}{8} {\Omega_3}
\, , \qquad 
\Phi_+=
\frac{a \bar b}{8} 
e^{- i J}
 \, .
\end{equation}
These are the six--dimensional analogues of   
(\ref{puremi2}), (\ref{purepl}), for  $c_-=-\frac{iab}{8} $, 
$c_+=\frac{a \bar b}{8}$. Since $\Omega$ is pure, as we saw, it defines 
an almost complex structure $I$ (the $(0,1)$ vectors $v_{(0,1)}$ 
being by definition the ones that satisfy $\iota_{v_{(0,1)}}\Omega=0$).
One can also see that compatibility $[\J_I, \J_J]=0$ 
of the corresponding generalized almost 
complex structures (\ref{gcscom}) imposes that 
\begin{equation}
    \label{eq:gIJ}
g_{mn}=I_m{}^p J_{pn}
\end{equation}
be symmetric; this is then, by definition, the metric given by the two 
pure spinors. This implies then that $J$ be of type $(1,1)$ with respect to
$I$. 
We recover the familiar statement that an SU(3) structure on $T$ gives 
rise  to a compatible 
pair of a purely complex and a purely symplectic (generalized almost complex) 
structures satisfying
\beq \label{compsu3}
J \wedge \Omega_3 =0 \ , \qquad  i \Omega_3 \wedge \bar \Omega_3= 
\frac{4}{3} \, J^3
\eeq
where the last condition comes from the fact that the 
pure spinors have the same norm, (\ref{eq:eqnorms}): 
$\langle \bar \Phi_+, \Phi_+ \rangle =\langle \bar \Phi_- , \Phi_- \rangle$. 

If, on the contrary, $\eta^2$ is orthogonal to $\eta^1$, they define a 
so--called {\it static 
 SU(2) structure} on $T$. This consists of a complex vector field without zeros
\beq\label{vwz}
\eta^{1\, \dagger}_- \gamma^m \eta^2_+= (v+iv')^m  \equiv \Omega_1^m \ ,
\eeq
a real 2-form $j$, and a holomorphic two--form $\Omega_2$. Equivalently, an  
SU(2) structure on $T$ is  defined by the intersection of two 
SU(3) structures. We can locally decompose the two SU(3) structures as
\beq
J^{1,2}=j\pm v \wedge v' \ , \qquad \Omega^{1,2}_3= \Omega_2 \wedge (v \pm i v') \, .
\eeq
One can also rewrite
\beq
\label{SU2st}
\eta^2_+= \frac12 b (v+i v')_m \gamma^m \eta_- \qquad \eta^1_+=a \eta^+ \, .   
\eeq
Inserting (\ref{SU2st}) in (\ref{pureforms}), we get
\begin{equation}
  \label{puresu2}
  {\rm static \, \, SU(2):} \qquad
\Phi_-=- \frac{ab}{8} \Omega_1  e^{-i\,j}  \ , \qquad
\Phi_+=-\frac{ia\bar b}{8}   \Omega_2 
e^{-i vv'} \ ,
\end{equation}
where we have  rescaled again the two spinors by two functions $a$ and $b$, 
with a redundancy for later convenience. 
The pure spinors have precisely the form (\ref{type}), for $B=0$. The type is $k=1$ for $\Phi_-$ 
and $k=2$ for $\Phi_+$.
Hence a static SU(2) structure on $T$  gives rise to a pair of  
compatible hybrid type 1 and type 2 generalized almost complex 
structures on \tts.

Note that in this paper  we are dealing with parallelizable manifolds, which have trivial structure group 
(on $T$ as well as on \tts). It may then be confusing to talk about SU(3) or SU(2) structures. The point is that
we are interested in the $\N=1$ vacua and therefore use  only two of the four globally defined Weyl spinors
that exist on a parallelizable manifold. So by  SU(3) structure solutions we denote those constructed out 
of  $\Phi_\pm$ of type 0 -- type 3, and similarly models will be called of SU(2) structure if the 
the corresponding  pure spinors are 
of type 1 -- type 2.

Generically the spinors $\eta^1$ and $\eta^2$ will neither be parallel nor orthogonal everywhere; 
the formulae  for the pure spinors in that case are more complicated \cite{jw}. 
  In particular, they can become parallel at some points,
in which case they do not define globally an SU(2) structure. All these cases, however, always define a global \stt\ structure on \tts. In this more general case (also called ``dynamic SU(2)" sometimes) we have a pair of compatible type 0--type 1 generalized almost 
complex structures
whose types can jump to type 2-type 1 or type 0-type 3 at special points. These are given by \cite{jw}
\beq \label{purestt}
{\rm{SU(3) \times SU(3):}} \qquad \Phi_-=-\frac{ab}{8} (v+iv') (k_\bot e^{-ij}+i k_\| \Omega_2)  \ , \qquad
\Phi_+=\frac{a \bar b}{8} e^{-ivv'} (\bar k_\l e^{-ij} -i \bar k_\bot \Omega_2) 
\eeq
where we are taking $\eta^1_+=a \, \eta_+$, $\eta^2_+=\frac12 b \, (k_\| \eta_+ + k_\bot (v+iv')_m \gamma^m \eta_-)$, with $|k_{\|}|^2 + |k_{\bot}|^2=1$.
As we shall see  in the next Section (see eqs. (\ref{fi-}) and (\ref{fi+})) , these more general forms of pure spinors are not consistent with any orientifold involution on our class of geometries, and thus we make no further comments on this case.

We are ready now to review the construction of the string vacua.

\section{ Constructing $\N=1$ vacua}
\label{N1vacua}

In this paper we are look for four--dimensional  $\N=1$ Minkowski vacua. They correspond to 
ten--dimensional backgrounds given by a  
warped product of four--dimensional Minkowski space  and an  
internal six--dimensional compact space
\beq \label{10dmetric1}
ds^2= e^{2A} \eta_{\mu \nu} dx^\mu dx^\nu + ds_6^2  \, .
\eeq

Our strategy for constructing such  vacua is to examine the ten--dimensional supersymmetry conditions and the equations of motion and Bianchi identities for the fluxes. For metrics of the type (\ref{10dmetric1}) , this is equivalent to solving the full set of ten--dimensional equations of motion \cite{dimitris}. We will first consider the differential-geometric conditions on the internal space and then pass to global conditions set by the compactness and the orientifold involutions.
  
\subsection{Local and global conditions}
\label{sec:necsuff}
The necessary and sufficient conditions for $\N=1$ supersymmetry when RR fields
are switched on were found in \cite{gmpt2}. The internal manifold must have \stt\  structure 
 on \tts\ and the 
pair of pure spinors $\Phi_{1,2}$ defining
the structure must satisfy\footnote{ In Appendix 
\ref{ap:N=1calc} we explain in detail  the intermediate steps 
in the calculations that 
lead to (\ref{int}), (\ref{nonint}). We have also changed notation in IIA relative to \cite{gmpt2}
so that the two equations have the same expression in both theories, see
(\ref{eq:redefA}).}
\bea
\label{int}
(d-H\wedge) (e^{2A-\phi}\Phi_1)&=&
0 \, , \\
(d-H\wedge) (e^{2A-\phi}\Phi_2)&=&
 \label{nonint} e^{2A-\phi} dA\wedge\bar\Phi_2 
+\frac i8 e^{3A} *\lambda(F)\ 
\eea 
and have norms $||\Phi_{1,2}||^2= \frac{e^{2A}}{8}$.
The $F$ that appears in these equations is a purely internal form, that is related to 
the total ten--dimensional RR field strength by
\begin{equation}
    \label{F10F6}
F^{(10)} = F+ \mathrm{vol}_4\wedge \lambda(*F)\ , \qquad 
F=\begin{array}{cr}\vspace{.2cm}
F_0+F_2+F_4+F_6   &\quad ({\rm IIA})\\
F_1+F_3+F_5 &\quad ({\rm IIB})
\end{array}
\end{equation}
where $\lambda$ is defined in  (\ref{defMukai}), and $*$ the 
six--dimensional Hodge dual.  
We are using the democratic formulation \cite{democratic}, in which $F^{(10)}$ is 
self--dual in ten-dimensions, but where $F$ need not be so in six. Notice that we have implicitly chosen 
the purely internal field strengths to be electric, and the ones with spacetime
indices to be magnetic; this is a convention we will stick to, in particular
below when talking about equations of motion for the fluxes versus Bianchi 
identities. Of course there is nothing special in this choice.
 For type IIA the pure spinors 
are
\beq \label{adef}
\Phi_1=\Phi_+ \ , \quad \Phi_2=\Phi_- \ , 
\eeq
and for type IIB
\beq \label{bdef}
\Phi_1=\Phi_- \ , \quad \Phi_2=\Phi_+ \ . 
\eeq
The pair of compatible even/odd pure spinors $\Phi_{\pm}$ can in general be written as (\ref{purestt}), 
but, as will become clear later on, in this paper
we will only need the particular cases (\ref{puresu3}) and  (\ref{puresu2}).

Equation (\ref{int}) says that the pure spinor that has the same parity as the RR fluxes, $\Phi_1$,
must be closed and thus any manifold suitable for $\N=1$ vacua should be 
a twisted generalized Calabi-Yau. This is only a necessary condition for 
having supersymmetric vacua. 
For example, all six--dimensional nilmanifolds 
are generalized Calabi-Yau \cite{CG} but we will see that very few of them can be promoted to supergravity vacua. 

The second equation relates the non-integrability of the second generalized complex structure to the RR fluxes.
Splitting it into real  and imaginary part we get
\bea
(d-H\wedge) (e^{A-\phi} \mathrm{Re} \Phi_2) &=& 0 \label{Re} \ , \\
(d-H\wedge) (e^{3A-\phi} \mathrm{Im} \Phi_2) &=& \frac{1}{8} e^{4A} *\lambda(F) \label{Im} \ .
\eea
From these we see that actually only the imaginary part of $\Phi_2$ is  not closed. 
Interestingly, it has been 
shown in \cite{ms, martucci} that the equations (\ref{int}), (\ref{Re}) and (\ref{Im})
can be rederived from demanding a generalized calibration condition
for D--branes, i.~e.~the existence and stability of supersymmetric D--branes on
backgrounds with fluxes. 
It was also noticed in \cite{WJ} that these equations
can be interpreted as conditions for a constrained critical point of the Hitchin functional for the pure spinor $\Phi_2$.

The ``limiting case'' in which $F=0$ is strictly speaking not 
covered by these
equations. In that case, the supersymmetry jumps to (at least) $\N=2$. This is
because the RR fields are the only term in the supersymmetry transformations
mixing $\epsilon_1$ and $\epsilon_2$; if one sets the RR to zero, then, one may as
well take different $\zeta$ in $\epsilon_1$ and the one in $\epsilon_2$ 
(see (\ref{eq:spinor})). (See also \cite{jw}).) Another way to see this
enhancement of supersymmetry is 
to remark that there are then two closed compatible pure
spinors; this implies (as we saw in Section 3) that there are two generalized
complex structures $\J_{a,b}$ which commute and are integrable.
This case was called {\it generalized K\"ahler} in \cite{gualtieri} and
proven to define a $(2,2)$ world--sheet model. This implies $\N=2$ spacetime
supersymmetry only if there exists a spectral flow for left-- and right--movers, 
but one can see that this condition is the vanishing of the right hand
side of $(d-H\wedge)\Phi_\pm$ in (\ref{eq:twistedint}).

In order to have a valid solution, we should
additionally impose  Bianchi identities and
the equations of motion for the fluxes \footnote{The equation of motion for $H$ comes 
from a reduction of the Chern-Simons coupling $B \wedge F_n \wedge \lambda(F_{8-n})$  (which can be written in spite RR flux being self-dual).}  
\bea \label{bianchi}
&& (d - H \wedge) F=\delta(\mathrm{source}) \ , 
\qquad (d+H \wedge ) (e^{4A} *F)=0 \ ,\\
\label{eqsH}
&&  dH=0 \ , \qquad d (e^{4A- 2 \phi} * H) = \mp e^{4A} F_n \wedge *F_{n+2} \ ,
\eea
where we have used the warped metric (\ref{10dmetric1}),  $\delta(\mathrm{source})$ 
is the charge density of the corresponding space--filling source 
(D--branes and O--planes) and the upper (lower) sign  corresponds to  
IIA (IIB). The electric sources are not present, because they would 
be points in spacetime, and hence break the assumption of
 four--dimensional Poincar\'e symmetry. More details on this are given in Appendix 
\ref{ap:N=1calc}.

Note that the equation of motion for $F$ is automatically satisfied. Indeed eq. (\ref{Im}) implies 
\beq \label{eeom}
\, \lambda [(d-H\wedge)  (e^{3A-\phi} \mathrm{Im} \Phi_2)] =   \, 
\mp (d+H\wedge) [e^{3A-\phi} \lambda ( \mathrm{Im} \Phi_2)] 
= \mp \frac{1}{8}  e^{4A} * F
\eeq
where the upper (lower) signs  correspond to  IIA (IIB) and come from commuting $\lambda$ with $*$ 
and $(d-H\wedge)$. From (\ref{eeom}) it follows  that the equation of 
motion for $F$ is automatic: $e^{4A} * F$ is $d+H\wedge$ exact, and hence also 
$d+H$ closed.

There is a no--go theorem that rules out vacua in which the 
four--dimensional space is Minkowski and the  internal compact manifold
has non-zero background fluxes and no sources. 
This is normally obtained  from the 
four--dimensional components of 
Einstein equation \cite{dWSHD,MN}: the RR and NS fluxes that preserve Poincar\'e invariance 
contribute a positive tension term to the energy momentum tensor.
However it also follows  from the Bianchi identities 
(\ref{bianchi}), as shown in 
\cite{GMPW} for the specific situation of flux $H$ on  
internal manifolds with $G$--structure, and as we are going  to see shortly in 
full generality.  
The Bianchi identities are more restrictive than 
four-dimensional Einstein equation,
since the latter is just a scalar equation, while the former are a form 
equation and give constraints for
every allowed cycle on which we can wrap supersymmetric sources.
The scalar component of the Bianchi identity is equivalent to the trace of the 
four-dimensional Einstein equation.

From equation (\ref{Im}) it follows 
that the scalar part of the Bianchi identity has a definite sign. 
To isolate it from (\ref{Im}) we can compute the paring with 
$e^{3A-\phi} \mathrm{Im} \Phi_2$. This form calibrates the cycle wrapped by 
a spacetime--filling brane or an orientifold, as
derived in the context of generalized complex geometry in \cite{ms}.
Using the adjunction property 
\begin{equation}
    \label{eq:adj}
\int \langle A,(d-H\wedge)B\rangle =
\int\langle (d-H\wedge)A,B\rangle    
\end{equation}
 for the differential $d-H$, one derives
\beq \label{tadpolesinglet}
\int \langle  (d-H \wedge) F, \ e^{3A-\phi}\mathrm{Im} \Phi_2  \rangle
=\int\langle F,\ (d-H\wedge) e^{3A-\phi}\mathrm{Im} \Phi_2\rangle
=\frac18 \int e^{4A} \,\langle F,\ *\lambda(F)\rangle
\eeq
which has always {\it negative} sign (see footnote \ref{fo:star}).  One can think of the left hand side
as of an  effective overall charge, since it is equal to 
$\int e^{3A-\phi}\langle {\rm source}, \mathrm{Im}\Phi_2 \rangle$.
Hence, to have a compactification consistent
with Gauss' law, there has to be some source of negative charge;
the only one available is an orientifold.
This is true no matter how much torsion (or, as it is often said, ``metric fluxes'') the manifold has.

We insist on the fact that it is the total effective charge  that 
has definite sign. If there is more than one contribution there is no 
general statement one can make for the sign of each contribution.
As we will see, there are examples where all individual contributions
have  the same sign (and therefore are all sourced by orientifold plane
charges) and examples where the individual contributions
have opposite signs.  
 In any case, once again, eqn.~(\ref{tadpolesinglet}) says that at least one
orientifold projection is always necessary.

\subsection{Orientifolds}

The no-go theorem we showed in the previous Section implies that in order 
to have compactifications to Minkowski 
the contribution of the
internal background fluxes to the energy-momentum tensor
must be cancelled by local sources of negative tension.

The theorem does not assume anything about the differential properties of the
internal manifold,
and therefore applies also to generalized Calabi-Yau. 
Then  any compactification in the
presence of RR and/or NS fluxes  needs sources of negative tension, 
like orientifold planes.

A (single) orientifold projection consists of modding out the theory by
 $\Omega_{\rm WS}(-)^{F_L} \sigma$ (for O3/O7 and O6) 
or $\Omega_{\rm WS} \sigma$ 
(for O5/O9 and O4/O8) where $\Omega_\mathrm{WS}$ is the reflection in 
the worldsheet, $F_L$ is fermion number for the left--movers 
and $\sigma$ a space--time involution. 
On Calabi-Yau manifolds $\sigma$ is an involutive symmetry of the 
manifold. For Type IIA this symmetry must be 
antiholomorphic, ($\sigma I = -I$), while in IIB it is holomorphic ($\sigma I = I$).  

For SU(3) structure manifolds, which still admit a nowhere vanishing 
fundamental form, the isometric involution $\sigma$ can then be found by imposing 
the same requirements as for the Calabi Yau \cite{BG}\footnote{In the context of RR
flux compactifications, one does not 
have control over the usual world--sheet definition of orientifolds.} (see also
\cite{GL}). This leads to the action on the
pure spinors
\begin{center}
\begin{tabular}{lccc}
IIA: &  & $\sigma \O_3 = \mp \bar \O_3$ & $\sigma e^{-i J} =  e^{i J}$ \ , \\
IIB: &  & $\sigma \O_3 = \mp \O_3$ & $\sigma e^{-i J} = e^{- i J}$ \ . 
\end{tabular}
\end{center}
The $\pm$ correspond to different O--planes (O3 vs O5, for example). 

In the general
\stt\ case there is not necessarily an almost complex structure, and the action
of the involution is determined by asking that it must  
%
reduce to the given action in the particular
case of a pure SU(3) structure. The involution should therefore act on the
pure
\stt\ spinors (\ref{purestt}) in the form showed in Table \ref{ta:sigma}  \cite{BG}.

\begin{table}[h]
\begin{center}
\begin{tabular}{c|c|c|c}
O3/O7 & O5/O9 & O6 &O4/O8 \\
\hline
$\sigma (\Phi_{+}) = -\lambda (\overline{ \Phi}_+)$ &
$\sigma (\Phi_{+}) = \lambda (\overline{\Phi}_+)$ &
$\sigma (\Phi_{+}) = \lambda (\Phi_+)$  &
$\sigma (\Phi_{+}) = -  \lambda (\Phi_+)$ \\
$\sigma (\Phi_{-}) = \lambda (\Phi_-)$ &
$\sigma (\Phi_{-}) = -\lambda (\Phi_-)$ &
$\sigma (\Phi_{-}) = \lambda (\overline{\Phi}_-)$ &
$\sigma (\Phi_{-}) = - \lambda (\overline{\Phi}_-)$
\end{tabular}
\label{ta:sigma} 
\caption{Action of the involution $\sigma$ on the pure spinors.}
\end{center}
\end{table}

In the next Sections we will look at orientifolds of nilmanifolds and solvmanifolds.
In these cases all orientifold planes can be allowed except for 
O3's, which only appear in the untwisted case\footnote{An O3 would
correspond to
an involution $\sigma (e^a)=-e^a$ for all $a=1,..,6$, which is incompatible
with
(\ref{strdef}) unless all structure constants are zero.}.

Notice also that a generic \stt\ spinor  (\ref{purestt}) with both $k_\| \neq0, k_\bot \neq
0$ is not consistent with
any orientifold involution\footnote{In this argument we neglected the possibility of an SU(2) rotation acting $j$, ${\rm Re}\, \Omega_2$ and $\Im \Omega_2$. As shown afterwards in \cite{koerber-tsimpis}, the consistency conditions are less rigid.}. To see this, take for example  an O5, where
$\sigma$ acts on the \stt\ spinors by
\bea \label{fi-}
\sigma\left((v+iv') (k_\perp e^{-ij}+ik_\| \Omega_2)\right) &=& -(v+iv')
(k_\perp e^{ij} -ik_\| \Omega_2)  \ , \\
\sigma \left(e^{-ivv'} (\bar k_\| e^{-ij} -i \bar k_\perp \Omega_2)\right)
&=& e^{-ivv'} (k_\| e^{-ij} -i k_\perp \bar \Omega_2) \ .
\label{fi+} \eea
Eq.(\ref{fi-}) implies $\sigma(k_{\perp} j)= - k_{\perp} j, \ \sigma( k_\|
\Omega_2) = -k_\| \Omega_2$, while eq.(\ref{fi+}) enforces $\sigma(\bar
k_{\|} j) = k_\| j$ and $ \sigma(\bar k_{\perp} \Omega_2) = - k_\perp \bar
\Omega_2$.
These two are inconsistent unless  $k_\perp =0$ ($\sigma(j)=j, \sigma
(\Omega_2)= - \Omega_2$ and
$k_\|$ real) or $k_\|=0$ ($\sigma(j)=-j, \sigma(\Omega_2)= - \bar \Omega_2$
and
$k_\perp$ real), which correspond to the pure SU(3) or static SU(2)  conditions,
respectively.

The O4/O8 projections are even more restrictive, since they are only  possible in SU(2) structure. 

The orientifold projections  fix  the relation between $a$ and $b$, and thus the norm
of the pure spinors.
More precisely, $a=b$ for O5/O9, $a=ib$ for O3/O7, and  $a=i \bar{b}$ for  O6, O4 and O8\footnote{The
supersymmetry preserved by orientifolds in type IIA  is generically $a=\bar{b} e^{i\theta}$. The phase
$\theta$ determines the location of the orientifold planes. In the O6 case,  ${\I}(e^{i\theta}
\Omega_3)=0$
at the orientifold. Moreover we can absorb the phase of $b$  by
redefining $\Omega_3$. Here, we take $\theta=\pi/2$ and $b$ real. 
The O4 orientifold is located at ${\R} (e^{i\theta} \Omega_1)=0$, while
the O8 is located at ${\R} (e^{i\theta} \Omega_1 j^2)=0$. The phase
of $b$ can be absorbed in $\Omega_2$. We take $\theta=\pi/2$ and
$b$ pure imaginary.
Something similar
happens with the relation between
$a$ and $b$ for O5 and O7 in a static SU(2) structure and the phase of
$\Omega_2$.
The relation $a=b$ for O5, for example, corresponds to fixing $
(\Omega_2)_{e^1_- e^2_-}$ being real. A relative phase between $a$ and $b$
would rotate $\Omega_2$.}.
Note that all orientifold projections and $\N =1$ supersymmetry require $|a|=|b|= e^{A/2}$. 

In the Table below we  explicitly give the pure spinors for each orientifold projection and their
transformation properties under the involution.

\begin{table}[h]
{\small \begin{center}
{\bf IIA} \\
\medskip
\begin{tabular}{c|c|c}
 & O6 &O4/O8 \\
\hline 
type 0   & 
$\sigma(i e^{B-ij}) = ie^{-B+ij}$  &  \\
type 1 & 
$ \sigma(i \Omega_1  e^{B-ij}) = -i\overline{\Omega}_1 e^{-B-ij}$ & 
$ \sigma( - i \Omega_1  e^{B-ij}) = -i\overline{\Omega}_1 e^{-B-ij}$  \\
type 2 & 
       $ \sigma(\Omega_2  e^{B-ij}) = -\Omega_2 e^{-B+ij}$
&  $ \sigma( \Omega_2  e^{B-ij}) =    \Omega_2 e^{-B+ij}$  \\
type 3 &  
  $ \sigma(\Omega_3  e^{B}) = -\overline{\Omega}_3 e^{-B}$  & \\
\end{tabular} 
\end{center}}
\bigskip

{\small \begin{center}
{\bf IIB}\\
\medskip
\begin{tabular}{c|c|c}
& O3/O7 & O5/O9  \\
\hline 
type 0  & $\sigma(i e^{B-ij}) = ie^{-B-ij}$ & $\sigma(e^{B-ij}) = e^{-B-ij}$  \\
type 1 & $ \sigma(i \Omega_1  e^{B-ij}) = i \Omega_1 e^{-B+ij}$ & 
$ \sigma(\Omega_1  e^{B-ij}) = -\Omega_1 e^{-B+ij}$  \\
type 2 & $ \sigma(\Omega_2  e^{B-ij}) = -\overline{\Omega}_2 e^{-B-ij}$ &
       $ \sigma(i \Omega_2 e^{B-ij}) = i\overline{\Omega}_2 e^{-B-ij}$   \\
type 3 &  $ \sigma(\Omega_3 e^{B}) = -\Omega_3 e^{-B}$ &
          $ \sigma(i\Omega_3  e^{B}) = i\Omega_3 e^{-B}$ \\
\end{tabular}
\caption{\label{ta:pure} Pure spinors for each
orientifold projection. The  pure spinors in the supersymmetry 
equations (\ref{int}), (\ref{nonint}) are the ones on the left hand side (after
multiplication by $ e^A/8$); the table also gives the action of $\sigma$ on them.}  
\end{center}}
\end{table}


\section{Twisted tori and new $\N=1$ Minkowski vacua}

In this Section we analyse in detail the possibilty of constructing $\N=1$ flux vacua compactifying on nilmanifolds or
compact solvmanifolds. As already mentioned in the previous Sections, one of the motivations for considering 
nilmanifolds is that they are all generalised Calabi-Yau \cite{CG}. More precisely six--dimensional
nilmanifolds are  not necessarily complex or symplectic, but they all admit at least one closed pure spinor of a certain type.
Notice also that nilmanifolds are all non Ricci--flat. For solvmanifolds it is not possible to make such a general statement,
but generically they are 
non Ricci--flat. Thus in the absence of fluxes they are not good string backgrounds since they 
do not solve the 10-dimensional supergravity equations of motion.
It is then natural to ask whether it is possible to promote some of them to a string vacuum by turning on fluxes. 

\subsection{General ideas}

We outline here the general strategy and results, while in the next subsections we present the 
solutions. In Appendix \ref{app:eqs} we give  the detailed form of the supersymmetry equations for the  various types of pure spinors and orientifolds. 

For every manifold,
the allowed orientifold planes are those compatible with
the structure constants of the manifold. Under the orientifold action $\sigma$, the globally defined one-forms $e^a$  divide into two sets, 
$e^\alpha_+$, $e^i_-$, according to their transformation properties $\sigma(e^\alpha_+)=e^\alpha_+$, $\sigma(e^i_-)=-e^i_-$. 
In this notation the structure constants can only be of the type $f^+_{++},f^+_{--}, f^-_{+-}$ in order 
to satisfy (\ref{strdef}). The list of orientifolds allowed for every algebra is given in 
Tables \ref{ta:nil} and \ref{ta:solv}.

Then we have to find a pair of compatible 
pure spinors, $(\Phi_+, \Phi_-)$  that are compatible with the orientifold projection 
and solve eqs. (\ref{int}), (\ref{nonint}).
Notice that twisted tori have trivial structure group, so that there are 
4 globally defined nowhere vanishing spinors (in six dimensions), or equivalently six globally defined nowhere 
vanishing vectors. 
We can therefore form many different pairs
$\Phi_+$, $\Phi_-$, such that the structure group of \tts\ is also trivial.  However if we are interested in
$\N=1$ vacua, all we need is to specify  two internal spinors (which can coincide), 
or in other words the \stt\ structure of \tts.  This is also reflected at the level of four-dimensional effective theories.
The reduction of type II supergravity on these manifolds 
should yield  four--dimensional
$\N=8$ gauged effective actions, whose vacua preserve different number
of supersymmetries from $\N=8$ (no twisting) to $\N=1$ according
to the background fluxes or sources one introduces. 
The amount of supersymmetry preserved by the vacuum determines also the number of relevant
pure spinor pairs one has to consider.

Once the supersymmetry variations are satisfied, we still have to solve the Bianchi identities for the fluxes (the equations 
of motion for the fluxes are implied by supersymmetry as shown in Section \ref{N1vacua}).

A delicate issue in finding a compact solution is the role of the warp factor. In that respect we find it more
practical to proceed in two steps. We first look for solutions in the limit of constant warp factor. In this limit the Bianchi identity is not satisfied as a local equation. The topological content of the
BI is simply that the sources are exact in $(d-H\wedge)$--cohomology.
This could be answered by just looking at the hypothetical sources
(orientifold planes allowed by the manifold, plus possibly some branes) and at the complex defined by the structure constants and $H\wedge$, obviously; but it would have nothing to say on the
pure spinors (and hence, metric and B--field) that we actually have.
We decided to impose a stronger condition: in the large volume limit, 
we will pretend that the branes --- and, more controversially, orientifold --- charges get smeared, so that the sources are 
invariant forms in the internal manifolds. Now they have a chance
of being equal to $(d-H\wedge)F$, and this is what we will check. 
More precisely, we demand
\beq \label{intBianchi}
 (d-H\wedge) F_{p-3}\equiv c_i \eta^i = Q_i  (\mathrm{source}) \ {\rm vol}^i 
\eeq
where $c_i$ are constants, $\eta^i$ invariant forms, $Q_i$ is the charge of the 
source (D--branes have $Q>0$, O--planes $Q<0$) and $\rm{vol}$ is 
the volume form along $\eta^i$ whose sign is by convention such that 
$\langle {\rm vol}^i, \mathrm{Im}(\Phi_2)\rangle >0$; see (\ref{tadpolesinglet}). We want ${\rm vol}^i$ to be dual to cycles, so each of them 
should be closed. 
Moreover, ${\rm vol}^i$ have to be decomposable (that is, 
wedges of one--forms) in order to be dual to planes.
In other words, in the large volume
limit, we replace the delta function of the sources by a 
constant contribution.  
Note that in 
the large volume limit the (constant) warp factors on the left and right hand side of (\ref{Im}) cancel ($\Phi_2$ is proportional to $e^A$), while there is an overall factor of $e^{-\phi}$.
In the constant warp factor, constant dilaton solution we take $e^{\phi}=\gs$.

The procedure  will become clearer in examples. Its validity is by no means proven 
in this paper, but it is implicitly followed in much of the literature on the 
subject. 
If the ``global" solution exists, the second non--trivial step is to promote it 
to a ``localized" one. If the sources we had to introduce are
all parallel and along the directions $e^{\alpha}$, it is enough to rescale\footnote{We could have started with a general rescaling 
$e^\alpha_+ = e^{pA} \tilde e^\alpha_+,  \  e^i_- = e^{qA} \tilde e^i_-$. It is not
hard to show that eqs. (\ref{int}), (\ref{nonint}) impose $p=-q=1$ for all orientifolds.}
\beq \label{rescale}
e^\alpha_+ = e^{A} \tilde e^\alpha_+ \, \qquad   e^i_- = e^{-A} \tilde e^i_- \  .
\eeq
This indeed works for the cases where there is a single contribution to (\ref{intBianchi}). 

In the case of multiple intersecting sources, we can use a similar trick, by introducing
one function $A_i$ for each source in (\ref{intBianchi}) such that each vielbein
rescales by an $e^{\pm A_i}$ according to whether it is parallel 
or perpendicular to the source, namely 
\beq \label{rescaleint}
e^m= e^{\sum_i (-1)^{sign_i(m)} A_i}  \ \tilde e^m 
\eeq
where $sign_i(m)$ is $+1$ if the source is along $e^m$, and $-1$ if it is orthogonal \footnote{We could have started again
with more general rescalings, but  eqs. (\ref{int}), (\ref{nonint}) impose (\ref{rescaleint}).}.
The total (four-dimensional) warp factor in this case is given by $e^{2 \sum A_i}$. This trick works well 
for partially smeared intersecting branes in flat space, i.e. when the functions $A_i$ 
depend only on the coordinates orthogonal to all sources.
We call this type of solutions partially local.

The natural question that arises proceeding this way  is whether any 
``global" (unwarped) solution can be promoted to a local (or partially local) solution 
by rescaling. 
It turns out that not all global solutions can be lifted to warped ones by the
 simple rescalings defined above. This does not mean, of course, 
that there are no localized solutions corresponding to the global ones we are 
finding; only that one obvious strategy does not work.
We will discuss it in more detail in Section \ref{sec:globallocal}.

The results of our systematic search are summarized in Table \ref{results}. 
We  explored all possible orientifolds of 
nilmanifolds and compact solvmanifold (see Tables \ref{ta:nil} and  \ref{ta:solv} in Appendix \ref{app:Nil}), namely  
O5 and O7  in IIB, and O6, O4 and O8 in IIA \footnote{More precisely, we looked at all involutions that
act by a sign on the basis in which the classification was given. All other
involutions we have found can be related to these via changes of coordinates; we do not exclude the possibility of having missed other possibilities}.
In all cases we 
considered the two possible pairs of pure spinors allowed by the orientifolding, 
type0-type3 and type1-type2, corresponding
to SU(3) and static SU(2) structures.
We  found that very few manifolds admit solutions, even global ones. Table \ref{results} contains
the list of all the solutions for both nilmanifolds (labelled by $n$) and solvmanifolds (labelled by $s$). 
As already mentioned there are two types of solutions: those that can be made
local by introducing a warp factor and those for which we could not.
We call the latter  global (or smeared) and denote them by an asterisk in
Table \ref{results}.

For nilmanifolds there is only one global solution, 3.14 in IIB. It  
requires two intersecting O5--planes,
but cannot be made partially local by the 
rescalings (\ref{rescaleint}). It corresponds to  Model 1 in Section \ref{global}. All 
other solutions 
for the nilmanifolds can be made local.  These  are 
actually related by T--duality to warped $T^6$ solutions with self-dual 
three-form flux, O3--planes and
D3--branes (usually referred to as ``type B") as discussed in \cite{Schulz, Kachru}. 
For all the T--dual models $dF_3$ is along the Poincar\'e dual of the 
orientifold and thus requires only one orientifold projection. We will describe 
some examples in Section \ref{tdualex}.

Among the solvmanifolds in Table \ref{ta:solv}, five have a truly six--dimensional 
algebra while the others
are defined by direct sums of lower dimensional solvable algebras 
with either trivial directions
or other lower dimensional solvable/nilpotent algebras.
There is no solution for any of the six--dimensional algebras, 
neither global or local. There are instead global solutions for one 
of the direct sums, namely the algebra 2.5.
The corresponding models (2, 3 and 4) all require two sources of opposite charge to cancel
the effective charge of the fluxes. The same as in Model 1, the
sources wrap intersecting cycles and this makes the introduction of the
warping hard.

Before moving to the explicit examples, we would like to make some comments about the supersymmetry of
the models built out of the algebra $s$ 2.5. The associated manifold admits a flat metric.
Flatness implies trivial
holonomy and therefore the existence of a basis of covariantly constant spinors. 
So far this is just like on $T^6$. However, unlike on $T^6$, some of these spinors will be 
covariantly constant but not explicitly constant. The associated pure spinors 
would then  be non--constant. In most of this paper (except when we 
try to promote global solutions to local ones) we consider forms which are
left--invariant, and hence with constant coefficients in the left--invariant basis
$e^a$. For this reason, we will not see all the possible pure spinors on that 
manifold. 
Due to the fact that $s$ 2.5 is flat, there are also models that
do not have fluxes, of the type discussed already in \cite{SS}
(for some realizations in seven dimensions see \cite{Dalpre}), and in which therefore the orientifold projections are 
strictly speaking not needed and their charge 
can be cancelled locally by D7--branes. Hence the amount of 
supersymmetry we see
in these fluxless  models  
is artificially restricted by the non--left--invariance of some
of the spinors. The  fluxless models are not in the Table 3  and
are discussed in \ref{sec:fluxless}.

\begin{table}[h]
\begin{center}
{\bf IIA} \\
\medskip
\begin{tabular}{|c|l|c|c|c|}
\hline
& algebras & \multicolumn{1}{|c|}{O4} &  \multicolumn{2}{|c|}{O6}  \\
\hline
&  & t:12 & t:30 &  t:12 \\
\hline
$n$ 3.5 & (0,0,0,12,13,23) && 456 &\\
$n$ 5.1& (0,0,0,0,0,12+34) & 6 & & \\
$n$ 5.2& (0,0,0,0,0,12) & 6 & &  \\
\hline
$s$  2.5 & (25,-15,$\pm$45,$\mp$35,0,0) &  &\minicent{2.2}{\vspace{.1cm}
(136 + 246)$^*$ \\ (146 + 236)$^*$\vspace{.1cm}
}
&\minicent{2.2}{(136 + 246)$^*$ \\ (146 + 236)$^*$} \\
\hline
\end{tabular}

\vspace{1cm}

{\bf IIB} \\
\medskip

\begin{tabular}{|c|l|c|c|}
\hline
& algebras &\multicolumn{2}{|c|}{O5}  \\
\hline
&  &   t:30 & t:12\\
\hline
$n$ 3.14& (0,0,0,12,23,14-35)  & (45 + 26)$^*$&  \\
$n$ 4.4 & (0,0,0,0,12,14+23)   & 56 & 56 \\
$n$ 4.5 &(0,0,0,0,12,34)  & 56 & 56 \\
$n$ 4.6 & (0,0,0,0,12,13)  & 56 & 56  \\
$n$ 4.7 & (0,0,0,0,13 + 42,14+23)    & 56 & 56 \\
$n$ 5.1& (0,0,0,0,0,12+34)  & 56 & 56  \\
\hline
$s$  2.5 & (25,- 15,$\pm$45,$\mp$35,0,0) &  \minicent{2.2}{\vspace{.1cm}(13 + 24)$^*$ \\ (14 + 23)$^*$\vspace{.1cm}}   & \minicent{2}{\vspace{.1cm}(13 + 24)$^*$ \\ (14 + 23)$^*$\vspace{.1cm}} 
\\
\hline
\end{tabular}
\caption{\label{results} { Summary of solutions. For every solution we give 
the type of the pure spinors and
the cycles where sources should be wrapped to obey BI.
 The asterisk means that the solution is not related by T--duality to O3 solutions on $T^6$.  We have not included in this table fluxless solutions (all the solutions with O7-projections, as well as some extra solutions with O6 projections). }}
\end{center}
\end{table}%

\subsection{Global non T--dual solutions}
\label{global}

In this Section we present the global new  solutions that are not related by T--duality to type B
solutions on $T^6$. 
Even if we were not able to find their local versions 
(it requires intersecting orientifolds and/or branes), it is 
worth considering such solutions because, to the best of our knowledge, they are  the only examples of Minkowski
vacua with internal compact solvmanifolds not obtainable from the standard $T^6$ by T--duality. 

In order to keep the discussion as light as possible, we put the supersymmetry equations and the general form 
of the pure spinors in Appendix \ref{app:eqs}. Here we simply give the form of the solutions. 
We organize them by models according to the algebras defining the manifolds.
Some solutions come in copies related by symmetries of the structure constants, like for example an exchange
 between 1 and 2 in the algebra $s$ 2.5. For these cases, we give 
only one solution.  
The algebra $s$ 2.5 admits solutions with orientifold projections of different dimensionalities, 
while the others admit only one type of orientifold projection.

\subsubsection{Model 1 ($n$ 3.14):  O5 orientifolds for  SU(3) structure}
\label{sec05}

\noindent

The supersymmetry condition  (\ref{int}) requires  that every 
type 3--type 0 solution in IIB theory should have an 
integrable  complex structure. From the equations (\ref{03O5}) we see that the pure spinor associated to such a complex structure is given by $e^A \Omega_3$. The compatible pure spinor corresponds to a
non integrable purely symplectic structure  defined by $J$. 
The three-form $F_3$ is the only non-trivial RR flux allowed by 
SU(3) structure together with the O5 projection, consistently with the results of \cite{gmpt}
(there can also be an exact H-field, $H=dB$, carrying no flux).
As usual, the equation for $F_3$ implies that its equation of motion,  (\ref{bianchi}), is
automatically satisfied. The Bianchi identity, on the contrary, has to be imposed. 
For the $F_3$ given by (\ref{03O5}), (\ref{bianchi}) reads
\bea \label{BianchiO5}
\gs dF_3=2 i \del \bar \del (e^{-2A} J) = g_s (\delta(D5)-\delta(O5))  \ , 
\eea 
where we have used the primitivity of $dJ$, guaranteed by the condition $dJ^2=0$.
Further  details about Bianchi identities 
for the case of O5 wrapping $T^2$ fibrations over a four--dimensional base are given in 
\cite{Schulz}.

As explained at the beginning of this Section, we search for solutions with constant warp factor, i.e. pairs $\tilde J, \tilde \Omega$ that satisfy (\ref{03O5}) for
$\phi$ and $A$ constant, and that satisfy the integrated Bianchi identities (\ref{intBianchi}) with
O5--planes and possibly  D5--branes as sources.

We found only one solution of this type: the model is 3.14 in Table \ref{ta:nil} and has structure 
constants (0,0,0,12,23,14-35). This corresponds to $M_6$ being an iterated fibration 
\[
\xymatrix{ 
S^1_{\{6\}}\ \ar@{^{(}->}[r] &M_6\ar[d]\\
T^2_{\{4,5\}}\ \ar@{^{(}->}[r] &M_5\ar[d]\\
& T^3_{\{1,2,3\}}
}
\]
where the indices denote the directions that span the various tori. 
We choose the 
orientifold to be along the directions 4 and 5. The holomorphic 3-form, the symplectic two-form and the RR flux are \footnote{This is not the most general solution. We have already imposed the consistency of the pure spinors with the second orientifold projection 
required by tadpole cancellation as discussed later.} (we omit tildes to simplify
notation, but all $e$'s are unscaled):
\bea \label{mod1}
 \Omega_3&=&(e^1 - i e^3) \wedge (e^2 +  i \tau e^6) \wedge (e^4 + i e^5)\, , \nn \\
J 
&=& - t_1 e^1 \wedge e^3 +  t_2 \tau_r  e^2 \wedge e^6 +  t_3 e^4 \wedge e^5 \, ,\nn\\ 
\gs F_3&=& -(\tau_i e^2 - |\tau|^2 e^6) \wedge \left(t_2 (e^1 \wedge e^4-e^3 \wedge e^5) + 
\frac{t_3}{\tau_r} (e^1 \wedge e^5+ e^3 \wedge e^4) \right) \, .\label{eq:model1}
\eea
Here $\tau=\tau_r + i \tau_i$ is the only complex structure modulus left, and $t_1,t_2,t_3$ are the surviving 
K\"ahler moduli. 
The metric defined by $J$ and $\Omega_3$ is positive definite if $t_i>0$. 
Notice that we have omitted a modulus
corresponding to the overall volume, that would have appeared
in the $\Omega$ and $J$ above multiplied by a power of $1/2$ and
$1/3$ respectively. Such a parameter would not affect the analysis
below, and in particular nothing would fix it, and we would be
able to make it large, consistently with the use of supergravity
and with the approximation of using of constant coefficients and smeared source, as discussed above.

To see that we need a second orientifold plane, it is enough to 
look at the derivative of $F_3$
\bea \label{mod1F3}
dF_3 &=& 
-2 \, \frac{|\tau|^2}{\gs} \left( \frac{t_3}{\tau_r} e^1 \wedge e^2 \wedge e^3 \wedge e^6 - 
 t_2 \, e^1 \wedge e^3 \wedge e^4 \wedge e^5 \right) \nn \\
\label{dfmodel1}& =& -2 \, \frac{|\tau|^2}{\gs} 
 \left(\frac{t_3}{\tau_r^2 t_1 t_2} \  {\rm vol1}_{-} +  \frac{t_2}{ t_1 t_3}\  \rm{vol2}_{-}
 \right) 
 \, ,
\eea
where $\rm{vol 1}_-$, $\rm{vol 2}_-$ denote the volume forms orthogonal to 45 and 26, 
normalized so that 
$\langle {\rm vol}i_-,{\rm Im}\Phi_+\rangle={\rm vol}i_-\wedge J=+{\rm vol}$ (see
comment after (\ref{intBianchi}) and (\ref{tadpolesinglet})).
Comparing to (\ref{intBianchi}) we see that  
both contributions have to equal the charges of orientifold planes 
wrapping  45 and 26 since the $t_i$ cannot be negative 
(a positive definite metric requires $t_i >0$). These two orientifold planes are
obtained by a projection of the theory by the group generated by 
$\{\Omega_{\rm{WS}} \, \sigma_1,\sigma_1\sigma_2\}$, with $\sigma_{1,2}$ the 
reflections along
$1236$ and $1345$ respectively.  
Both projections are allowed by the structure constants. The integrated Bianchi 
identities 
then fix $t_i$ in terms of $\tau$.
In the absence of branes, we get 
$t_1=\frac{1}{16 \,  \gs}\frac{|\tau|^2}{|\tau_r|}, 
t_2=4 \, \sqrt{\gs} \frac{1 }{|\tau|}, t_3= 4 \, \sqrt{\gs} \frac{|\tau_r|}{|\tau|}$, 
where $\tau$ is free.

Unfortunately this ``global" solution needs intersecting orientifolds, and 
this makes it difficult to introduce
the warp factor. Even the possibility of "smearing the orientifolds" 
in  the 26 and 45 directions 
and performing the rescalings (\ref{rescaleint}) 
is excluded since it is not compatible with the local Bianchi 
identities. We return to this issue
in Section \ref{sec:globallocal}.


The model is non T--dualizable to an O3 because $\del_4$ and $\del_5$ are not
isometries of the metric. (The general criterion for a direction $x$ 
to be an isometry is that it should not appear as a lower index in a structure 
constant, $f^a{}_{xb} = 0$ for all $a,b$.) 
Another way to see that we do not get this model by T--duality of a $T^6$ is that this manifold has $b_1=3$,
 while by performing two T--dualities on a $T^6$ with an $H$ flux that has at most 
one leg on the T--dual directions
we get models with $b_1\ge4$.

To conclude our discussion of this model, notice that the solution (\ref{mod1}), (\ref{mod1F3}) seems to have an apparent isometry in the direction 6. So it is natural to ask if we may T--dualize and what the result of the dualization would be
\footnote{In order to perform T--duality 
we need to  choose the modulus $\tau$ to be real. Otherwise, we are forced to redefine $\tilde e^2=e^2 - \tau_i e^6$ and 
$\tilde e^6=e^6$. The resulting change  in the structure constants would lead to the loss of the isometry}.
One would naively get an O6 along 456 and an O4 along 2, on a manifold
whose corresponding algebra would be (after a redefinition) $n$ 4.6. The projections for these O--planes are
$\Omega_{\rm WS} (-)^{F_L} \sigma_{123}$ and
 $\Omega \sigma_{13456}$ respectively. The product of the two is 
$(-)^{F_L}\sigma_{1456}$. This is not an usual orbifold, in that
it treats left-- and right--moving fermions asymmetrically. 
The non--geometric construction of \cite{hmw} is a resolution of
a similar orbifold. One would expect therefore to need some
conceptual change to the framework used in this paper (which is 
fully geometric) to describe such a configuration, which we expect
to be a valid string--theory background because it arises as T--dual
to a valid vacuum. If one tries to proceed in spite of this, one
finds a model with pure spinors of type 1--type 2, that indeed 
have the appropriate $F$ to account for sources along the O6 and
O4 detailed above. 
However, on the fixed locus of $(-)^{F_L}\sigma_{1456}$ one expects a source for $H$, and this is not 
the case. It is interesting, nevertheless, that the geometrical approach ``almost works" also in this case; it seems to indicate
that the pure spinor approach might be modified to account for some
non--geometric vacua as well. Another hint in the same 
direction will be described in Section \ref{sec:ng}.

This is the only solution for O5--planes and SU(3) structure.
Before moving on to the other solutions (on solvmanifolds), it may be useful to show how
the other  constructions of this type fail.
There are  17 six--dimensional nilmanifolds
with some nonzero structure constants that admit a closed type 3 
spinor $\tilde \Omega_3$ (see Fig. \ref{fig:nil}).
For all these, one can find an integrable complex structure 
compatible with at least one of the possible O5 projections, i.e. a closed $\tilde \Omega_3$ of the form
(\ref{Om}) but 
only 8 out of the 17 admit a compatible $\tilde J$ that satisfies  $d \tilde J^2=0$.
The last equation in (\ref{03O5}) defines $F_3$. 
For for all these model $dF_3$ is purely along 
$e^1_+ \wedge e^2_+ \wedge e^i_- \wedge e^j_-$ and it has the sign of 
a D--brane charge. So an extra orientifold is needed in the directions
(+,+,-,-).  
Such orientifold planes would be supersymmetric
with the original one (since they would have four mutually orthogonal directions), 
but in all cases the structure constants are such that
they do not allow this second projection. 
Therefore, there is no possible solution to the Bianchi identities. 
Notice that 3.14 is the only example that does not 
show this feature.

From the general form of $\Omega_3$ and $J$ we see that for models with SU(3) 
structure and O5--planes there are 5 complex structure moduli:
the constants $\tau^+, \tau^1,.., \tau^4$ in the expression in (\ref{Om}),
and 5 real K\"ahler moduli, the  $t_i$ and $b$ in (\ref{Je}). 
Demanding integrability of the complex structure fixes one, two or at best 
three of them. Similarly the equation $d \tilde J^2=0$, whenever it has 
solutions, fixes at most two (real) K\"ahler moduli.


\medskip
\noindent
\subsubsection{Model 2 ($s$ 2.5): O5 orientifolds for static SU(2)}

This is the first model of a series that uses the same group, corresponding to the algebra $s$ 2.5. We know that
this group can be made compact by a choice of a $\Gamma$ because of the criterion  
of ref. \cite{saito}, as reviewed in Section 2. A priori, it is guaranteed that this 
lattice exists, but not that it is sent into itself by the orientifold action. 
However  by writing down an explicit representation for the algebra 
we can explicitly give the lattice (with integer coefficients) and check that it is left invariant 
by all the orientifold projections we consider in this paper.

For one possible quotient $G/\Gamma$, the topology of the manifold is $S^1_{\{6\}}
\times M_5$,
where $M^5$ is a $T^2_{\{1,2\}}\times T^2_{\{3,4\}}$--fibration 
over $S^1_{\{5\}}$; the fibration is defined by each of the two $T^2$ 
identified with itself up to a $\pi/2$ rotation after $x^5\to x^5+1$. 
This can be thought of as an orbifold of $T^6$ in which one quotients
by a shift by one quarter of the period in direction 5 along with the
rotation just described in directions 1234.

\medskip
\noindent
We will start with a static SU(2) structure in IIB. 
The supersymmetry condition  (\ref{int}) requires  that the manifold should have 
a hybrid  integrable structure with one complex dimension and four (real) symplectic ones.

Model 2 corresponds to 2.5 in Table \ref{ta:solv} with $\alpha=1$\footnote{This algebra was also obtained by Villadoro and Zwirner from a gauged supergravity analysis based on \cite{VZ, Dere, VZ2}: it realizes Scherk-Schwarz compactifications from 5 to 4
dimensions that break $\N=4$ to $\N=2$ (which becomes $\N=1$ after the orientifold),
corresponding to a twist in the diagonal subgroup of $SO(2) \times SO(2) \subset SO(5) \sim
USp(4)$, and to a consistent ${\cal N}=4$ gauging. We thank them for pointing it to us.}: (25,-15,45,-35,0,0)
and an orientifold in the directions 13.
The pure spinors  (\ref{puresu2}) compatible with the orientifold projection 
are built out of
\bea \label{solmodel1}
\Omega_2&\equiv& z^1 \wedge z^2 = (e^1 + i (-\tau^2_2 e^2 + \tau^1_2 e^4 + \tau^1_3 e^5)) \wedge
(e^3 + i (\tau^2_2 e^4 + \tau^2_3 e^5 + (2 \frac{b}{t_2} \tau^2_2  + \frac{t_1}{t_2} \tau^1_2)e^2)) \nn \\ 
\Omega_1&\equiv& z^3= \tau^-_3 e^5 + \tau^-_4 e^6 \nn \\
j&=& \frac{i}{2} \left( t_1  z^1 \wedge \bar z^1 + t_2 z^2 \wedge \bar z^2 + b ( z^1 \wedge \bar z^2 - \bar z^1 \wedge z^2) \right)  
\eea
where $\tau^i_j, t_1, t_2$ and $b$ are real, while $\tau^-_{3,4}$ are complex.
There is no  NS flux in the solution. The RR three-form flux  has a long and not very illuminating expression, so we only give it for a  special choice of moduli below.  On the contrary  its exterior derivative is quite simple
\beq \label{tad25O521}
dF_3= -2 \frac{|\tau^-_4|^2}{\gs} \frac{t_1 (\tau^1_2)^2 + t_2 ((\tau^2_2)^2-1)+2 b \tau^1_2 \tau^2_2  }{t_2\Im (\tau^-_3 \bar \tau^-_4)} (e^1 \wedge e^3 \wedge e^5 \wedge e^6 - e^2 \wedge e^4 \wedge e^5\wedge e^6) \ .
\eeq
The two  contributions need to be matched by sources wrapping the directions 13, as well as
24. We can a priori have orientifolds in both directions\footnote{In order to apply consistently a second ${\mathbb Z}_2$ projection in 1356 (to have an orientifold fixed plane in 24) 
the moduli would need to be restricted further. The particular solution in (\ref{parto5}) 
allows for this second ${\mathbb Z}_2$.}. 
Wedging with $\Im \Phi_+$ as in (\ref{tadpolesinglet}) to obtain the singlet in the flux effective charge, we get indeed a positive quantity, which means that the solution needs orientifold planes. 
As for the individual contributions in (\ref{tad25O521}), 
it is hard to see
what their sign is for a general choice of moduli such that the metric 
defined by
$\Phi_+, \Phi_-$ is positive definite. (It is easy to write down 
explicitly the inequalities that define the open set in which the signature
is positive by looking at its characteristic polynomial, but the expressions
are complicated and the correlation with the sign of the charges is not clear.)
However, we can make an easy choice
\beq \label{parto5}
\tau^1_3=\tau^2_2=\tau^2_3=0 \ , \qquad \tau^-_3=1 \ , \ \tau^-_4=i  \ , \ b=0 \ ,
\eeq
while we leave $\tau^1_2$ free, and $t_i>0$ for the positivity of the metric. In this case, 
the metric is diagonal:
\beq
g={\rm diag}(t_1,\ \frac{t_1^2}{t_2}(\tau_2^1)^2,\ t_2,\ t_1(\tau_2^1)^2, \ 1, \ 1) \ .
\eeq
and  is clearly positive definite. For this choice, the flux and its tadpole read
\bea \label{BImodel2}
F_3&=& -\frac{1}{\gs}(1-\frac{t_1}{t_2}(\tau^1_2)^2) (e^1 \wedge e^4 \wedge e^6 + e^2 \wedge e^3 \wedge e^6) \nn \\ 
dF_3&=& -\frac{2}{\gs} (1-\frac{t_1}{t_2}(\tau^1_2)^2)  (e^1 \wedge e^3 \wedge e^5 \wedge e^6 - e^2 \wedge e^4 \wedge e^5\wedge e^6)
\eea
Wedging the individual terms with $\Im \Phi_+$ we get that both are proportional to the volume,
with a constant of proportionality equal to  
 $\frac{2}{\gs t_1 t_2} (1-A)$ for the first term, and
$-\frac{2}{\gs t_1 t_2} \frac{1-A}A$ for the second, where 
$A=\frac{t_1}{t_2}(\tau_2^1)^2$ . While the sign of each depends on 
the sign of $1-A$, we see clearly that they have opposite signs, and one of them 
is matched by O--plane charge (the first one, if $1-A$ is positive) while the 
other is cancelled by D--brane charge. We also see  that in any case the total 
charge of the sources should be negative.  Imposing charge quantization will 
this time fix the ratio $t_1/t_2$.


Notice also that for $\frac{t_1}{t_2} (\tau_2^1)^2=1$ the manifold is flat 
with the metric given above. In this case, there is no  $F_3$ flux, and the O5 charge has to be cancelled
by D5--branes on top of it. We will say more about this in 
Section \ref{sec:fluxless}.

\subsubsection{Model 3 ($s$ 2.5): O6 orientifold with SU(3) structure}
The same algebra we used for Model 2 admits a solution with O6--planes in 136.
This model is actually T--dual to the previous model, but we include
it because it is the only SU(3)--structure O6 we have found. (This model has also been found in \cite{Camara}.) 
 
The pure spinors are given in terms of the forms
\bea \label{Om6}
\Omega_3&=&  (e^{2} + 
 i (\tau^{1}_1 e^{1} +\tau^{1}_2 e^{3} +   \tau^{1}_3 e^{6} )) \wedge (e^{4} + 
 i (-\tau^{1}_1 e^{3} +\tau^{2}_3 e^{6} + (  \frac{t_1}{t_2} \tau^1_2
  -2  \frac{b_3}{t_2} \tau^{1}_1 ) e^{1} )   ) \wedge 
 (e^{5} + i \tau^3_3 e^{6})  \nn\\
 & \equiv& z^1 \wedge z^2 \wedge 
 z^3  \nn \\
J&=& \frac{i}{2}  \sum_{i=1}^{3}  t_{i} \ z^{i} \wedge  {\bar z}^{i}  +  \I  \left(  b_1 \  z^{2} \wedge  {\bar z}^{3} + b_2   \  z^{3} \wedge {\bar  z}^{1} + b_3 \  z^1 \wedge {\bar z}^2  \right)
 \ , 
\eea
where $b_1 = - \frac{1}{\tau_3^3} ( \tau^1_3 b_3 + \tau^2_3 t_2)$ and 
 $b_2 =  \frac{1}{\tau_3^3}( \tau^1_3 t_1 + \tau^2_3 b_3)$.  
 The $H$-flux is zero, as it should be for SU(3) structure.
 As in the previous example
 the general solution for the  RR two-form is too long to be shown. Its exterior derivative is 
 \beq
 d F_2 = -\frac{2 (\tau^3_3)^2 }{\gs} \frac{(\tau^1_2)^2 t_1 + ((\tau^1_1)^2 -1) t_2 -2 \tau^1_1  \tau^1_2 b_3  }{t_2 ((\tau^1_3)^2 t_1 + (\tau^2_3)^2 t_2 - (\tau^3_3)^2 t_3 + 2 \tau^1_3 \tau^2_3 b_3)} ( e^2 \wedge e^4 \wedge e^5 - e^1 \wedge e^3 \wedge e^5) \ .
 \eeq
 This should be matched by sources wrapping cycles in the directions 136 and 246. 
 The net effective charge of the fluxes is obtained wedging with ${\rm Im} \Omega_3$ and it is easy to check that it has to be cancelled by orientifolds planes. In this case we can also determine
 the sign of the individual contributions and it turns out that they have opposite sign, i.e. one need orientifolds, while the other needs D--branes. Which one is which depends on 
 the relative values of the moduli. Note however that if we needed orientifolds
 wrapped in 246, we would need to constrain the moduli further, such that
 the pure spinors transform appropriately under a reflection in 135. This would require $\tau^1_3 = \tau^2_3 = 0$. 
 
 We can also find a particularly simple solution 
\beq 
 \tau^1_1 = \tau^2_3 = \tau^1_3 =\tau^2_2 = 0, \quad  \tau^3_3 =1 \, \qquad  \qquad  b_1 =b_2 =b_3 =0
  \eeq
leaving $\tau^1_2$ free. The metric is diagonal
\beq  
g={\rm diag}(\frac{t_1^2}{ t_2}(\tau^1_2)^2 , t_1, t_1 (\tau^1_2)^2, t_2, t_3,t_3) \, ,
 \eeq
 and we leave $t_i$ free. Here $F_2$ has a simple expression
 \beq 
 F_2 = - \frac{1}{\gs }\frac{t_1 (\tau^1_2)^2 -t_2}{ t_2 t_3} ( e^1 \wedge e^4 + e^2 \wedge e^3) \, .
 \eeq
For $(\tau^1_2)^2 t_1 =t_2$ (and any value of $t_3$) the flux is zero and the manifold is flat.


We obtain a similar solution for O6 wrapping 146 (which, in order to cancel tadpoles, needs 
D--branes wrapping 236).

\subsubsection{Model 4 ($s$ 2.5): O6 orientifolds for static SU(2) structure  }

The solvable algebra 2.5 admits also a type 1-type 2 solution with an O6 along the directions 136.
The pure spinors are built out of
\bea
\Omega_1 &\equiv& z^3=\tau^-_3 e^5 + i \tau^+_3 e^6 \nn \\
\Omega_2&\equiv& z^1 \wedge z^2= ( \tau^1_1  e^1 + \tau^1_2  e^3 ) \wedge 
 ( \tau^2_1 e^2 + \frac{\tau^1_2}{\tau^1_1} \tau^2_1 e^4 + \tau^2_3 e^5)  \nn \\
j &=&  \frac{i}{2} \left( t_1  z^1 \wedge \bar z^1 + t_2 z^2 \wedge \bar z^2  \right) 
\eea
where $\tau^{\pm}_3$ are real and  $\tau^i_j$ are complex.
The NSNS flux and the exterior derivative of the RR 2-form flux are 
\bea
H&=& h_{12} e^1 \wedge e^2 \wedge e^6 - h_{21} e^3 \wedge e^4 \wedge e^6+
  h_{32} e^1 \wedge e^5 \wedge e^6 + \nn \\
&& \frac{1}{2} \left( \frac{\tau^1_2}{\tau^1_1} h_{12} -  \frac{\tau^1_1}{\tau^1_2} h_{21}\right) 
(e^1 \wedge e^4 \wedge e^6 - e^2 \wedge e^3 \wedge e^6) -  \nn \\
&& \frac{1}{2} 
\left(\frac{\tau^1_2 \tau^2_3}{\tau^1_1 \tau^2_1} h_{12} + \frac{\tau^1_1 \tau^2_3}{\tau^1_2 \tau^2_1}
h_{21}  - 2 \frac{\tau^1_2 }{\tau^1_1 } h_{32}  \right)
e^3 \wedge e^5 \wedge e^6  \nn \\
dF_2&=&2\frac{ |\tau^1_1|^2 t_1 - |\tau^1_2|^2 t_2}{\gs} \,  \frac{ {\Im} (\tau^1_1 \bar \tau^1_2)}{ |\tau^1_1|^2 \tau^+_3} \, 
(e^1 \wedge e^3 \wedge e^5 - e^2 \wedge e^4 \wedge e^5)
\eea 
Since $F_0=0$, $dF_2$ has to be fully cancelled by sources wrapping 
the cycles 246 and 136. It is not hard to see that when wedging the individual terms 
in $dF_2$ with ${\Im} \Phi_-$, they have opposite signs. One is therefore cancelled by 
D--branes, while the other by O--plane charges.  
One more time, the net
charge of the sources needs to be negative, verifying the no--go theorem.
We give again a particular solution for which $H$ is real and 
the metric is diagonal, and the fluxes have short expressions: 
\beq
\tau^1_1=\tau^2_1=1 \ , \quad \tau^1_2= i \ , \quad \tau^2_3=0\ , 
\quad h_{21}=-h_{12}\ , \quad h_{32}=0 \ ,\label{ez4}
\eeq
in which case the metric reads
\beq
g={\rm diag}(t_1,t_2, t_1,t_2,(\tau_3^-)^2,(\tau_3^+)^2) \ .
\eeq
For the solution, the NS and RR fluxes are
\bea
H&=& h_{12} (e^1 \wedge e^2 \wedge e^6 + e^3 \wedge e^4 \wedge e^6) \ , \nn \\
F_2&=& \frac{1}{\gs} \left( \frac{h_{12} }{\tau^+_3} \left( e^1 \wedge e^2 + e^3 \wedge e^4 \right)
- \frac{t_1 -  t_2}{\tau^-_3}  \left( e^1 \wedge e^4 + e^2 \wedge e^3 \right) \right)  \ .
\eea
The exterior derivative of $F_2$ is
\beq
dF_2 = - 2\frac{t_1 - t_2}{\gs \, \tau^-_3} \,  (e^1 \wedge e^3 \wedge e^5 - e^2 \wedge e^4 \wedge e^5) \ .
\eeq
The sign of each contribution depends on the sign of $t_1-t_2$. If 
$t_2>t_1$, then the source is matched by O6--planes wrapping 246, and D6--branes 
wrapping 136, and the 
converse for $t_2<t_1$.
The case $t_1=t_2$ corresponds again to a flat metric. We also have
a D4--charge induced by $H$ along the direction 5, proportional to $h_{12}^2/(g_s (\tau^+_3)^2 t_1 t_2)$.

We obtain a similar solution for O6 wrapping 146 (which needs D6--branes in 236 to 
match the effective flux charge).

Finally, a puzzle seems to arise about applying T--duality to this model. 
Even in the particular case (\ref{ez4}), the T--dual algebra would have structure
constants (25,-15,45,-35,0,12+34). The algebra is obviously solvable, yet we could not find it (or anything related by change of coordinates)  in the 
classification \cite{Mubara5}.  Although it gives rise to a  model which  seems to be 
perfectly sensible; we do not write  down the solution.\footnote{For $h_{12}=0$ this puzzle does not arise.  Since $H$ vanishes, the T--duality does not change the topology of the  internal space, and  the dual manifold is still given by $s$ 2.5. The T--dual  solution has  $SU(3)$ structure  with 05--planes. This solution was missing in the earlier versions and is discussed in \cite{david}. We have updated Table 3 to include it, and do not discuss it further.}

\subsubsection{Fluxless models}
\label{sec:fluxless}

As we have already remarked, some of the solvable algebras admit a flat metric. 
It is hence possible to have compactifications without any flux at all. 
We have already seen that each of the models found on the algebra $s$ 2.5
 included a flat limit possibility. There are a few other such 
examples that one can build. As we reviewed earlier, on a flat manifold there
is a complete basis of covariantly constant spinors. Since in this paper
we are considering left--invariant forms only (for global solutions) we
should restrict our attention to spinors which are also constant. For the 
algebra $s$ 2.5, it
turns out that there is a two--dimensional space of constant, covariantly constant O(6) spinors
of a given chirality. Now, consider one of them,  $\eta^1$, and build from it a ten--dimensional
supersymmetry parameter $\epsilon^1$ (see the decomposition (\ref{eq:spinor})). 
The orientifold projection will determine 
now a second supersymmetry parameter $\epsilon^2$, with an internal spinor
$\eta^2$. This spinor, however, will not necessarily be covariantly constant, 
even if a basis of such spinors exists at every point, because it might
lack the appropriate explicit coordinate dependence to cancel with the 
spin connection term. 
Hence in our analysis of fluxless solutions with orientifolds on the algebra s 2.5 
we can find anywhere from $\N=2$
to no supersymmetry at all. 

Some supersymmetric examples we have seen already in the previous 
subsections (all based on the algebra $s 2.5$), arising at particular points
in the moduli space of solutions where the flux goes to zero, and the manifold 
becomes flat. There are
a few more O6 solutions with no flux (for example, $s 2.5$ has a solution with 
no flux  with an orientifold in 125, which is not obtained as a limit of a 
solution with flux).
$s 2.5$, however, is not the only algebra whose associated manifold
is Ricci-flat
with the metric equal to the identity in the basis $e^a$. 
The algebras  2.4 and 4.1 of Table \ref{ta:solv} also share this property. 
  These  are built out of 
the only three--dimensional compact solvable algebra: (23,-13,0) 
(which gives rise to a
Ricci--flat manifold for the identity metric).  2.4 consists of two copies of
this algebra, while 4.1 is a direct sum of this algebra plus 3 
trivial generators. 
However, none of the possible compatible pairs of closed pure spinors for 4.1
transforms in the right way under the allowed O5, O6 or O7 projections.  
On the contrary,
2.4 admits no O5 solution (and obviously no O6, since the structure 
constants do not
allow for an involution corresponding to an O6), but it does admit an 
SU(3)  solution with O7--planes wrapped in 1236.

The O7 projection is special though, since
for  the SU(3) case, the 
constant warping solution requires that $F=0$. Hence
a flat metric is the only possibility. 
This implies that none of the nilmanifolds can be a SU(3) solution with O7--planes, while
solvmanifolds can. Since there is no flux,  
strictly speaking there is no need for O7 to cancel tadpoles. If we do quotient by an O7 projection, we
can then just add 4 D7's on top to cancel the tadpole locally, so that 
no RR field--strength is required. 
  

We give here the O7 solution for
the algebra 2.5, for $\alpha=1$\footnote{The family of algebras in 2.5 gives rise for any $\alpha\in \Bbb Z$ to Ricci--flat compact solvmanifolds for the identity metric. However, only $\alpha=\pm1$ admits O7 solutions, where the orientifold planes are wrapped either in 1256 or 3456.}, 
and orientifolds wrapping 1256. The pure spinors for the SU(3) case are built out of
\bea
\Omega_3 &\equiv& z^1 \wedge z^2 \wedge z^3= (e^5 + \tau^1_2 (e^2 + i e^1))
\wedge (e^6 -  \tau^1_2  \frac{b}{t_2} (e^2 + i e^1))  \wedge (e^3 + i e^4) \nn \\
J &=&  \frac{i}{2} \left( t_{1} \ z^{1}\wedge  {\bar z}^{1} +  t_2 \ z^{2} \wedge  {\bar z}^{2}+ b \  z^{1} \wedge  {\bar z}^{2} - \bar b \ {\bar z}^{1} \wedge  z^{2} + t_3 \  z^3 \wedge  {\bar z}^{3 } \right) \ ,
\eea
where $\tau^1_2$ and $b$ are complex, while $t_i$ are real.
$J$ and $\Omega_3$ are closed, and compatible with an O7 projection reflecting 
the directions 3 and 4. 
As usual, not any set of moduli gives rise to a positive definite metric. We need 
to take  $t_3 > 0$, and $t_1 t_2 - |b|^2>0$. Once $t_3$ is greater than zero, 
the latter is implied by  the
normalization condition  (\ref{eq:eqnorms}), which 
requires $t_3 (t_1 t_2 -|b|^2)=1$. Finally, to see that this model is 
$\N=2$ as predicted by the
general discussion above, one notices that sending $e^2+i e^1\to e^2-i e^1$ in 
both $z^1$ and $z^2$, conjugating $z^3$, 
and changing $\tau_2^1\to \bar\tau_2^1$, one gets
a new pair without affecting the metric.

We will now also show that there are O7's with SU(2) structure on the same manifold, $s 2.5$. 
Take as before $\alpha=1$, and the orientifold wrapping 1256.
As in the type 3 -- type 0 case, the constant warping solution has no flux 
(a priori the type 1 -- type 2 case allows for $H$, but demanding $H$ to be 
 closed, compatible with the orientifold projection and the pure spinors to be 
 $d-H$ closed  sets $H=0$ for this algebra).
 Therefore, the pure spinors have to be closed. They are  given by those in 
Table \ref{ta:pure} with 
 \bea
 \Omega_1 & \equiv& z^3 = \tau^+_3  e^{5} + \tau^+_4 e^6 \ ,  \nn \\
\Omega_2&\equiv & z^1 \wedge z^2=  ( e^{3} + i (\tau^1_1 e^{1} +\tau^1_2 e^{2} +
\tau^1_3 e^{5} )) \wedge ( e^{4} +  i (-\tau^1_2 e^{1} + \tau^1_1 e^2 + \tau^2_3 e^5))  \ ,\nn \\
j &=&\frac{ i}{2}  t  \left(z^1 \wedge \bar z^1 +  z^2 \wedge \bar z^2  \right) \ ,
 \eea
where $\tau_{3,4}^+$ are complex, while all the rest are real. Choosing 
for simplicity
$\tau^+_3$ real, $\tau^+_4$ pure imaginary, the metric defined by the
pure spinors is positive definite for any $t>0$. The normalization condition
(\ref{eq:eqnorms}), which for the general O7 pure spinors
given in (\ref{12O7}) requires $(t_1 t_2 - |b|^2)=1$, fixes in this case $t=1$.  

Once again, this example has $\N=2$ supersymmetry: one can obtain a new
pair which yields the same metric by sending $\tau_i^j\to -\tau_i^j$.

To see that these examples are not the only ones, one can for example
T--dualize along the direction 6. This gives rise to two fluxless solutions 
(a type 3-0 and a type 1-2) again on the algebra s 2.5 with O6 orientifolds wrapping 125
\footnote{Starting from the similar solution with O7--planes in 3456, we get by T--duality other O6 fluxless solutions with orientifolds in 345.}. We
 do not give their explicit form here. Just note that unlike the other fluxless solutions
 with O6 orientifolds on $s$ 2.5, these do not arise as special points in moduli space of
 solutions with flux. They also have no RR field strength
switched on, and need branes parallel to the orientifold planes to cancel their
charge, as usual.

\subsection{From global to local}
 \label{sec:globallocal}

We consider now the problems that may arise when trying to promote a global solution
into a local one. We illustrate these with the case of O5 orientifolds in SU(3) structure, and a single type of source. We also discuss briefly the case of multiple sources.

For an SU(3) structure and a single warp factor $e^{2A}$,
if a (unrescaled) pair $\tilde J, \tilde \Omega_3$
is a solution of the O5 equations 
in the limit of constant warping, then the rescaled $\Omega_3$ would solve
\beq
d (e^A \Omega_3) = 0
\eeq
for any function $A$, with the one-forms rescaled
as in (\ref{rescale}).
On the other hand, given $d \tilde J^2=0$, the
 condition $dJ^2=0$ is not automatic, since $(J_{--})^2$ and $J_{++}J_{--}$
scale differently with $e^A$. In order for an ``unwarped solution" to be promoted to
a full solution, we need the stronger requirement
\beq \label{req1}
d(\tilde J_{++} \wedge \tilde J_{--})=0 \ ,
\eeq
(as well as $d (\tilde J_{--})^2=0$).
Inserting the expression for $F_3$ in terms of $dJ$ (eq. (\ref{03O5})) in the Bianchi
identity imposes the extra constraint
\beq \label{req2}
d \tilde J_{--}=0 \ .
\eeq
If this is not satisfied, then the Bianchi identity for $F_3$ gets contributions along
$e^+ e^- e^- e^-$ directions, which cannot be cancelled by any supersymmetric source.
Imposing these strong requirements, the Bianchi identity for $F_3$ (\ref{BianchiO5}) reduces to
\beq \label{b05}
\gs dF_3= \tilde \nabla^2_- (e^{-4A})
\widetilde{\rm{vol}} _-+  2 i  \del \bar \del  (\tilde J_{++})
=  g_s \sum_i Q_i \frac{\delta(x_--x_-^i)}{\sqrt{\tilde g}_-}  \ \widetilde{\rm{vol}}_-
\eeq
where $\tilde \nabla^2_-$ denotes a Laplacian along the unwarped $\tilde e^-$ directions,
$\sum_i Q_i = 2N_{D5}-32$ is twice the charge of $N_{D5}$ branes and
16 O5--planes (the factor of 2 arises because the orientifold has half the
volume of the original torus)
and
\bea
\widetilde{\rm{ vol}}_-&=& \sqrt{\tilde g_-} \, \tilde e^{1}_-\wedge \tilde e^{2}_-\wedge \tilde e^{3}_-\wedge \tilde
e^{4}_- =   (t_1 t_2 -|b|^2)  (\tau^2_r \tau^3_r - \tau^1_r \tau^4_r) \
\tilde e^{1}_-\wedge \tilde e^{2}_-\wedge \tilde e^{3}_-\wedge \tilde
e^{4}_-  \nn \\
&=& \frac{1}{2} (\tilde J_{--})^2
\eea
is the unwarped volume along the minus directions, with $\tau^i_r=\R \tau^i$. The total six--dimensional volume
(including warp factor) is given by
\bea \label{volO5}
\rm{vol}&=& \sqrt{\tilde g} \ e^1_- \wedge e^2_- \wedge e^3_- \wedge e^4_- \wedge
e^1_+ \wedge e^2_+ =
\rm{vol}_- \wedge \rm{vol}_+ \nn \\
&=&\frac{1}{2} J_{--}^2 \, J_{++}= \frac{1}{6} J^3=\frac{i}{8} \Omega
\wedge \bar \Omega = -8 i e^{-2A} \left<  \Phi_\pm, \bar \Phi_\pm \right>
\eea
where in the last two equalities we have used the normalization condition
(\ref{compsu3}) for the pure spinors (\ref{puresu3}).

All the solutions obtained by T--duality from a solution on a conformal $T^6$ and imaginary
self-dual three-form flux satisfy the requirements (\ref{req1}), (\ref{req2}) as we will
see more explicitly in Section \ref{tdualex}.

All the non--T--dual solutions need intersecting sources.
Let us discuss now the possibility of partially localizing these solutions
by the help of two functions, $A_1$ and $A_2$, depending only on the coordinates
orthogonal to all sources. This implies that we are smearing the sources
in the Neumann-Dirichlet directions.
This trick succesfully describes  partially smeared
intersecting branes in flat space. It has chances of succeeding also in
the simplest case, namely in the models associated to the algebra 2.5.
Unfortunately it does not, as these models need sources of different charges.
One possible reason for this failure is that it would have been unphysical 
anyway to smear an orientifold source. However, it is instructive to consider why
this failure occurs technically.
Let us show this very briefly for Model 2, for the simple choice of moduli (\ref{parto5}).

Model 2 needs intersecting sources wrapping the directions 13, and 24. Lets us call the corresponding warp
factors $e^{2A_1}$ and $e^{2A_2}$. We define $e^1=e^{A_1-A_2} \tilde e^1$, $e^2=e^{-A_1+A_2} \tilde e^2$
and so on, as in eq. (\ref{rescaleint}). $A_1$ and $A_2$ are functions of 5 and 6 only.

We want to see whether $ \Omega_1, j,  \Omega_2$ given in (\ref{solmodel1}), which are solutions
in the constant warping limit, can be promoted to solutions when $A_1$ are $A_2$ depend on 5 and 6. All the equations among the list in  (\ref{12O5}) that impose closure of forms are satisfied if we implement these rescalings.
Note that for this we need $j=j_{+-}, \ \Im \O_2=(\Im\O_2)_{+-}$ with respect to both
projections in 2456 and 1356, and so
the equations are not satisfied for the generic solution: we need to restrict the moduli
such that the pure spinors are compatible with both projections.
The equation that determines the 3-form flux is
\beq
\gs e^{4(A_1+A_2)} * F_3 = (d(e^{2(A_1+A_2)} \R \O_2)=d(e^{4A_1} \tilde e^1 \wedge \tilde e^3 +
\tau^2  e^{4A_2} \tilde e^2 \wedge \tilde e^4)
\eeq
were  $\tau\equiv\tau^1_2$. 
From this we get the BI (cf. its large volume limit, eq. (\ref{BImodel2}))
\bea
dF_3&=&\frac{1}{\gs} \left[ - df   (\tilde e^1 \tilde e^4   \tilde e^6
+\tilde e^2  \tilde e^3   \tilde e^6) + (\tau^2 \nabla^2_{56} e^{-4A_1} + 2f)  \tilde e^2  \tilde e^4  \tilde e^5 \tilde e^6 + (\nabla^2_{56} e^{-4A_2} - 2f) \tilde e^1 \tilde e^3  \tilde e^5  \tilde e^6 \right] \nn \\
&=& \delta({\rm source}) = Q_{1i} \, \delta(x^{5,6}-x_{1i}^{5,6}) \, \tau^2 \tilde e^2  \tilde e^4  \tilde e^5 \tilde e^6  +Q_{2i}\, \delta(x^{5,6}-x_{2i}^{5,6}) \,  \tilde e^1  \tilde e^3  \tilde e^5 \tilde e^6
\eea
where   $\nabla^2_{56}$ denotes a Laplacian in the 56 directions, and we have defined
$f = e^{-4A_2} -  \tau^2 e^{-4A_1}$.
We see that the first term is $+ - - -$, and is not cancelled by any other term,
nor we can wrap a supersymmetric source on those cycles. It should therefore be zero,
which implies that $f$ is constant.  The two warp factors are therefore the same, up to an additive constant. But on the other hand, we see that the effective charges have opposite signs, and therefore there is no solution, unless $f=0$. In this case there is no effective flux, and
the solution corresponds to (partially smeared) intersecting sources in flat space.

All the solutions based on the algebra $s$ 2.5 have this feature, namely the effective charges
have opposite signs.
Model 1, on the contrary, needed two orientifolds to cancel tadpoles. But for that model the
situation is much worse, and there are all sort of $(+ - - -)$ uncancelled terms in the Bianchi identities.

None of the non T--dual solutions found is therefore localizable (not even partially)
by the rescalings (\ref{rescaleint}). We stress one more time that this does not mean
that there is no way of localizing the solutions we found, but it just means that the strategy
that works in flat space does not work for nil- and solvmanifolds.

\section{T--duality, GCG and string vacua}

T--duality has been an important tool in producing new vacua, and this is exactly the way the nilmanifolds first entered the scene \cite{Kachru}. In our approach the T--dual solutions are special, not due to the way they are found (in this sense they are not different from the rest), but because they are particularly nice --- they have a single source term in the BI, and can be fully localized. The complete list of these solutions can be found in Table \ref{results}. We will discuss here some examples. We use this occasion to give a discussion of how  T--duality acts on pure spinors that goes beyond the immediate application to nil(solv)manifolds.

\subsection{Pure spinors and T--duality}
\label{sec:T}

T--duality on pure spinors is a Hodge star on the T--dual directions. In this
Section, we are going to see this in the simple cases we will need in this 
paper, as well as (for three T--dualities) in greater generality, revisiting 
the duality for
SU(3) structure manifolds \cite{fmt} and giving some hints about the so--called
non--geometric cases. 

We will start with a single T--duality (in the direction $x^1$, say). 
An obvious way to compute its 
action on the pure spinors is by going to the bispinor picture and
computing the action on the spinors. Let us first see an example in flat space. 
It is easy to see that the action of a single T--duality leaves $\eta^1_+$ invariant, 
while multiplying $\eta^2_+$ by the gamma matrix in the dualized direction \cite{hassan}; by
comparing with (\ref{pureforms}), one gets that $\Phi$ is multiplied from the right:
\bea
&& \Phi_-=
(dx^1+i dx^2)\wedge (dx^3+i dx^4)\wedge (dx^5+i dx^6)\ , \nn \\
&& \sla\Phi_-\buildrel{T_1}\over\longrightarrow \sla\Phi_- \gamma^1= 
(1-i\gamma^{12})
(\gamma^3+i \gamma^4)(\gamma^5+i \gamma^6) \ ,
\label{eq:Tcliff}
\eea
from which we conclude 
\[
(dx^1+i dx^2)\wedge (dx^3+i dx^4)\wedge (dx^5+i dx^6)
\buildrel{T_1}\over\longrightarrow
 (1- i dx^1\wedge dx^2) \wedge (dx^3+i dx^4)\wedge (dx^5+i dx^6)\ .
\]
One can actually avoid going back and forth from forms to bispinors 
by mapping back left multiplication using (\ref{eq:gammamap}).
Obviously this gives the same result as above. 

We would now like to see what happens for more general $S^1$ fibrations. The 
method above can become confusing: for example, the manifold is changed by 
T--duality, and one has to understand on which manifold the $\gamma$'s live.
Here we will present an alternative method, which is a little more precise, and 
that in the end shows what the rule (\ref{eq:Tcliff}) really means.
\begin{enumerate}
\item Consider a manifold $M$ which is $S^1$ fibred. (We will use $S^1$ for simplicity.)
Compute the annihilator of the initial pure spinor $\Phi$. This is
a subbundle $L_\Phi$ of dimension six of $T\oplus T^*$ on $M$. 
\item T--duality can now be thought of as reexpressing $L_\Phi$ as
a new bundle $\tilde L$ on the dual $S^1$ fibration $\tilde M$. This is defined
by dualizing the fibre (in the case of $S^1$, just inverting its radius), 
and exchanging the components of $H$ with one leg in the fibre with the
Chern class of the fibration; operationally, this will be equation 
(\ref{eq:xch}) below.
\item Finally, interpret $\tilde L$ on $\tilde M$ as the annihilator of
a new pure spinor $\tilde \Phi$. This actually only determines $\tilde \Phi$
up to pointwise rescaling. The pure spinor equations 
(\ref{int}),(\ref{nonint}), however, can be used
to fix this ambiguity. 
\end{enumerate}

Let us isolate the coordinate $x^1$ on the fibre from the remaining ones
$y^m$. Also, similarly to \cite{fmt}, write metric and B--field as
\begin{equation}
  \label{eq:s1fibration}
ds^2_M = \sigma^2 (dx^1 + \lambda)^2 + ds^2_{\mathrm{base}}\ ; \qquad
B=b_2 + b_1 \wedge (dx^1 + \frac12 \lambda)\ .    
\end{equation}
The expression for the metric is the usual KK one; the thing to be noticed 
is that
in $B$, it proves useful to put a seemingly strange $\frac12$ in front of the
connection one--form $\lambda$. $b_2=\frac12 b_{mn} dy^m\wedge dy^n$ and $b_1=
b_m dy^m$ are forms on the base. 

The virtue of this definition is that T--duality can be expressed now as 
\cite{fmt}
\begin{equation}
  \label{eq:xch}
  \lambda\leftrightarrow b_1 \ , \qquad \sigma \leftrightarrow \frac1\sigma\ .
\end{equation}
(We will also use $\ \widetilde{}\ $ to denote variables on the dual manifold $\tilde M$. 
For example, in this case, $b_1=\tilde\lambda$, $\lambda=\tilde  b_1$, $\tilde \sigma=
\frac 1\sigma$.) 
The second relation is the well--known inversion of the radius; the 
first implies,
by taking $d$ on both sides (nothing depends on $x^1$), that 
$\iota_{\del_{x^1}} H \leftrightarrow c_1$ \cite{ag,duff,bem}. 

We will now carry out the method itemized above for a pure spinor of 
the form 
\[
\Phi= (e^1 + i e^2) \wedge \phi\ .
\]
While $e^1= \sigma(dx^1+ \lambda)$, we do 
not require that $e^2$ is one element of the vielbein on the base. 
In fact, to fix ideas, one can take $\phi= (e^3+ i e^4) \wedge
\exp(i e^5\wedge e^6)$, and again all the $e^a$, $a\neq 1$ appearing in this 
expression do not form a vielbein on the base. 

The annihilator $L_\Phi$ is in this case given by 
\[
L_\Phi=\{(e^1+i e^2), \ \ E_2^i \, \del'_i - \frac i \sigma \del_1, \ \ 
(e^3+ i e^4) , \ \ (E_3+i E_4)^i \del'_i, \ \ 
E_5^i \, \del'_i -i e^6,\ \  E_6^i \, \del'_i +i e^5\}\ .
\]
Here, the index $i$ runs on the base; 
$\del'_i \equiv \del_i - \lambda_i \del_1$; and $E_a^i$ is a basis
of vectors dual to the basis of one--forms defined by the $e^a_i$ since 
(as in Section 2) $E_a^n e^b_n=\delta_a{}^b$. 
 
It turns out, actually, that applying the method directly to $\Phi$ gives a 
$\tilde \Phi$ which depends explicitly on all the forms on the base defined
above ($\lambda, b_1, b_2$). This gets even less pleasant when considering
higher dimensional torus fibres. A better result can be
obtained by starting not from $\Phi$ itself, but from $e^B \Phi$ (compare
(\ref{eq:eb})).
We need not do any extra work to derive the annihilator of the latter: it is
enough to apply an appropriate transformation to the derivatives: $\del_i \to
e^B \del_i \, e^{-B}$, $\del_1 \to e^B \del_1 e^{-B}$. This amounts to 
substituting
\bea
\label{eq:eB}
(L_\Phi \to L_{e^B\Phi}:) && \del'_i \to \hat\del_i = \del_i 
- (b_{ij}+b_{(i} \lambda_{j)})dy^j -(b_i dx^1+\lambda_i \del_1)\ , \nn \\
&& \del_1\to \del_1+b_1\ .
\eea 
Notice that the only element of $L_\Phi$ that 
actually contains $\del_1$ explicitly is the second.
 
Now we can apply step 2: this means $\del_1\leftrightarrow dx^1$ (or, in other words, 
$\del_1=\widetilde{dx^1}$ and $dx^1 =\tilde\del_1$) as well as
(\ref{eq:xch}). In performing this step, we are not as much changing as simply reinterpreting
the various forms, vectors and functions on $M$ with others on $\tilde M$. This step is remarkably easy because of the happy circumstance that
$\hat\del_i$ in (\ref{eq:eB}) is {\it invariant}: it has the same expression on both $M$ and $\tilde M$, 
$\hat\del_i=\widetilde{\hat \del}_i$. Also, the second element now
contains $\frac1\sigma(\del_1+ b_1)=\tilde\sigma(\widetilde{dx^1} +\tilde\lambda) = \tilde e^1$. 
After writing $\tilde L$, it is also easy (since the 
$\hat\del_i$'s are invariant, again) to backtrack and retransform $e^B$ out,
getting $\tilde\del'_i=\del_i-\tilde \lambda\tilde\del_1$ everywhere. We obtain
\[
L_{\tilde\Phi}=\{(\frac1{\tilde\sigma}\tilde\del_1 +i e^2),\ \  E_2^i \tilde\del'_i - i\tilde \sigma \tilde e^1, \ \ (e^3+ i e^4) ,\ \  (E_3+i E_4)^i \tilde\del'_i, \ \
E_5^i \tilde\del'_i -i e^6, \ \ E_6^i \tilde\del'_i +i e^5\}\ .
\]
It is clear that the last four elements have come just for the ride, getting
back unchanged, and for that reason we could have considered any $\phi$ 
in $\Phi=(e^1+ i e^2)\phi$. We have chosen a definite example $\phi= 
(e^3+ i e^4)\wedge\exp(ie^5\wedge e^6)$ for sake of illustration. 
The resulting pure spinor can now be read off the annihilator we have computed:
\begin{equation}
  \label{eq:sTd}
  e^B\Phi=e^B (e^1+i e^2)\wedge\phi\buildrel T_1 \over \longrightarrow 
 e^{\tilde B} \exp(-i \tilde e^1\wedge e^2) \wedge \phi \ . 
\end{equation}
The result is  
very simple: apart from the initial $e^{\tilde B}$, ($\tilde B= b_2+ \tilde b_1(\widetilde{dx^1} + \frac12
\tilde \lambda)$), one can perform the T--duality as if it were in flat
space, by treating the $e$'s as $dx$'s, and apply separately the exchange
$\iota_{\del_{x^1}} H \leftrightarrow c_1$. That exchange, in our nilmanifold
case, reads
\begin{equation}
  \label{eq:fH}
  f^1{}_{ab}\buildrel T_1 \over \longleftrightarrow H_{1ab} \ .
\end{equation}
The reason for this simplicity ought to be clear. The real content
of (\ref{eq:xch}) is (\ref{eq:fH}), which is its topological part: the
curvature, in other words, not the connections. By using diffeomorphisms on 
one side and B--field gauge transformations on the other,
we can always choose the connection to be for example zero in a certain 
region over the base, and concentrate its curvature somewhere else.
But T--duality is local on the base. Hence it should be possible to ignore
all connections and remember them only when writing down the global structure
of the T--dual manifold, which is given by (\ref{eq:fH}). This is practically
what we do in the rest of the paper.

There is actually a shorter way of seeing that the Clifford multiplication was
essentially the right T--duality operation. By composing the various steps in this subsection 
(B--transform, T--duality and - B--transform), we have 
\begin{equation}
    \label{eq:fullexchange}
    \del_1    \longrightarrow  \frac1\sigma e^1 \ , \qquad 
    e^1 \longrightarrow \frac1\sigma \del_1 \ , \qquad 
\del'_i \longrightarrow \del'_i \ , \qquad
dx^i \longrightarrow dx^i \ .
\end{equation}
This transformation can be reproduced (denoting collectively, as usual, 
$\del_1,\del'_i, e^1,dx^i$  by $\Gamma_\Lambda$) 
$\Gamma_\Lambda\to U \Gamma_\Lambda U^{-1}$, where $U= \del_1 - e^1$ (up to an overall sign which changes nothing in all 
the annihilators). This is essentially the rule (\ref{eq:Tcliff}), which we
now understand in this way: we multiply the pure spinors by $U$, and then 
put tilde's on everything. In other words, this rule gives the functional 
dependence of the transformed pure spinor on $\sigma$, $b_1$, $\lambda$ and
the other parameters in the metric. 

We can then see another feature of the result: a single T--duality acts on the 
type of the pure spinor (which is  its lowest form degree: 
see eqn.~(\ref{type})) by changing it by 
$\pm1$. This is because the operator $U$ is linear in the $\Gamma_\Lambda$'s. 

As we have already remarked applying the T--duality rules
to $e^B \Phi$ (as opposed to $\Phi$) was essential in order to reconstruct
$e^{\tilde B}$ on the other side (rather than having some components of it). This is
related to the fact that it is $d+H\wedge$, rather than $d$, that acts
on $\Phi$ in the pure spinor equations coming from supersymmetry, 
(\ref{int}),(\ref{nonint}). Indeed we can  always write locally
$d+H\wedge = e^{-B} d e^{B}$. This way, $d(e^B \Phi)=0$ in one
theory will be mapped to $d(e^B \tilde\Phi)=0$ in the other theory. 
We will come back on this idea in the following Section.

Finally, let us remark on a point that we have ignored so far. 
The T--dual pure spinor in (\ref{eq:sTd}) might have been, a priori,
multiplied by an arbitrary function, without any change in its
annihilator. As noticed in item 3.~above, however, the normalization
condition $||\Phi||^2=e^{2A}/8$ (imposed right below (\ref{int}) and (\ref{nonint})) fixes this ambiguity. The 
factor $e^{2A}$ does not
transform under T--duality, being a component of the spacetime metric.
Hence, remembering (\ref{defMukai}), that condition says that $\Phi\wedge\lambda(\bar\Phi)$ should
be proportional to the volume form ${\rm vol}$ and that $\tilde\Phi\wedge\lambda(\bar{\tilde\Phi})$ should be proportional
to the dual volume form $\widetilde{\rm vol}$, with the same proportionality factor. This eliminates the 
possibility of a rescaling by an arbitrary function, and fixes
the normalization as in (\ref{eq:sTd}). 
The fact that we used the supersymmetry equations to fix this factor should 
not come as a surprise. This is the requirement that supersymmetric vacua
be sent to other supersymmetric vacua, and it should be thought of as a
``square root'' of the way Buscher rules were derived in the first place, 
namely by requiring that vacua be sent into vacua. 

In fact, one can even think of the T--duality action on the pure spinors
as {\it implying} the Buscher rules, if one considers a pair of pure spinors rather than 
a single $\Phi$ as we did here. The reason is that a pair of compatible pure 
spinors determines a metric and B--field (see Section \ref{sec:comp}). 
Consider the metric $M= \mathcal{I}\J_a \J_b$.
 In our case, $\J_a=\J_+$ and
$\J_b=\J_-$. Using that the $\J$'s are both hermitian (hence 
$\J^t\mathcal{I}=-\mathcal{I}\J$) and that they commute, we can rewrite 
this as either
\begin{equation}
    \label{eq:M}
    M= - \J_+^t\mathcal{I}\J_-=-\J_-^t\mathcal{I}\J_+\ ;
\end{equation}
if one now transforms 
\begin{equation}
    \label{eq:TcJ}
\J_\pm\to (O^t)^{-1} \J_\mp O^t\ ,    
\end{equation}
for $O\in \mathrm{O}(6,6)$, one finds that
\begin{equation}
    \label{eq:TM}
M\to -O \J_-^t O^{-1} \mathcal{I}(O^t)^{-1} \J_+ O^t= -O \J_-^t\mathcal{I}\J_+
O^t=OMO^t
\end{equation}
where in the first equality we have used 
that $O\in\mathrm{O}(6,6)$ (hence $O\mathcal{I}O^t=\mathcal{I}$).
(\ref{eq:TM}) is 
the appropriate transformation rule for $M$, as found in the T--duality
literature (see for example \cite{gpr}) long before GCG was defined. If
$O={{a\ b}\choose{c\ d}}$, with $a$, $b$, $c$, $d$ $6\times6$ matrices, 
$E=g+B$ transforms as $E\to (aE+b)(cE+d)^{-1}$. 
For an 
even element of O$(6,6)$, we would have needed $\J_\pm\to 
(O^t)^{-1} \J_\pm O^t$ instead of (\ref{eq:TcJ}).  The same result was found 
in the context of mirror symmetry for tori in \cite{kapor}, which remarkably 
enough predates the definition of generalized complex structures. 

\subsection{Spinoff: applications to mirror symmetry}

We can use similar techniques to revisit a question considered in  \cite{fmt}:
the transformation law under ``mirror symmetry'' of intrinsic torsions for
SU(3) structures.

Calabi--Yau three--folds are $T^3$--fibred, and mirror symmetry amounts to 
T--duality along those fibres \cite{syz}. The reasoning behind these statements
comes from considering moduli spaces of  D--branes, and there is no
reason those statements should remain true more generally. However, if 
a manifold happens to be $T^3$--fibred, one can define a mirror by dualizing
that fibre. This by itself is no profound definition; the surprise in 
\cite{fmt} was that, however, various quantities measuring the failure of
$M$ and $\tilde M$ to be Calabi--Yau (called intrinsic torsions) transform
in a way which was more covariant than a priori expected. Specifically,
one could  summarize their transformation law without reference to the 
$T^3$ fibration structure on each side. The intrinsic torsions for manifolds
with SU(3) structure are differential forms $W_i$ (to be reviewed shortly) 
in the representations
$1,8,6,\bar 6, 3, \bar 3$; the slogan in \cite{fmt} was that 
$8+1 \leftrightarrow  6 + \bar 3$. 

Rather than reviewing how this came about in \cite{fmt}, we are now going
to show how generalized complex geometry helps rederive those results
in a much shorter way, which makes them completely natural and expected.
(The relevance of pure spinors was anticipated in \cite{fmt}, but not fully
put to fruition.) The reason we are including this discussion here is
that it is an easy application of the method described in the previous
subsection and of some ideas in \cite{gmpt2}. 

Let us start by reviewing what are the objects we want to transform.
Given an SU(3) structure defined by a pair $(J,\Omega)$, we want to 
give a measure of its failure to be a Calabi--Yau. For the latter, 
we know that they are both covariantly constant, $\nabla J= \nabla\Omega=0$. 
The condition $dJ=d\Omega=0$ might seem to be weaker, but one can show with
some SU(3) group theory that it is actually equivalent. Hence we can 
use the differential forms $dJ$ and $d\Omega$ to classify SU(3) structures
which are not Calabi--Yau. 

It is customary to break up these differential forms 
in SU(3) representations. The aim 
of \cite{fmt} was to compute the transformation laws of the different SU(3) 
representation appearing in $dJ$ and $d\Omega$. Here we will use a different 
approach which is more suitable to the general context of SU(3)$\times$SU(3) 
structures.

First of all, it is natural to use $e^{i J}$ rather than $J$. 
Second, rather than using 
SU(3) representations, one should use a decomposition which is more natural
for SU(3)$\times$SU(3) structures. This has been suggested
in \cite{gmpt2}: we review it here. One should use the ``pure Hodge diamond''
(which here we specialize to the SU(3) case) 
\begin{equation}
  \label{eq:hodgeSU3}
  \begin{array}{c}\vspace{.1cm}
e^{iJ} \\ \vspace{.1cm}
e^{iJ}\gamma^i  \hspace{1cm} \gamma^{\bar i}e^{iJ} \\ 
\Omega\gamma^{\bar i} \hspace{1cm} \gamma^{\bar i} e^{iJ}\gamma^i \hspace{1cm} 
\gamma^i \bar \Omega\\
\Omega \hspace{1.2cm}\gamma^{\bar i}\Omega\gamma^{\bar j} 
\hspace{1cm} \gamma^i\bar\Omega\gamma^j 
\hspace{1.2cm}\bar\Omega\\
 \gamma^{\bar i} \Omega\hspace{1cm} \gamma^i e^{-iJ}\gamma^{\bar j} 
\hspace{1cm}\bar\Omega\gamma^i\\
\gamma^i e^{-iJ} \hspace{1cm}  e^{-iJ}\gamma^{\bar i}\\
 e^{-iJ}\\
  \end{array}\ ;
\end{equation}
we have written here the bispinors corresponding to the differential forms.
Remember that in the main text we are deliberately confusing forms and
bispinors to avoid cluttering the equations. 
The gamma matrices acting from the left and from the right were 
divided in Section \ref{sec:metric} into four bundles of 3 dimensions 
each, $L_{\pm\pm}$. Explicitly, 
\begin{eqnarray}
    \label{eq:corners}
\buildrel\to\over\gamma{}^i= P^i{}_n(dx^n+ iJ^{np}\del_p)\  \ , & 
\qquad & \buildrel\leftarrow\over\gamma{}^{\bar i}= 
(-)^p\bar P^{\bar i}{}_n(dx^n+iJ^{np}\del_p) \ ,\\
\buildrel\leftarrow\over\gamma{}^i= P^i{}_n(dx^n- iJ^{np}\del_p)\ , & \qquad &
\buildrel\to\over\gamma{}^{\bar i}= (-)^p
\bar P^{\bar i}{}_n(dx^n-iJ^{np}\del_p) \ ,
\end{eqnarray}
where $p$ is the degree of the form these are acting on;  
the indices $i,\bar i$ are holomorphic, and $n, p$ are real; and $P=\frac12(1-iI)$ is, as earlier, the holomorphic projector
for the almost complex structure $I$ defined by $\Omega$. 
(When $I$ is not integrable, complex coordinates  $dz^i$ might not exist.) We have used 
(\ref{eq:gammamap}) as well as
$g^{mn}=I^m{}_p J^{pn}$.
These act in the four possible directions
on the bispinors (or forms) in (\ref{eq:hodgeSU3}): for example, the sections 
$\buildrel\to\over\gamma{}^i$ of $L_{++}$ act on the 
diamond (\ref{eq:hodgeSU3}) by going up one position 
and right one position. This is consistent with the fact that they 
annihilate $e^{iJ}$ and 
$\Omega$. (In \cite{gmpt2} $L_{++}$ was called $L_\nearrow$, for this reason.)
The four groups of gamma matrices have been carefully placed in the four
corners so as to correspond to the four directions of their action in the
Hodge diamond. 

One can now define intrinsic torsions as follows
\bea
&(d+H\wedge)\O =
W^{00} e^{iJ} + W^{11}_{\bar i j} (\gamma^{\bar i} e^{iJ} \gamma^j) 
+ W^{22}_{i\bar j} (\gamma^i
e^{-iJ} \gamma^{\bar j}) + W^{33} e^{-iJ}+
W^{20}_{\bar i} (\Omega \gamma^{\bar i}) 
+ W^{31}_{\bar i} (\gamma^{\bar i} \Omega)&  \hspace{1cm}\\
&(d+H\wedge)e^{iJ} =
W^{30} \Omega + W^{21}_{\bar i \bar j} (\gamma^{\bar i} \Omega \gamma^{\bar j}) 
+ W^{12}_{ij} (\gamma^i\bar\Omega{}\gamma^j) + W^{03}\bar\Omega{}+
 W^{10}_i (e^{iJ}\gamma^i) 
+ W^{01}_{\bar i} (\gamma^{\bar i} e^{iJ}) \  .&
\eea
It turns out that these are the only torsions present (for example, 
$(d+H\wedge)\Omega$ does not have any piece proportional to $\gamma^a
\bar\Omega$, even if it would be allowed by parity).
The torsions $W^{ab}$ are denoted according to the position of the 
corresponding form in
the Hodge diamond, starting from 00 on the top, 10 and 01 for the first 
row, etc. One can obtain more explicit expressions for the $W^{ab}$  by
using the pairing (\ref{defMukai}). 
One has for example
$W^{00}=-8i\langle e^{-iJ}, (d+H\wedge)\Omega\rangle$, 
or $W^{12}_{ij}=2i
\langle \gamma_i\Omega\gamma_j, (d+H\wedge) e^{iJ}\rangle$ 
(we used (\ref{trace}) and we normalized the spinors to 1, thus forgetting
factors of $e^A$).
The following relations hold among the simplest intrinsic torsions: $W^{03}= W^{33}$, 
$W^{30}=-\overline{W^{00}}$.

 SU(3)$\times$SU(3) intrinsic torsions transform better than the  usual SU(3) torsions.
In order to see that, let us
 derive the transformation law of $e^{iJ}$ and $\Omega$.
Explicitly, these forms are taken to be 
\begin{equation}
\label{eq:JOE}
J=-V_{i\alpha} dy^i\wedge e^\alpha\ , \qquad 
\Omega=E^1\wedge E^2\wedge E^3\ , \quad 
\Big(E^a= i e^a_i dy^i + V^a_\alpha (dx^\alpha+\lambda^\alpha)\Big)
\end{equation}
$E^a$ being the $(1,0)$ vielbein;  $V^a_\alpha$ is a
vielbein for the three fibre directions. For more details on the setup, see \cite{fmt}.
The computation is now a variation on the one we saw in the previous Section, only
this time with three T--dualities rather than one. Not surprisingly, the 
result is that 
\begin{equation}
  \label{eq:eijO}
  e^B \Omega \leftrightarrow |g_f| e^{\tilde B} e^{-i\tilde J}\ .
\end{equation}
As in the previous subsection, the presence of $e^B$ on both sides 
is essential in getting a result that does not depend on the explicit 
splitting among base and fibres (the analogue of eq.~(\ref{eq:s1fibration})).
This result was argued in \cite{fmt} in a much clumsier way. 
It also appeared in the context of
Calabi--Yau manifolds: for example in \cite{lyz, gm} it was used to
show that the D--term BPS conditions for B--branes \cite{mmms}, which 
read $\mathrm{Im}(e^{i \theta} e^{B+F} e^{iJ})$, are mapped by mirror symmetry
to the stability conditions for A--branes, $\mathrm{Im}(e^{i\theta}e^{F}\Omega)$. 
A similar mapping was argued in \cite{dfr} 
for the expression of the central charge for B--branes, 
$\int_B e^{B+iJ} ch(E)\sqrt{\frac{Td(M)}{Td(B)}}$
and for A--branes, $\int_A \Omega$. Notice that in both these examples 
the exchange seems to be more $e^{B+iJ}\leftrightarrow \Omega$, as opposed
to (\ref{eq:eijO}). However, in the Calabi--Yau case the B--field is 
$(1,1)$, and hence $e^B\Omega=\Omega$.

Coming back to our endeavor of T--dualizing the intrinsic torsions, 
notice that $(d+H\wedge)= e^{-B}d e^{B}$, 
at least locally in the base. (This is no loss of generality: both 
T--duality and the intrinsic torsions are certainly local in the base.)
It is now easy to compute explicitly the action of T--duality on any intrinsic
torsion. Let us consider for example $W^{00}$:
\bea
    \label{eq:w00w30}
\frac i8 W^{00}=\langle e^{-iJ}, (d+H\wedge)\Omega\rangle&=& \langle e^{B-iJ},
d(e^B \Omega)\rangle
\longrightarrow  \nn \\
 \langle e^B\Omega, d(e^B e^{-iJ})\rangle &=& 
\langle \Omega,(d+H\wedge)e^{-iJ}\rangle=-\frac i8\overline{W^{30}}\ .    
\eea
We have used that $d$ does not transform, as it only contains
derivatives along the base. The factor $|g_f|$, which 
drops out here, turns out to be important for
example for the exchange  $W^{10}\leftrightarrow -\overline{W^{20}}$. The computation
of the duals for the other $W^{ij}$ is a bit more subtle than the 
formal manipulation in (\ref{eq:w00w30}). The reason is that one has to decide
whether to transform the $\gamma$'s or not. The answer is clear in the 
framework of Section \ref{sec:T}. We saw there that the transformation law
for the pure spinor $\Phi$ is defined by the one for the annihilator $L_\Phi$. 
The $\gamma$'s in this Section are nothing but the four common annihilators
(of $\Phi_+$ and $\Phi_-$, of $\Phi_+$ and $\bar \Phi_-$, and so on) 
$L_{\pm\pm}$. So they do transform too: specifically, the ones acting from 
the left are being swapped as 
\begin{equation}
    \label{eq:exgammas}
\buildrel\to\over\gamma{}^i\leftrightarrow \buildrel\to\over\gamma{}^{\bar i}\ .  \end{equation}
This gives for example to $W^{31}_{\bar i}\leftrightarrow W^{32}_i$. This 
exchange makes sense once one remembers that 
none of these intrinsic torsions depend on the coordinates on the fibre: we
are swapping forms on the basis ($W^{31}_{\alpha}\leftrightarrow 
W^{32}_\alpha$), and multiplying by the appropriate projector at the end. More 
details can be found in \cite{fmt}. 

The general rule is now simple: $W^{ij}
\leftrightarrow -\overline{W^{3-i,j}}$. This 
amounts to a reflection of the Hodge diamond, which does indeed look 
like a mirror map.  Turning in particular 
 to the exchange $W^{00}\leftrightarrow -\overline{W^{30}}$, 
we recall the earlier remarked relation, $W^{30}=-\overline{W^{00}}$. Hence
$W^{00}$ is invariant. In fact, this 
had been derived in \cite{fmt}, but it is not captured by the
slogan $8+1\leftrightarrow 6 +\bar 3$. All the 
transformation rules derived in this way coincide with those in \cite{fmt}.

The present formulation presenting mirror map as the reflection of the Hodge diamond applies to more general \stt\ structures.  Indeed, also for this general case  the pure Hodge diamond (\ref{eq:hodge}) can be introduced,  and the method explained in subsection  \ref{sec:T} can be applied to pure spinors of any type, not just 0 and 3.  The \stt\ framework  also makes it much clearer that morally T--duality  is simply a  multiplication from the left by the product of the transverse gamma matrices $\Gamma^\perp$, as one sees  from (\ref{eq:exgammas}). Indeed, taking $\Phi_\pm=\eta^1_+
\otimes\eta^{2\,\dagger}_\pm$ (even when $\eta^1=\eta^2$) it is natural 
to multiply one spinor by the  transverse gammas and not the others: it turns out to be just the usual
transformation law $\epsilon^1\to \Gamma^\perp \epsilon^1$, $\epsilon^2\to
\epsilon^2$ specialized to the decomposition (\ref{eq:spinor}).  

\subsection{Non--geometric cases?}
\label{sec:ng}
We have also tried to extend (rather formally) 
the method described in subsection
\ref{sec:T} to the so--called non--geometric T--duals. 
When $H$ has more than one leg along the fibre to be T--dualized, the 
expression one gets from the standard Buscher rules for the dual metric and 
B--field becomes well--defined only up to a T--duality--valued monodromy: 
the metric and B--field are not well defined separately \footnote{
Incidentally, the simple exercise of checking the topology of the twisted tori (see Section 2) can serve also as an illustration of the non-geometricity. Going back to the three-dimensional nilpotent Heisenberg algebra $(0,0, N \times 12)$, it is not hard to see that  the corresponding nilmanifold can be produced by 
T--dualizing an ordinary $T^3$ with the NS flux given by 
$H =Ne^1 \wedge e^2 \wedge e^3$. This configuration naively allows for 
two T--dualities. Indeed,
choosing the original NS two-form as $B_2 = N x^1 e^2 \wedge e^3$, and the metric after the first T--duality as  (\ref{heis}) $dx_1^2+ dx_2^2+(dx_3+ N x_1 dx_2)^2,$
we do expect to be able to perform the second T--duality along direction 2.
Yet due to the twisting discussed earlier $ (x^1,x^2,x^3)\simeq 
(x^1+1,x^2,x^3-N x^2)$,  $\partial_2$ is no longer a well-defined vector: 
under $x^1\to x^1+1$, $\partial_2\to \partial_2+N\partial_3$.
Making it well--defined by e.g.~changing the metric to 
$dx_1^2+ dx_2^2+(dx_3 - N x_2 dx_1)^2$  makes the direction 2 cease being an 
isometry. Hence the simple claim that the non--geometricity is associated 
with inability to perform consecutive T--dualities in a background that 
naively has more than one isometry.}. 

The only difference with respect to subsection \ref{sec:T} is that 
the B--field also includes
components with two legs along the fibre:
\begin{equation}
  \label{eq:Bgeom}
  B= b_2 + b_{\alpha i} (dx^\alpha + \frac12 \lambda^\alpha) \, dy^{i} + 
  \frac12 B_{\alpha\beta} e^\alpha e^\beta \ , \quad e^\alpha= dx^\alpha + 
\lambda^\alpha \ .
\end{equation}
The Buscher rules now read
\begin{equation}
  \label{eq:nongbus}
  B_{\alpha\beta} \to \hat B^{\alpha\beta}\equiv 
-\left(\frac1{h+B} B \frac1{h-B}\right)^{\alpha\beta}\ ,
\qquad
V_{\alpha i} \leftrightarrow 
\hat V^\alpha{}_i\equiv 
\left( \frac1{h+B} \right)^{\alpha\beta} V_{\beta i}\ , 
\qquad 
\lambda^\alpha{}_i\leftrightarrow b_{\alpha i}\ , 
\end{equation}
where  
$h_{\alpha\beta}= V_\alpha{}^a V_\beta{}^b \delta_{a b}$ is the
metric on the fibre.
Notice that $\hat V$ is a vielbein for the dual metric, which is defined as
$\hat h^{\alpha\beta}= (\frac 1{h+B})^{\alpha\gamma} h_{\gamma\delta}   (\frac 1{h-B})^{\delta\beta}$. 

Consider a pair of type 0--type 3 pure spinors. Applying the 
same procedure as in subsection \ref{sec:T}, one gets for the dual of the
odd spinor 
\begin{equation}
  \label{eq:tsu3}
 e^B \Omega \to |g_f| \exp[\tilde B -i \tilde J - \frac12 \tilde B_{\alpha\beta} 
 (\tilde e^\alpha + i \tilde V^\alpha{}_i dy^i) 
 (\tilde e^\beta + i \tilde V^\beta{}_j dy^j)] \ .
\end{equation}
In the equation above, the tilde denotes quantities defined on the T-dual 
manifold.
We have managed to get rid of all the hats on the rhs, 
by using $\hat V_{a\alpha} 
\hat B^{\alpha\beta} \hat V_{b \beta}= - V_a^\alpha B_{\alpha\beta} V_b^\beta$ and
$\hat V_{a\alpha}=(\delta_\alpha{}^\beta-B_{\alpha\gamma}h^{\gamma\beta})V_{a\beta}$, and then to recombine all the new non--geometrical contributions
 (containing $\hat B$) into a single square. Notice also that, by the 
definition of
 the $(1,0)$ vielbein above (\ref{eq:eijO}), this new non--geometrical term is
 a $(2,0)$ form. 

The fact that one can still write pure spinors associated with a 
non--geometrical 
background suggests that the approach of the present paper might be applied in
some way to those cases as well, although we will not pursue this here any
further. 

We will present now some details about the computation that
leads to (\ref{eq:tsu3}), using again the strategy
outlined in section \ref{sec:T}. 

Taking $\Omega$ as in (\ref{eq:JOE}), one can compute 
its annihilator as usual. It is, however, better to start from 
the annihilator of $e^B \Omega$, just like in section \ref{sec:T}. 
This is most readily obtained by using that 
\[
\begin{array}{c}\vspace{.2cm}
e^B\del_i e^{-B} = \del_i -(b_{ik} +b_{\alpha [i}\lambda^\alpha_{k]}
+ B_{\alpha \beta} \lambda^\alpha_i \lambda^\beta_k ) dy^k -
(b_{\alpha i} + B_{\alpha \beta} \lambda^\beta_i ) dx^\alpha\ ,  \\
e^B \del_\alpha e^{-B} = \del_\alpha+ b_{\alpha i } dy^i - B_{\alpha\beta} e^\beta\ ,
\end{array}
\]
so that, similarly to (\ref{eq:eB}), one has
\bea
\label{eq:hatdeli}
(L_\Phi \to L_{e^B\Phi}:) && \del'_i\equiv \del_i -\lambda_i^\alpha\del_\alpha{} 
\to \hat \del_i \equiv \del_i -(b_{ik} + b_{\alpha(i} \lambda^\alpha_{k)} ) dy^k 
- b_{\alpha i} dx^\alpha- \lambda_i^\alpha \del_\alpha\ , \\
&&\del_\alpha{}\to  \del_\alpha+ b_{\alpha}  - B_{\alpha\beta} e^\beta\ ,
\eea 
with $b_{\alpha}=b_{\alpha i} dy^i$ one--forms. Hence the annihilator of $e^B \Omega$ reads:
\[ 
L_{e^B \Omega}=\Big\{ e_a^i \hat \del_i 
-i V_a^\alpha (\del_\alpha + b_\alpha -B_{\alpha\beta} e^\beta) \, ,\ i e^a_i dy^i + V^a_\alpha e^\alpha \ 
\Big\}\ .
\]
We can now apply the T--duality rules in (\ref{eq:nongbus}). Again (see comments after
(\ref{eq:eB})) this step consists of rewriting $L_{e^B \Omega}$ by reinterpreting forms
on $M$ as forms on $\tilde M$: for example, the transformation of $B_{\alpha\beta}$ 
in (\ref{eq:nongbus})  is to be read $\tilde B_{\alpha\beta} = \hat B^{\alpha\beta}\equiv
-((h+B)^{-1} B (h-B)^{-1})^{\alpha\beta}$ -- or, inverting, $B_{\alpha\beta}= 
\widetilde{\hat B^{\alpha\beta}}= 
-((\tilde h+ \tilde B)^{-1} \tilde B (\tilde h- \tilde B)^{-1})^{\alpha\beta}$. In other
words, a $\ \widetilde{}\ $ says that a certain quantity lives on $\tilde M$; a 
$\ \hat{} \ $ is simply
the operation of multiplying left and/or right (as appropriate)  
by $(h\pm B)^{-1}$. One can easily see that
$\hat \del_i$ in (\ref{eq:hatdeli}) is invariant under T-duality
 (just like it was
in section \ref{sec:T}) in the sense that $\hat\del_i=\widetilde{\hat\del_i }$. 
After having applied this transformation, one can undo a  $\tilde B$--transform to obtain 
\[ 
L_{\tilde \Phi} =\Big\{ 
e_a^i \widetilde{\del'_i} - i \widetilde{\hat V}_{a\alpha} 
(\tilde e^\alpha - \widetilde{\hat B^{\alpha\beta}}
(\tilde\del_\beta+ \tilde B_{\beta\gamma}\tilde e^\gamma))\, , 
\ i e^a_i dy^i + \widetilde{\hat V}^{a\alpha} 
(\tilde \del_\alpha+ \tilde B_{\alpha\beta}\tilde e^\beta) \Big \}\ .
\]
One can take then $\tilde \Phi$ to be 
\[
\exp[-i \widetilde{\hat V}_{\alpha i} \tilde e^\alpha dy^i 
-\frac12 \widetilde{\hat B^{\alpha\beta}} \widetilde{\hat V}_{\alpha i}
\widetilde{\hat V}_{\beta j }dy^i dy^j-\frac12\tilde B_{\alpha\beta} \tilde e^\alpha
\tilde e^\beta] \ .
\] 
This expression is more complicated than its counterpart in section \ref{sec:T} due
to the presence of hat symbols. Fortunately we can get rid of them by using 
the relations 
\[
\widetilde{\hat B^{\alpha\beta}}\widetilde{\hat V}_{\alpha i }
\widetilde{\hat V}_{\beta j}= -\tilde B_{\alpha\beta} \tilde V^\alpha_i \tilde V^\beta_j\ ,  
\qquad
\widetilde{\hat V}_{\alpha i}= \tilde V_{\alpha i} - 
\tilde B_{\alpha\beta} \tilde h^{\beta\gamma}  
\tilde V_{\gamma i}\ . 
\]
We get 
\begin{equation}
    \label{eq:tsu3final}
\tilde \Phi= \exp[ i \tilde V_{\alpha i } dy^i \tilde e^\alpha+\frac12 \tilde B_{\alpha\beta}
(\tilde V^\alpha_i dy^i + i \tilde e^\alpha) (\tilde V^\beta_j dy^j + i \tilde e^\beta)] \ .    
\end{equation}
 This is the result in (\ref{eq:tsu3}) where  we used 
the explicit expression 
for $J$ given in (\ref{eq:JOE}), and the fact  that the pure spinors 
connected by T--duality
are $e^B \Omega$ and $e^{\tilde B} \tilde\Phi$, with $\tilde\Phi$ as in (\ref{eq:tsu3final}).

\subsection{T--dual local solutions}
\label{tdualex}

In  this Section we will present some samples of the  solutions related by a sequence of  T--dualities to  an O3  compactification on  $T^6$  with a non trivial self-dual three-form flux and $F_5$ proportional to the warp factor (a type B solution). The complete list of these is given in Table \ref{results}.  The first examples of such local solutions appeared in \cite{Kachru} and
their localization is given in \cite{Schulz}. These are of the same type as the first 
model we discuss below, $n$ 4.4 with O5--planes and SU(3) structure. We will see here 
that all solutions of T--dual type have completely localized 
liftings.   In a decreasing order of completeness we will discuss three solutions of 
IIB with O5--planes, one with type 0-type 3 pure spinors and two with type 1-type 2 
(an $\N=2$ version without $H$ flux after T--duality and  an $\N=1$ with $H$ flux), 
and one solution of IIA with O6--planes with type 0-type 3 pure spinors.

\vskip 0.4cm

\noindent
{\sl (n 4.4)  with O5--planes, SU(3) structure.} $\,\,\,\,$ We start with the 
discussion of the IIB background involving pure spinors of type 0 and type 3. This is 
a standard items in the SU(3) structure classification. It has RR three-form flux and is typically labeled as type C. We consider the nilpotent algebra 4.4 in 
Table \ref{ta:nil} defined by the structure constants 
(0,0,0,0,12,14+23). The general equations for this case are collected in \ref{sec05A}.  Choosing to perform the orientifold projection  along the directions 5 and 6, we can 
 build $\Omega_3$ and $J$ (as in (\ref{Om}), (\ref{Je})), 
 \bea \label{OmTdual}
\Omega_3 &=& z^1\wedge z^2 \wedge z^3=  (  e^{1} + 
 i \tau^{1}   e^{2} +  i \tau^2    e^3) \wedge (  e^4 + 
 i \tau^{3}   e^2  + i \tau^4    e^3) \wedge 
 (  e^5 + i \tau^+   e^6)  \ ,\\\nonumber
J &=& \frac{i}{2} \left( t_1  z^1 \wedge \bar z^1 + t_2 z^2 \wedge \bar z^2 + 
b z^1 \wedge \bar z^2 - \bar b\bar z^1 \wedge z^2 
+t_3  z^3 \wedge \bar z^3\right)  
\eea
where the directions orthogonal to the orientifold are $e_-^{1,\ldots,4} = 
e^{1,\ldots,4}$ and $e_+^{1,2} = e^{5,6}$ correspond to the O5 location.
Overall there are 15 real free moduli.
Some of them are fixed by imposing the closure of $\Omega_3$ and $J^2$ 
as required by equations (\ref{03O5})
\bea \label{ruleclo}
&& \tau^2 (1+ \tau^3 \tau^+) + \tau^+ (1- \tau^1 \tau^4) = 0 \ , \nn \\
&& {\rm Re} \left( t_2 \tau^4 - i b \tau^2 \right) = 0 \ ,\nn \\
&& {\rm Re} \left(  b (1+ \tau^1 \bar \tau^4 - \tau^2 \bar \tau^3) 
+ i (t_1 \tau^1 \bar \tau^2 +  t_2 \tau^3 \bar \tau^4) \right) =0  \ ,
\eea
and the normalization condition  of the two pure spinors given by
$ t_3 (t_1 t_2 -|b|^2)=1$.
 
Notice that the further constraints that we normally impose in order to have a local
solution, namely $d \tilde J_{--}=0, d \tilde J_{++} \wedge \tilde J_{--}=0$, are 
satisfied without any extra conditions on the moduli. It turns out that this is always true for all the T--dualizable cases, so that any ``global" solution  can be turned into a good local one.

In order to check the tadpole condition, we have to compute the Hodge 
star of the supersymmetry equation for the RR three-form 
in  (\ref{03O5}). The metric appearing in the star is determined from 
the pure spinors as in Section \ref{Gentwist}. For simplicity we 
derive it here for a particular subset 
of the solutions of (\ref{ruleclo}) where all $\tau$'s real and
satisfy
\beq
\tau^3=0 \ , \quad \tau^1= -\frac{1}{\tau^4} \ , \quad \tau^2=-2 \tau^+ \ , 
\quad \rm{Im} b = t_2  \, \frac{\tau^4}{2 \tau^+} \ .
\eeq
Without specifying these further, we are not guaranteed to have a positive--definite metric on the internal space. Thus the $\tau$'s should be further constrained.
There are plenty of values of moduli that ensure this.   The only non-vanishing flux, 
$F_3$,  is given by
\bea\label{F3p}
\gs F_3&=&  t_3 \left[ -  \tilde e^1 \wedge \tilde e^2 \wedge \tilde e^5 +2 \, \frac{\tau^+ }{ \tau^4}
 \tilde e^2 \wedge \tilde e^4 \wedge \tilde e^5 -2 \tau^4
\tilde e^1 \wedge \tilde e^3 \wedge \tilde e^5 + \right. \nn \\
 &&  \left.   (\tau^+)^2 \left( \tilde e^1 \wedge \tilde e^4 \wedge \tilde e^6
 + \tilde e^2 \wedge \tilde e^3 \wedge \tilde e^6+  4 \tilde e^3 \wedge \tilde e^4 \wedge \tilde e^5 \right)\right] + e^{2A} *_4 d(e^{-4A})  \nn 
 \eea
where $*_4$ is the star on the base. Its exterior derivative 
\beq
dF_3= \frac{1}{\gs} \left( 6\, t_3 (\tau^+)^2 + \frac{1}{t_3}  \tilde \nabla^2_- (e^{-4A}) \right)
\tilde e^1 \wedge \tilde e^2 \wedge \tilde e^3 \wedge \tilde e^4 
\eeq
has only components in the four directions transverse to the orientifold.
The constant piece in the equation above 
gives the contribution that must by cancelled by sources. 
Wedging it with $J$, we have
\beq
dF_3^{\rm{const}} \wedge J=\frac{6}{\gs} t_3^2 (\tau^+)^3  e^1 \wedge e^2 \wedge e^3 \wedge e^4 \wedge e^5 \wedge e^6 =  \frac{4}{\gs} t_3^2 J^3 
\eeq
and we can check that its sign is consistent with the no-go theorem discussed in Section \ref{N1vacua}:
the RR--fields have to be sourced by O5--planes.

As already mentioned this solution, 
like all the other global ones that can be 
warped, is mapped by two T--dualities 
to a type B solution  with  O3 orientifolds transverse
to a six--dimensional flat torus. 
The two T--duality directions correspond to the position of the O5 in the 
internal manifold, in this case  5 and 6. Since there is no B-field, 
5 and 6 are indeed isometries of the metric.

As discussed in Section \ref{sec:T}, the effect of T--duality is to
exchange torsion and NS fluxes (\ref{eq:fH}).  The effect on the vielbeine  
is a rescaling and the 
disappearance of the connection pieces. 
 To be more precise, the actual metric vielbeine are on both sides
\bea
& e_g^5=e^A \sqrt{t_3} \tilde e^5=e^A \sqrt{t_3} (dx^5+ x^1 dx^2)
 \ , \qquad \qquad \qquad & e^5_{g_D} =\frac{e^{-A}}{\sqrt{t_3} } \tilde e^5_{D}= \frac{e^{-A}}{\sqrt{t_3} } dx^5  \\
& e_g^6=e^A \sqrt{t_3}  |\tau^+| \tilde e^6=e^A \sqrt{t_3} (dx^6+ x^1 dx^4+x^2 dx^3)
 \ , \quad & e^6_{g_D} =\frac{e^{-A}}{{\sqrt{t_3}} |\tau^+|} \tilde e^6_{D}= \frac{e^{-A}}{{\sqrt{t_3}} |\tau^+|} dx^6 \nn
 \eea 
 where $e^5_g, e^6_g$ are the vielbeine,  $e^{5,6}_{g_D}$ are the T--dual 
vielbeine, and $\tilde e^5$, $\tilde e^6$ are elements of the basis
of one--forms that  satisfy the differential  equations (\ref{strdef}). 

The T--dualized vielbeine have the right rescaling of an O3 solution, 
namely all $e$'s get  a factor of $e^{-A}$. 
However the function $e^{-4A}$ for an O5 depends on the 
transverse distance, $r$,  to the O--planes  as $1/r^2$
(see eq. (\ref{b05}), the Laplacian is in four transverse coordinates).
On the O3 side, this is what we expect from a ``smeared" warp factor, 
independent of the directions  5 and 6 
(an O3-type function $e^{-4A} \sim 1/r^4$ integrated over $x^5, x^6$
gives a dependence of the form $1/r^2$).
This feature is model-independent ---  the directions along which O5 is extended 
 after T--duality  will stay smeared in the resulting O3;  exactly the same happens 
with D5--branes, if the model has some.

 According to the rules in Section \ref{sec:T}, the pure spinors
are T--dualized by doing a Hodge star in the T--duality directions 56.
 In this case it is not hard to show
 that the T--dual $\Phi_\pm$ have exactly the same functional form 
 as the original ones except for an overall $i$. This agrees 
 with the fact that the supersymmetry preserved by an O3 is $a=ib$ while 
 for an O5 is $a=b$ and therefore  the O3 pure spinors should have 
relative $i$ with respect to those for the O5. 
But this also implies that the form of $\Omega_3$ and
 $J$ in terms of the vielbeine does not change after two T--dualities 
(as we expect from the discussion in 
 Section \ref{sec:T}, given that 56 appear  in $z^3$ only). 

Since T--duality acts on the RR fields also as the Hodge star in the direction 56, 
it is easy to see that the last term in the three-form flux (\ref{F3p}) turns into five-form flux
 \beq
\gs  F_5^D= e^{4A} * d(e^{-4A}) \ ,
 \eeq
where we have used $*_{4} *_2=*$, and $F_5^D$ denotes the 
5-form flux for the T--dual solution. This is  the five-form flux of  type B solution and is related via Hodge duality to the exterior derivative of the warp factor. All  other terms of $F_3$ are mapped into three-form flux. This is the RR part of 
the self-dual complex three-form of the type B solution. Its Hodge star 
is
\bea \label{F3d}
\gs *F_3^D&=&-*_2 dJ=   t_3 {\rm Re}(\tau^+) *_2 (- \tilde e^1 \wedge \tilde  e^2 \wedge \tilde  e^6 + \tilde  e^1 \wedge \tilde  e^4 \wedge \tilde  e^5 + \tilde  e^2 \wedge \tilde  e^3 \wedge \tilde  e^5)  \nn \\
&=&    t_3 \rm{Re}(\tau^+)  \left(  \tilde  e^1 \wedge \tilde  e^2 \wedge \tilde  e^5_D + \tilde  e^1 \wedge \tilde  e^4 \wedge \tilde  e^6_D + \tilde  e^2 \wedge \tilde e^3 \wedge \tilde e^6_D \right) \ .
 \eea 
Finally, the NS part comes from the torsion as in (\ref{eq:fH}) and is given by
 \beq
 H_3^D = \tilde e^1 \wedge   \tilde e^2 \wedge \tilde  e^5_{D} +  \tilde e^1 \wedge \tilde e^4 \wedge \tilde e^6_{D} +  \tilde e^2 \wedge \tilde e^3 \wedge \tilde e^6_{D} \ .
\eeq
We can therefore see that one of the type B requirements, namely 
$*F_3=e^{-\phi} H_3$ is satisfied, since 
$e^{-\phi_D}=e^{-\phi} \sqrt{|g_f|}=\frac{e^{-2A}}{\gs} e^{2A} t_3 \rm{Re} \tau^+$.
For a supersymmetric Type B solution the 3-form fluxes $F_3$ and $H_3$  
must also be  primitive, and have no (0,3) or
(3,0) component. From (\ref{F3d}) we see that $*F_3^D$ 
(and therefore of $F_3^D$)
is primitive if $dJ$ is primitive, which is one of the supersymmetry conditions
of the type C solution ($dJ^2=0$). Furthermore, $F_3$ will have no  
(3,0) and (0,3) components if $dJ$ does not have them. The compatibility 
condition $J \wedge \Omega_3=0$  implies that the SU(3) singlet component 
in $dJ$ is proportional to the singlet component in
$d \Omega_3$. Since the type C solution requires $d\Omega_3=0$, this singlet 
is zero,
and therefore $dJ$ and consequently $F_3^D$ have no singlet component.
The T--dual solution is therefore a supersymmetric type B solution.  

Once more, we refer to the Table \ref{results} for the full list of solutions of this type.  Basically for every
twisted torus with $b_1=4$ (i.e. with only two non-closed one-forms, 
$e^5$ and $e^6$),
admitting an orientifold in 56, there is such a solution. It is actually 
very easy to see that for the dual models $*F_3^D=H_3^D$ 
since (up to factors of $t$'s, $\tau$'s and $\gs$)
\beq
*F^D_3=-*_2 dJ=-*_2 d(e^5 e^6)=-*_2 (f^5_{ij} e^6 e^i e^j - f^6_{ij} e^5 e^i e^j )=
f^5_{ij} e^5 e^i e^j +   f^6_{ij} e^6 e^i e^j=H_3^D 
\eeq
Primitivity and absence of singlets are again guaranteed by primitivity 
and absence of singlets in $dJ$.

\vskip  0.4cm

\noindent
{\sl (n 4.6) with O5--planes, SU(2) structure, $\N=2$} $\,\,\,\,$  Our next two examples, both involving type 1 -- type 2 pure spinors, we will 
discuss from the ``other end''. Namely we will start from a type B solution 
on $T^6$ with O3--planes, perform two T--dualities (these will be of a different type than in the previous case) and arrive at a localized solution on a nilmanifold. We have chosen to present two different solutions with different amounts of supersymmetry on the same nilmanifold.

Let us start from a configuration given by
 \bea
J&=&  e^1\wedge  e^4- e^5 \wedge e^2+e^6 \wedge  e^3 = e^{-2A} \left( \tilde e^1\wedge \tilde e^4-\tilde e^5 \wedge \tilde e^2 +\tilde e^6 \wedge \tilde e^3\right) \nn \\
\Omega_3&=& -( e^1+i  e^4)\wedge ( e^5-i  e^2)\wedge ( e^6+i  e^3) \nn \\
\gs F_3 &=& \tilde e^2 \wedge \tilde e^4 \wedge \tilde e^5 +\tilde e^3 \wedge \tilde e^4 \wedge \tilde e^6 \nn \\
H_3 &=&  \tilde e^1 \wedge   \tilde e^3  \wedge \tilde e^6 + \tilde e^1\wedge \tilde e^2\wedge \tilde e^5 \nn \\
\gs F_5 &=& e^{4A} * d(e^{-4A})
\eea
It is not hard to check that it does correspond to a solution of B type, and has  $\N=2$ supersymmetry.
From $J$ and $\Omega_3$ we construct $\Phi_\pm$ for an O3 as in Table 2, which
transform under T--duality by a Hodge star on 56. 
Smearing  the warp factor in directions $56$,  we can perform two  T--dualities in these directions. The dual fields are
\bea
\Phi_-^D &=& e^A (   e^1+ i    e^4) \wedge e^{-i(  e^5 \wedge e^2 + e^6 \wedge e^3)} \nn \\
\Phi_+^D &=& i e^A   ( e^5+ i  e^2)  ( e^6- i  e^3) e^{-i e^1 \wedge e^4}   \nn \\
\gs F_3^D &=& -\tilde  e^2\wedge  \tilde  e^4\wedge  \tilde  e^6+ \tilde  e^3\wedge  \tilde  e^4\wedge  \tilde  e^5 +
e^{2A} *_4 d(e^{-4A}) \nn \\
H_3^D&=&0
\eea
where $e^5$ and $e^6$ are now T--dual vielbeine, ($e^{5,6}=e^{A} \tilde e^{5,6}$)
 and the structure constants are given by (0,0,0,0,12,13). The T--dual fields
 have the form of type 1--type 2 pure spinors for an O5, and solve our eqs. (\ref{12O5}). 
Bianchi identities read
\beq
dF_3^D=\frac{1}{\gs} (2 +  \tilde \nabla_-^2 (e^{-4A})) \tilde e^1\wedge  \tilde  e^2\wedge  \tilde  e^3 \wedge \tilde  e^4=
2 (\sum_{i=1}^{16} \delta(x-x^i) -\sum_{i=1}^N \delta(x-x^i)) \tilde  e^1\wedge  \tilde  e^2\wedge  \tilde  e^3 \wedge \tilde  e^4 \ ,
\eeq
where the first term in the last equality comes from the orientifold planes located at 
each of the 16 fixed points $x^i$, and the second term is the charge of $N$ D5-branes
located at $x^i$. In the absence of branes, tadpole cancellation fixes $\gs=1/16$. 
The constant contribution is the T--dual of the type B flux effective D3-charge 
$H_3 \wedge F_3= \frac{2}{\gs}  \tilde  e^1\wedge \tilde  e^2\wedge  \tilde  e^3 \wedge \tilde  e^4 \wedge 
\tilde  e^5 \wedge \tilde  e^6$,
and fixes $\gs$ to the same value in the absence of D3--branes. 
On the T--dual side, we can see that this solution is $\N=2$ because
the following pair $\Phi'_\pm$
\bea
\Phi'_- &=& e^A (   e^1- i    e^4) \wedge e^{-i(  e^6 \wedge e^3+ e^2 \wedge e^5)} \nn \\
\Phi'_+ &=& -i e^A e^{ i e^1 \wedge e^4}  ( e^6- i  e^3)  ( e^5- i  e^2)  
\eea
determines the  same metric as $\Phi^D_\pm$  and is also a solution to  the equations for the same flux $F_3^D$. 

Even if the solution involves pure spinors of type 1 and 2,  it is close to the previous type C example since only the RR three-form flux is non-vanishing. As for the previous case, there is a one-to-one correspondence between two solutions, i.~e.~to every supersymmetric type B solution there is a T--dual supersymmetric type C.
 
 \vskip 0.4cm

\noindent
{\sl (n 4.6) with O5--planes, SU(2) structure, $\N=1$} $\,\,\,\,$ Via T--duality we can also obtain ``static SU(2) structure'' (type 1--type 2 pure spinors but this time with $\N=1$ supersymmetry)  O5 solutions but with non zero $H$ flux.  As before, we start from  type B parent solution on $T^6$ with O3--planes
\bea
\label{t-solu}
J&=& \left(e^1 e^4+e^6 e^3-e^5 e^2\right) \nn \\
\Omega&=&  (e^1+ie^4)  (e^5-ie^2) (e^6+ie^3) \nn \\
\gs F_3 &=& \tilde e^1 \tilde e^4 \tilde e^5- \tilde e^1\tilde e^4 \tilde e^6 - 2  \tilde e^1 \tilde e^5 \tilde e^6-\tilde e^2 \tilde e^4 \tilde e^6+ \tilde e^2 \tilde e^5 \tilde e^6- \tilde e^3 \tilde e^4 \tilde e^5- \tilde e^3 \tilde e^5 \tilde e^6 \nn \\
H_3 &=& -\tilde e^1  \tilde e^2 \tilde e^4-\tilde e^1 \tilde e^2 \tilde e^6-\tilde e^1 \tilde e^3 \tilde e^4 -\tilde e^1 \tilde e^3 \tilde e^5- 2 \tilde e^2\tilde e^3 \tilde e^4 +\tilde e^2\tilde e^3 \tilde e^5 +\tilde e^2\tilde e^3 \tilde e^6\\
\gs F_5 &=& e^{4A} * d(e^{-4A})
\eea
which has $\N=1$ (we have omitted the wedges between the forms). Smearing again  the warp factor in directions $56$, and performing two T--dualities in these directions, we arrive at the dual configuration with the dual fields given by
\bea
\Phi_-^D &=& -
e^A 
( e^1+ i  e^4) e^{-i (- e^2 \wedge   e^5 +  e^3  \wedge e^6)}\nn \\
\Phi_+^D &=& i e^A ( e^5+ i  e^2) \wedge   ( e^6- i e^3) \wedge   e^{-i e^1 e^4}   \nn \\
\gs F_3^D &=& -\tilde e^1\tilde e^4 \tilde e^6-\tilde e^1\tilde e^4 \tilde e^5-\tilde e^2\tilde e^4 \tilde e^5+\tilde e^3\tilde e^4 \tilde e^6 
+ e^{2A} *_4 d(e^{-4A}) \nn \\
\gs F_1^D &=& e^{2A} (2 \tilde e^1- \tilde e^2+\tilde e^3) \nn \\
H_3^D&=&-\tilde e^1 \tilde e^2 \tilde e^4-\tilde e^1 \tilde e^3 \tilde e^4- 2 \tilde e^2\tilde e^3 \tilde e^4
\eea
The structure constants one gets from T--dualizing the $H$ flux in (\ref{t-solu}) are  (0,0,0,0,-13+23,-12+23). This set of structure constants is not to be found in the Table \ref{ta:nil}, yet after  doing some change of coordinates we arrive at more canonical  model given by (0,0,0,0,12,13). This solves our eqs. (\ref{12O5}), and the  equations of motion for the fluxes (\ref{eqsH}), as well as the Bianchi identities. For the latter, the flux charge is cancelled by the O5 charge together with additional 11 D5--branes. On the O3-side, we
need (besides the orientifold planes), 11 D3--branes to cancel the flux contribution $H\wedge F_3$.

\vskip 0.4cm

\noindent 
{\sl  (n 3.5) with O6--planes, SU(3) structure} $\,\,\,\,$ Finally turning to IIA we just quote one of the solutions that is related to a type B solution on $T^6$ with O3--planes by three T--dualities.  Considering the nilmanifold given by 
(0,0,0,12,13,23), i.~e.~3.5 in Table \ref{ta:nil}, with an O6 projection acting along 456 we can write down the following solution:
\bea
\Omega_3 &=& (e^1+ie^6) \wedge (e^2-ie^5) \wedge (e^3 - 2 i e^4) \nn \\
J&=& e^1 \wedge e^6 - e^2 \wedge e^5 - 2 e^3  \wedge e^4 \nn \\
\gs F_2&=&2(\tilde e^1 \wedge \tilde e^6- \tilde  e^2 \wedge \tilde  e^5 + \tilde  e^3 \wedge \tilde  e^4) - e^{A} *_3 d(e^{-4A}) \nn \\
dF_2&=& \frac{1}{\gs} (-6 +  \nabla^2_- 2 (e^{-4A})) \tilde e^1 \wedge \tilde e^2 \wedge \tilde e^3
\eea
A generic solution compatibile with supersymmetry and the orientifold projection would have 5 free moduli (closure of $\R \Omega$ imposes 3 equations for the nine real moduli $\tau^i_j$, while closure of $J$ imposes another equation). The solution we give above is not the most general one in that the 5 moduli have been fixed. 

Notice that there are no solutions for O6 and 
SU(2) structure obtained by T--duality. A priori one could get them, for 
example, starting from a complex structure that couples 5 and 6 
on a single holomorphic coordinate, and $J=56+...$, and doing T--duality 
in 456. It so happens, however, that there are none. 
Once more, for a full list of IIA solutions the Table \ref{results}  can be consulted.

\section{Further possibilities}

We have concentrated so far on four--dimensional Minkowski vacua partly due 
to the internal space having a generalized Calabi-Yau structure.  Extending 
our analysis to $AdS_4$ while involving very similar methodology requires 
in general some modification of the internal geometry. We are not going to do 
this here. Instead as we will see now one particular class of $AdS_4$ 
solutions does allow a GCY internal space, and this is exactly the case 
we are going to present here. In this Section we shall also discuss 
flat solvmanifolds and solutions on non--compact spaces.

\subsection{AdS vacua on SU(3) structure generalized Calabi-Yau manifolds}

The supersymmetry equations for 
the pure-spinor given in (\ref{int}) and (\ref{nonint})  
 were derived under the condition of having four
dimensional Poincar\'e invariance. However they can be easily
generalized to include a term related to a 
non--zero four--dimensional cosmological constant
\cite{gmpt2}. The contribution of the cosmological constant modifies
(\ref{int}) and (\ref{nonint}):
\bea
\label{muint}
(d-H\wedge) (e^{2A-\phi}\Phi_+)&=&
- 2\mu\,e^{A-\phi}\,\mathrm{Re}(\Phi_-) \, ,
\\
\label{munonint}
(d-H\wedge) (e^{2A-\phi}\Phi_-)&=&
-3i e^{A-\phi}  \,\mathrm{Im}(\bar \mu \Phi_+)
+ e^{2A-\phi} dA \wedge \bar\Phi_- + \frac i{8}e^{3A} *\lambda(F) \, 
\eea
where $\mu$ is related to the cosmological constant
$\Lambda$ as $\Lambda =- |\mu|^2$. These equations are derived in Appendix
 \ref{ap:N=1calc}; the change of variables (\ref{eq:redefA}) has been 
done in IIA.

As we can see, equation (\ref{muint}) does not imply any longer that the manifold
is generalized Calabi--Yau, as its counterpart with $\mu=0$, (\ref{int}), does. 
There is still some kind of geometrical interpretation, though. From (\ref{muint}), 
(\ref{munonint}) it follows that
\begin{equation}
    \label{eq:ghf}    
(d-H\wedge) {\rm Re}(i\,{\bar \mu} \,e^{2A-\phi}\Phi_+)=0=(d-H\wedge) {\rm Re}(
e^{A-\phi}\Phi_-)\ .
\end{equation}
An SU(3) structure manifold obeying the conditions
\begin{equation}
    \label{eq:hf}
    d{\rm Re}(e^{iJ})=0=d{\rm Re}(\Omega)
\end{equation}
is a {\it half--flat} manifold. 
In view of (\ref{eq:ghf}), one might want to call the general case (twisted) ``generalized
half--flat''. In particular, consider the SU(3) structure case. The phase of
$\mu$ does not appear in the cosmological constant, and we can tune it so 
that (\ref{eq:ghf}) implies (\ref{eq:hf}). A half--flat manifold has indeed been 
proposed in \cite{bobby} as the ten--dimensional realization of the four--dimensional AdS vacua
of \cite{K-taylor}. The appearance of half--flat geometries  looks curious 
in light of their role as mirrors of Calabi--Yau's with $H$ flux \cite{glmw};  this allows for 
an explicit construction  of half--flat manifolds starting from CY geometry
\cite{t}.

The general AdS solution does not involve therefore a GCY, and 
this is the reason we 
have not considered the AdS case so far in this paper.
However, it was shown in \cite{K-taylor}, that it is possible to have  
an  AdS solution of type IIA theory (in the large volume limit) on a Calabi--Yau --- specifically  
an (untwisted)  $T^6$. The
cosmological constant is 
generated by a combination of the singlet components of the  $H$ flux, 
$F_0$ and $F_4$. 
It is then natural to ask whether there exist other $AdS_4$ vacua where the 
internal manifold is a GCY and in particular a nilmanifold or a solvmanifold.
As we will see this is possible only in type IIA and for SU(3) structure and thus 
for type 0 -- type 3 pure spinors. 
$AdS_4$ solutions of IIA have been analyzed in detail 
in \cite{dimitris}. 
We will give here a parallel analysis in terms of the pure spinors. 

For type 0-type 3 pure spinors, eq. (\ref{muint}) gives 
the following one- and three-form
conditions (the five-form equation is implied by  the previous two):
\bea
\label{ads-even}
&& d(3A-\phi) = 0 \ ,\nonumber \\
&& dJ - iH = - 2 i {\hat \mu} e^{-A} {\rm Re}  {\hat \Omega}_3 \ ,
\eea
where we have defined ${\hat \Omega}_3 = - i e^{i(\alpha + \beta)} \Omega_3$,
and $\hat
\mu=\mu e^{-i(\alpha -\beta)}$ ($\alpha$ and $\beta$ being the phases
of $a$ and $b$). From the same equation one also gets 
that the difference $\alpha - \beta$ must be constant\footnote{In the O6 
equations in Section 5 we also give the equations in terms of $\hat{\Omega}_3$ 
but with the ``gauge  choice''
$\alpha+\beta=\pi/2$.}.

Turning to the equation for $\Phi_-$ and introducing ${\hat \mu} = m + i {\tilde m}$ we get the following four conditions:
\bea
\label{ads-odd}
&& 3 {\tilde m}  - e^{4A} *F_6  = 0 \ ,\nn \\
&& 3 m J + e^{4A}  *F_4 =0 \ ,\nn \\
&& \frac{3}{2} {\tilde m}  J^2 + e^{4A} *F_2 = - d(e^A  \rm{Im} \hat \Omega_3)
 \nn \\
&& \frac{1}{2}  m J^3 - e^{4A} *F_0   =  e^A H \wedge
\rm{Im}{\hat \Omega}_3 \ ,
\eea
where we have used (\ref{ads-even}) to simplify the system.  Clearly, the RR equations of motion $(d+ H)(e^{4A}*F)=0$ follow straightforwardly, while
the Bianchi identities are yet to be imposed.\footnote{Recall that
$F=(d-H)\wedge C - F_0 e^B$ and in configurations with non-vanishing $F_0$
the bare potential $B$ becomes important.}

The equation of motion for the NS flux (\ref{eqsH}) simplifies to 
\beq
3m (d -  J\wedge *d) {\rm Im} (e^A \hat \Omega) + 12  m {\tilde m} J \wedge J = -8m dA \wedge e^A \hat \Omega.
\eeq
It is satisfied by $dA=0$ and 
\beq
\label{ads-dO}
d \hat \Omega =  i \left( W_2^- \wedge J -  \frac{4 {\tilde m}}{3} e^{-A} J \wedge J \right) \, ,
\eeq
where $W_2^- $ is a real primitive (i.e. zero when wedged with $\hat \Omega$ or $J\wedge J$) form in the representation {\bf 8} of SU(3), and the second
term is consistent with the expression for $dJ$ in (\ref{ads-even}).  It
is not hard to check that (\ref{ads-dO}) also solves the $F_4$ Bianchi identity (which requires that $F_2 \wedge H \sim {\rm Re} \hat \Omega \wedge  (*d{\rm Im} \hat \Omega) =0$).

Finally we can  collect everything and write down the general supersymmetric solution of the equations of  motion:
\bea\label{ads-sol}
&3A = \phi = const \, , \qquad  
&dJ =  2 {\tilde m}e^{-A} {\rm Re} \hat \Omega \, , \qquad  H  =  2 {me^{-A}} \rm{Re} \hat \Omega \, , \nn \\
&F_0 =  5 m e^{-4A} \, , \qquad \qquad 
&F_2 =  -  e^{-4A} * \left( e^A d {{\rm Im} \hat \Omega_3}  + \frac{3 {\tilde m}}{2} J \wedge J \right)\, , \nn\\
&F_4 = 
\frac{3m}{2} e^{-4A} J\wedge J \, , \qquad &F_6=  - \frac{\tilde m}{2}  e^{-4A} J \wedge J  \wedge J \, .
\eea
The solution is subject to the last constraint coming from the Bianchi identity for $F_2$,
\begin{equation}
    \label{eq:bianchif2}
dF_2 - F_0 H = Q \delta_3 \, .
\end{equation} 
Here $\delta _3$
denotes the Poincar\'e dual to the possible localized codimension three
sources (O6--planes), and $Q$ is a numerical coefficient that may be zero.
This equation can be rewritten in a form that makes 
the constraints on the geometry obvious. Using the fact that $\hat \Omega_3$ is 
imaginary  anti-self-dual and
$d {\rm Re} {\hat \Omega}_3=0$, we obtain
\beq
\label{ads-omega}
( \Delta + 10 e^{-2A}  m^2 - 6 e^{-2A}  {\tilde m}^2) {\rm Re}  {\hat \Omega}_3 = - Q \delta_3
\eeq
where $\Delta = d\delta + \delta d$ is the Hodge operator.

Notice  that by taking $\hat \mu$ to be real, we find ourself in a situation 
where the twisted generalized half--flat manifold happens admits a GCY 
structure (albeit not twisted, in spite of having $H$-flux). Indeed, it 
is not hard to see that by taking $\tilde m = 0$, we find the conditions of  
closure of  $J$ (and thus  $d \Phi_+ = 0$) and $ {\rm Re} {\hat \Omega}_3 $, 
familiar from analyzing the Minkowski vacua. The flux $F_6$ now vanishes, 
and  the two-form flux has only the primitive component, while the remaining 
RR and NS fluxes all contain only singlets. The only new equation to
consider is (\ref{ads-omega}). Differently from the Minkowski
case this has two types of solutions, with or without O6--planes,
respectively.

\medskip
\noindent
$\bullet \,\,\,\,\,$ {\sl Backgrounds with O6--planes.} The simplest
geometrical situation corresponds to a closed ${\hat \Omega}_3$. This is 
consistent with the supersymmetry equations  provided that $F_2=F_6=0$ 
and the  tadpole is cancelled.  The dilaton and warp factor are fixed in the 
solution, the only non-vanishing fields are $H$, $F_0$ and 
$F_4$ and the only active component is the singlet, so the tadpole 
analysis is not hard: (\ref{eq:bianchif2}) becomes 
$F_0 H= -Q \delta_3$.
The obvious question to ask now
if we can find any geometry, other than the flat $T^6$, that admits {\it
two} closed pure spinors, and the right involution.
One such solution is given by the solvmanifold (25,-15,-45,35,0,0), the 2.5 of Table \ref{ta:solv} (taken here with $\alpha = -1$) . It is not hard to 
check that 
the real two-form $J=e^1 \wedge e^3-e^2 \wedge e^4+e^5 \wedge e^6$ and
holomorphic three-form $\hat{\Omega}_3=- (e^1+i e^3) \wedge (e^2-i e^4) \wedge (
e^6-ie^5)$ are both closed, while the corresponding pure spinors are
compatible. Finally, they transform in the correct way 
under the action of the involution 
corresponding to an O6--plane in the directions 345: $\sigma(e^{-iJ})= e^{iJ}$
and $\sigma (\hat{\Omega}_3) = - {\bar{\hat{\Omega}}}_3$. We can
therefore build an $AdS_4$ solution
\`a la \cite{K-taylor} on this twisted torus; see 
also \cite{bobby} for a similar analysis.  From the other side  
it is  $ {\rm Re}  {\hat \Omega}_3$ that gets balanced by the  localized 
sources. Notice that $ {\rm Re}  {\hat \Omega}_3$  contains  a part aligned 
with the O6--plane source (- - -) but also pieces orthogonal to it, in (+ + -). 
The projection into the top forms is easy and tells us that all the 
contributions come with the same sign, and thus in the solution multiple 
intersecting O6--planes 
are needed, wrapping the directions 126, 346, 325 and 145. The pure spinors
above transform as they should also under these additional projections.

\medskip
\noindent
$\bullet \,\,\,\,\,$ {\sl Backgrounds without O6--planes.} Eq. (\ref
{ads-omega}) allows for another type of solutions --- if both terms in the
left hand side of the second equation are non-trivial, a cancellation
between the two is possible and so is a compactification to $AdS_4$ without
O--planes. This in principle increases the number of possibilities --- looking
at e.g. 26 symplectic nilmanifolds we do not have to worry about the
compatibility of the structure constants and the involution. In practice
finding closed $J$ and $ {\rm Re}  {\hat \Omega}_3 $ is already very hard. 
So is solving (\ref{ads-omega}). 
Our search for solutions for this situation has not been as exhaustive as for
the Minkowski vacua, and so far we have not found any solutions to
(\ref{ads-omega}) of this type.

\subsection{Flat manifolds }
\label{flat}

Let us consider again the manifold $s$ 2.5 defined by (25,-15,-45,35,0,0). 
As already mentioned, it has closed compatible forms 
$J=e^1 \wedge e^3-e^2 \wedge e^4+e^5 \wedge e^6$ and
$\hat{\Omega}_3=- (e^1+i e^3) \wedge (e^2-i e^4) \wedge (
e^6-ie^5)$. Moreover,   we can check explicitly that the manifold is flat. 
 As we mentioned before, this should not come as a surprise, since 
all Ricci--flat homogeneous
spaces are flat \cite{flat}. 

As we just saw Ricci--flat solvmanifolds with
closed $J$  and $\Omega$ may lead to solutions relevant for $AdS_4$ vacua.
By virtue of being Ricci flat, these manifolds also would allow for
Minkowski vacua (of course, without any RR fluxes and thus hard tadpole
conditions), albeit with $\N=2$ supersymmetry.
From the four--dimensional effective action point of view (e.g. minimization of $\N=4$
superpotential) these are as hard to get as any others.  Yet while being
compact they easily circumvents the no-go theorems due to the absence of RR
fluxes.

A classification of lower dimensional ($d=3,4,5$) compact flat solvmanifolds exists \cite{Morgan}. 
Not counting  straight tori $T^d$, there are 6 compact flat solvmanifolds in 
three dimensions, 20 in four, and 62 with the first Betti number higher than one in
five. These numbers appear to be higher than one would naively expect --- we had seen there exist only 
two compact tree-dimensional solvable (not nilpotent) algebras, one four--dimensional one, 
four five-dimensional ones and eight six--dimensional ones. 
Thus the number of flat $d$--dimensional solvmanifolds is much higher
than that of $d$-dimensional compact algebras. So many of such manifolds
are obtained as cosets of higher dimensional algebras. Due to this fact, 
even if there is a full classification, 
it is not given in terms of structure constants and it is not obvious that 
these admit involutions compatible with O6
planes and are thus suitable for compactifiations. The example mentioned here 
does correspond to one of the five-dimensional compact solvable 
algebras 
(trivially extended by $S^1$). It is also not hard to check that the thee-dimensional 
compact algebra $E_2$ is flat as well.

\subsection{Solvmanifolds beyond our simple search}
\label{sec:scope}

As we remarked in Section \ref{sec:solvm}, while our search  for vacua was systematic for internal spaces given by nilmanifolds, this was not the case for solvmanifolds. The compactness criterion of  \cite{saito}  requires sometimes that a particularly nice basis for the Lie algebra be found, whereas
 for    assembling  the Table \ref{ta:solv}  we just applied the criterion to the bases provided by the classification of solvmanifolds \cite{Tarko,Mubara5}. Since our search was not systematic, we did not consider all possible compact solvmanifolds. In this section, we will consider in more detail an example of solvmanifolds that goes beyond our simple search. This example was already considered in the literature \cite{Camara}; on this space, a supersymmetric Minkowski solution can be found. 
This is a solution of Type IIA with type 0-3 pure spinors, where the solvmanifold is
defined by  (0,13,12,0,-16,-15).\footnote{By coordinate
redefinitions this can be made (25,15,45,35,0,0) and differs from the
flat compact case by signs only. Those signs are such that in the former
the adjoint representation on the nilradical contains  two copies of $E_{1,1}$  (see discussion around
eq.(\ref{e11})) while the flat compact case it contains two copies of $E_2$. In a previous version of this paper, it was erroneously claimed that the example in this section was noncompact. This was due to the mistake commented on in footnote \ref{foot:mistake}.} To establish compactness, we can apply Saito's criterion, as described in Section \ref{sec:solvm}. First of all notice that the nilradical is given by the span of $\{e^2,\ldots,e^6\}$. 
Following the construction in \cite[Version 1]{haque-shiu-underwood-vanriet}, we then 
define the basis $\eta^1= \frac1{\sqrt{3}}e^1$, $\eta^2=\sqrt{3}e^2$, $\eta^3 = e^3$, $\eta^4=\frac1{\sqrt{3}}e^4$, $\eta^5=\sqrt{3}e^5$, $\eta^6 =- e^6$. In this basis, the group $P(G)$ is given by elements of the form
\begin{equation}\label{eq:M-cfi}
	\exp \left[ a_1 \left(  \begin{array}{ccccc}
		0 & 3 & 0 & 0 & 0 \\
		1 & 0 & 0 & 0 & 0 \\
		0 & 0 & 0 & 0 & 0 \\
		0 & 0 & 0 & 0 & 3 \\
		0 & 0 & 0 & 1 & 0 
	\end{array}
	\right) \right]
	=
	\left( \begin{array}{ccccc}
		 \cosh(\sqrt{3} a_1) &\sqrt{3}\sinh(\sqrt{3} a_1) & 0 & 0 &0 \\
		 \frac1{\sqrt{3}}\sinh(\sqrt{3} a_1)&\cosh(\sqrt{3} a_1) &0 & 0&0 \\
		0&0 &1 &0 &0 \\
		0&0&0 &\cosh(\sqrt{3} a_1) & \sqrt{3}\sinh(\sqrt{3} a_1)\\
		0 &0 &0 &\frac1{\sqrt{3}}\sinh(\sqrt{3} a_1) & \cosh(\sqrt{3} a_1)
	\end{array}\right)\ .
\end{equation}
This matrix is integer--valued for $a_1=\alpha$, where $\alpha$ is defined by  $\cosh(\sqrt{3}\alpha)=2$. Hence $P(G)/[P(G)]^{\Bbb Z}={\Bbb R}/{\Bbb Z}$ is compact, and Saito's criterion tells us that the space can be compactified. 

We can make the construction a little more explicit (again following \cite[Version 1]{haque-shiu-underwood-vanriet}), and identify the basis of globally-defined one-forms. To do this, we can realize the forms 
$\eta^i$ defined earlier as 
\begin{equation}\label{eq:etadx}
	\eta^i= (M^{-1}(x^1))^{ij}d x^j\ ,
\end{equation}
where $M$ is the matrix in  (\ref{eq:M-cfi}), and we identify $x^1=a_1$. The cocompact subgroup $\Gamma$ is then defined by the five translations $T_i: x^j\to \delta^j{}_i x^j$, $i=2,\ldots,6$, and by the operation
\begin{equation}\label{eq:T1}
	T_1: \left\{ \begin{array}{c}\vspace{.3cm}
		x^1 \to x^1 + \alpha\\
		x^i \to (M(\alpha))^{ij} x^j\, \quad i=2,\ldots,6\ .
	\end{array}\right.
\end{equation}
The forms $\eta^i$ are well--defined on the compact space defined in this way. Let us see why. A way to understand the transformation (\ref{eq:T1}) is to say that there is a transition function that glues the coordinates $\tilde x^i$ after one monodromy  $x^1 \to x^1  + \alpha $ to the coordinates $x^i$ before the monodromy, according to the law $\tilde x^i =(M(\alpha))^{ij} x^j$. One--forms will similarly be glued according to $d\tilde x^i =(M(\alpha))^{ij} dx^j$. The forms $\eta^i$ will then be glued according to $\tilde \eta^i=(M^{-1}(x^1+\alpha))^{ij} d \tilde x^j=(M^{-1}(x^1))^{ij}(M^{-1}(\alpha))^{jk} d \tilde x^k=(M^{-1}(x^1))^{ij}(M^{-1}(\alpha))^{jk} (M(\alpha))^{kl} d x^l=(M^{-1}(x^1))^{ij}dx^j=\eta^i$. So each of the forms $\eta^i$ is actually glued to itself under monodromy. This means that each of them is globally well--defined, and then also that each member of our original basis $e^i$ is well--defined.

We can now give the SU(3) structure data of the supersymmetric solution. We have
\bea
\label{non-comp}
\Omega &=& \prod_i (e^i + i \tau^i \hat e^i) \ ,\nn \\
J &=& \frac{i}{2} \frac{t_i}{\tau^i}  \, z^i \wedge \bar z^i \ ,
\eea
where  $e^i=(e^4,e^5,e^6)$,  $\hat e^i=(e^1,e^2,e^3)$ and $z^i = (e^i + i
\tau^i \hat e^i)$. $\tau_i$ and $t^i$ are real parameters. 

It is not hard to see that ${\rm Re} \Omega$ is closed, and so is $J$ under
the condition
\beq
t_2 =t_3  \ .
\eeq
The two-form calculated  by taking the Hodge star of $d {\rm Im} \Omega$ (see (\ref{03O6}))  is
$$ F_2=-\frac{1}{\gs \, t_1}(\tau^2+ \tau^3) (e^2 \wedge e^5- e^3 \wedge e^6)$$ and
the Bianchi identity gives $dF_2=-\frac{2}{\gs \, t_1} (\tau^2 + \tau^3) \times
(e^1\wedge e^3 \wedge e^5- e^1 \wedge e^2 \wedge e^6)$.  There are
two contributions to the tadpole, and in order to see whether 
they should be cancelled by D6--branes or O6--planes it is easier to
project to the singlet component (the top-form) by wedging
$dF_2$ with ${\rm Im} \Omega$. Using the fact that $t_1 t_2 t_3=1$, one can see that both sources are due to O6--planes. These planes suggest two orientifold projections, and one can then check that the SU(3) structure forms transform in the correct way under both, without any further restriction on the parameters.

\section{Discussion}

In this paper we have found new $\N=1$ vacua of type II theories by solving directly 
the ten--dimensional supersymmetry conditions for all compact cosets $G/\Gamma$, with $G$
generated by a six-dimensional solvable algebra. 
We believe these are
the first supergravity geometrical Minkowski vacua which are neither Calabi--Yau nor related to them by any duality. Solutions
that are not Calabi--Yau but are connected to them via dualities 
are much more common (see for example \cite{Kachru,Schulz,fy,bbfty,
das-tatar}). Another
class of non--Calabi--Yau vacua has been found by four--dimensional
methods in \cite{ckt}, but they involve small cycles, possibly
with high curvature, so they cannot be found in ten--dimensional 
supergravity alone.)

The reason we have chosen to work in ten dimensions is essentially twofold. First, working 
with four--dimensional effective theories as it is usually done cannot
presently address non--trivial warp factors.
Second, geometrical methods are available
which make the construction rather systematic. 
Showing this was one of the purposes of this paper.   Indeed, the construction 
of backgrounds is rather algorithmic. First one finds a closed pure spinor (see
eq.~(\ref{int}))
defining a generalized CY structure on a compact six--dimensional manifold (in this paper,
a solvmanifold $G/\Gamma$). For nilmanifolds, this step has been to some extent already carried out 
for us in \cite{CG}. Examples in which 
(\ref{int}) and (\ref{Re}) are satisfied in the case of solvmanifolds have recently been considered in \cite{tove,debto}. Then one
proceeds to build the most general pure spinor compatible with it, and such 
that 
eq.~(\ref{nonint}) is satisfied.  While its real part (\ref{Re}) is closed, the
imaginary part (\ref{Im}) defines the RR field strength. Imposing the Bianchi identity
on the latter is then the hardest part. 
 
The only situations in which we have managed to find complete localized solutions are 
those where the fluxes present are such that $(d-H\wedge)F$ contains only a single 
component. This is the case for solutions connected by T--duality to the IIB 
compactifications on a conformally flat $T^6$ with imaginary self--dual three--form flux. 
Examples of the large volume limits of T--dual solutions and the $AdS_4$ solutions have 
been constructed in the last few years. For the non--T--dual solutions, there are 
mutiple orthogonal sources (depending on the model, these can be either only O--planes or 
combinations of planes and D--branes).  Fully localized versions of these solutions are 
then notoriously very hard to find. 

It is curious that three of the five solutions that we find (and almost all of the fluxless solutions) are associated with 
the same manifold (2.5 of Table \ref{ta:solv}), that happens to admit a flat 
metric (even if in the multi-source solutions discussed here the metric is not flat). 
This is hardly a coincidence.  Since the manifold is of trivial holonomy, it admits 
eight covariantly constant spinors - just like $T^6$! 
Yet it supports nontrivial flux configurations which are not supported on the latter. 
The basic reason seems to be that on $T^6$ it is impossible to find a pair of compatible 
(left--invariant) pure spinors, where only one is not closed; while they 
do exist on $s$ 2.5 due to the 
non trivial twisting. Similarly, because of the non trivial twisting, some of the covariantly constant spinors are not constant, as well as some of the closed forms do not have constant coefficients (and are therefore
not left--invariant). 

Let us also make some remarks on the geometry. The proof of the 
existence of closed pure spinors in all six--dimensional nilmanifolds \cite{CG} has been one motivation for this work.
However, the supersymmetry equations lead  to two major modifications: in the presence of 
$H$ flux, we need a twisted closed rather than a closed pure spinor.  In addition to this, 
when passing from global to local solutions, taking into account the orientifold 
involutions, we weight different internal directions by different powers of the dilaton and
of the warp factor.  Hence, the GCY will not be left--invariant (that is, with constant
coefficients in the basis $e^a$ of forms provided by the nilpotent or solvable algebra). 
In this paper, we have chosen to look for GCY structures which are left--invariant, hoping
they will be one day completed to warped supersymmetric solutions. It might be, however,
that there exist solutions for which the GCY structure is not left--invariant in any limit. 

In scanning the possible solvmanifolds we have been systematic. 
We have, however, restricted our attention to those that, as well as being compact, 
have the following properties: 1)  the solvable group $G$ of which they are a 
quotient $G/\Gamma$ is algebraic; 2) they have a discrete isotropy group. 
The first requirement is needed in the criterion \cite{saito} 
that we have used to check that $G$
admits a subgroup $\Gamma$ so that $G/\Gamma$ is compact. That criterion only applies
to groups with that property (explained in Section \ref{sec:compact}). This could
be circumvented by using the criterion \cite{auslander}, which needs no such hypothesis. 
As for the second property, not all six--dimensional solvmanifolds are of the 
type $G/\Gamma$  where $G$ is six--dimensional.  Indeed, one can take a $6+d$ dimensional 
group $G$ with $\Gamma$ having in addition to six discrete dimensions $d$ continuous ones. 
The quotient would still be a six--dimensional compact manifold.  To the best 
of our knowledge, no models of this type have been discussed so far. 
Unfortunately a more systematic study of these would be  complicated due to 
the absence of classification of the higher dimensional solvable algebras. The 
relative abundace of compact flat manifolds  which are predominatly of this 
type \cite{Morgan} is some measure of the wealth of this unexplored class of 
manifolds. 

Another possible lack of generality is in the choice of the involutions used
for the orientifold projections. We have considered involutions that act as reflections
in the basis in which the algebras were given. There might be more general involutions. There certainly are some
obvious ones: for example, in the algebra in which the only non--trivial structure
constant is $f^6{}_{12}$ (5.2 in Table \ref{ta:nil}), one can exchange the two coordinates
$x^3$ and $x^4$. This is obviously compatible with the structure constants, but it is
related by a change
of coordinates to a sign flip on the coordinate $x^3$, and therefore physically equivalent.
We do not 
know whether there exist involutions that cannot be related by changes of coordinates
to the ones we considered.

Finally, it would be interesting to extend some of these considerations to the so--called
``non--geometric'' compactifications. In particular, from the present point of view
it is natural to consider the doubled formalism in \cite{hull} as a certain class
of twelve--dimensional cosets (whose structure constants should be invariant under
O$(6,6)$). The constraints we solved here for the manifold to be compact might
be relevant to those cases as well.

\vskip 1cm
\noindent
{\bf Acknowledgments:} We would like to thank O.~Aharony, M.~Berkooz, I.~Brunner, D.~Bump, A. ~Dabholkar, J.-P.~Derendinger,  T.~Grimm, C.~Hull, S.~Kachru, J.~Shelton,  W.~Taylor, D.~Tsimpis, G.~Villadoro,  D.~Waldram, B.~Wecht and F.~Zwirner for helpful discussions.  
We also thank D.~Andriot and C.~Bock for many comments that were very helpful for the revision (v3).
RM and AT would like to thank KITP, UC Santa Barbara, and RM -- the 
Weizmann Institute for hospitality during the course of this work. This work is supported in part by RTN contracts  MRTN-CT-2004-005104 and  MRTN-CT-2004-512194 and by ANR grants BLAN06-3-137168 (MG and RM) and BLAN05-0079-01 (MP); AT is supported by the DOE 
under contract DEAC03-76SF00515 and by the NSF under contract 9870115.

\appendix
\section{Pure spinor equations for $\N=1$ vacua}
\label{ap:N=1calc}
In this Appendix we sketch for completeness the computation that 
reformulates the supersymmetry equations as the pure spinor 
equations (\ref{int}), (\ref{nonint}). The equations are from \cite{gmpt2}.

We will show the IIB computations; the IIA case is almost identical. In fact, the
equations given in \cite{gmpt2} for type IIA become exactly the same as the ones
for IIB in terms of new fields
\begin{equation}
    \label{eq:redefA}
F_A'=-\lambda(F_A)\ \quad \Big(\Rightarrow i*\lambda(F'_A)=-i * F_A
\Big)\ ,\qquad \mu'=-\mu\ , \qquad H'=-H\ .
\end{equation}
where $\mu$ appears in the cosmological constant, see (\ref{eq:ksp}) below, but so 
that its sign is immaterial.

This computation might have been done by decomposing $\Phi=\sum_d \phi_d$, 
remembering that each $\phi_d$ is a spinor bilinear because of the Fierz 
identities 
(\ref{fierz}), and  computing the exterior derivative of each bilinear. 
Happily, this is not necessary: 
one can always work at the level of bispinors, without ever 
applying (\ref{fierz}).

First of all, since we are looking for vacua, the spacetime has 
maximal symmetry (it is AdS$_4$ or Minkowski$_4$); and, since we
are looking for $\N=1$ solutions, we can only use one spacetime spinor
$\zeta_+$ (and its Majorana conjugate $\zeta_-$). 
The ten--dimensional Majorana supersymmetry parameters
$\epsilon_{1,2}$ then decompose, without loss of generality, as
\begin{equation}
  \label{eq:spinor}
  \begin{array}{c}
\epsilon^1= \zeta_+\otimes \eta^1_+ + \zeta_-\otimes \eta^1_- \\
\epsilon^2= \zeta_+\otimes \eta^2_+ + \zeta_-\otimes \eta^2_- \\
  \end{array}\ .
\end{equation}
It will be convenient to choose a basis in which the internal gamma
matrices $\gamma_m$\footnote{Indices $M$ are ten--dimensional, 
$m$ are internal and $\mu$ are along the spacetime.} 
are all purely imaginary. Due to the symmetry, we can choose a basis
of four--dimensional spinors $\zeta$ that obey 
\begin{equation}
    \label{eq:ksp}
D_\mu \zeta_-= \frac12 \mu \gamma_\mu \zeta_+    
\end{equation}
so that the cosmological constant $\Lambda=-|\mu|^2$. 

One now starts from the gravitino variations \cite{democratic}:
\begin{equation}
  \label{eq:gravB}
\begin{array}{c}
\left(D_M - \frac14 H_M\right) \epsilon^1 +
\frac {e^\phi}{16} (\sla F_h +\sla F_a) \Gamma_M \epsilon^2 =0\\
\left(D_M + \frac14 H_M\right) \epsilon^2 +
\frac {e^\phi}{16} (-\sla F_h +\sla F_a) \Gamma_M \epsilon^1 =0
\end{array}  
\end{equation}
(the slash is ten--dimensional here; $H_M\equiv H_{MNP}\Gamma^{NP}/2$; 
$F_h= F_1+F_5+F_9$ and 
$F_a=F_3+F_7$ are somehow misnomers, in that they are not really hermitian 
and anti--hermitian). By using the self--duality of $F^{(10)}$ 
(more precisely, $F^{10}_n= (-)^{\mathrm{Int}[n/2]}*_{10} F_{10-n}^{10}$), 
the general splitting
(\ref{F10F6}) and the formula
\begin{equation}
  \label{eq:star}
  \sla C\ \gamma =i\ \slashh{\ \lambda(*C)}
\end{equation}
(remember that $\lambda$ is the transposition as defined in (\ref{defMukai}), 
$\slas{\lambda(C_k)}= (-)^{\mathrm{Int}[k/2]}\sla C_k$)
one can reexpress $(\pm\sla F_h + \sla F_a)$ in (\ref{eq:gravB}) in terms
of the internal $F$ alone. This gives
\begin{equation}
  \label{eq:intgravB}
  \left(D_m-\frac14 H_m\right) \eta^1_+ +
\frac{e^\phi}8 \sla F \gamma_m \eta^2_+=0\ ; \qquad 
\fbox{$\eta^1 \to \eta^2 \  ,  \ \ \sla F \to -\sla F^\dagger \ , \ \
H\to -H$ }\ .
\end{equation}
(The equation for $\eta^2$ is obtained by the boxed rule, which is valid 
for  the external gravitino and for the modified dilatino as well.) 
For $M=\mu$, one also has a contribution from (\ref{eq:ksp}): 
\begin{equation}
  \label{eq:extgravB}
\frac12 \mu  \eta^1_-+ 
\frac12 \,e^{A}\, \sla \,\del A\, \eta_+^1 -\frac18 e^{A+\phi} \sla F\eta^2_+=0 \ .
\end{equation}
Finally we have to look at the dilatino; but we prefer using the combination
$\Gamma^M \delta \psi_M - \delta \lambda$, which gives $(\sla D - 
\sla \,\del  \phi) \epsilon^1 -\frac14 \sla H \epsilon^1 $. In our case this gives
\begin{equation}
  \label{eq:moddilB}
2\mu\,e^{-A} \eta^1_- + \sla D\eta^1_+ + 
\Big(\sla \,\del (2 A - \phi) -\frac14 \sla H \Big)\eta^1_+=0\ .
\end{equation}

\subsection{$d\Phi_+$}
Let us now compute the exterior derivatives of the pure spinors defined
in (\ref{pureforms}). 
\begin{eqnarray}
 \nonumber
2\slas{\ \, d \Phi_+}&=&\{\gamma^m, D_m (\eta^1_+\eta_+^{2\,\dagger})\}= 
\sla D \eta^1_+ \eta_+^{2\,\dagger}
+ \gamma^m \eta^1_+ D_m\eta_+^{2\,\dagger} +D_m\eta^1_+ \eta_+^{2\,\dagger}\gamma^m
+\eta^1_+ \sla D \eta_+^{2\,\dagger} =\\
&&\nonumber
\Big(-2\mu\,e^{-A} \eta^1_- 
-\sla \,\del ( 2A - \phi)\eta^1_+ 
+\frac 14\, \sla H \eta^1_+\Big)\eta_+^{2\,\dagger}
+ \gamma_m \eta^1_+ \Big(\eta_+^{2\,\dagger} \frac {H_m }4 
+\frac18 \eta_+^{1\,\dagger}\gamma^m \,\sla F\Big) \\
&&\nonumber
+ \Big(\frac{H_m}4\eta^1_+ -\frac 18 \,\sla F\gamma_m \eta^2_+\Big)
\eta_+^{2\,\dagger}\gamma^m 
+\eta^1_+\Big(-2\bar\mu\,e^{-A} \eta_-^{2\,\dagger}
- \eta_+^{2\,\dagger}\,\sla \,\del(2A-\phi) +\frac14 
\eta_+^{2\,\dagger}\sla H\Big)=\\
&&\label{Fpart}
-4\mathrm{Re}\Big(\bar\mu\,e^{-A}\, \sla \Phi_-\Big)
-\{\sla \,\del(2A-\phi), \sla \Phi_+\} +\frac14 \Big[\{ \sla H, \sla \Phi_+\}
+ \gamma_m \,\sla \Phi_+ H^m +H_m \,\sla \Phi_+ \gamma^m \Big] \\
&&\nonumber
\frac{e^\phi}8 \gamma_m \eta^1_+\eta_+^{1\,\dagger} \gamma^m \sla F -
\frac{e^\phi}8 \sla F \gamma_m \eta^2_+\eta_+^{2\,\dagger}\gamma^m  \ .
\end{eqnarray}
The $H$ part reconstructs $H\wedge$. To see this, it proves very 
useful to use the Clifford$(d,d)$ techniques reviewed in Section 
\ref{Gentwist}, in particular (\ref{eq:gammamap}). 
Defining $\lambda^m\equiv dx^m\wedge\ , \ \iota_m\equiv \iota_{\del_m}$, 
we have, for any even form $C_\mathrm{ev}$\footnote{\label{foot:H}
The right actions 
in this computation and in (\ref{eq:Hwodd}) are subtle: one has signs from 
(\ref{eq:gammamap}), and from the fact that a right action inverts the
order. Also, an expression like $(\lambda\iota^2)^{mnp}$ does not
suffer from ordering ambiguities because 
it is multiplied by an antisymmetric form.}:
\begin{eqnarray}
  \label{eq:Hweven}
&&\{\sla H,\sla C_{\mathrm{ev}}\}+ 
\gamma^m \,\sla C_{\mathrm{ev}} H_m +
H_m \,\sla C_{\mathrm{ev}} \gamma^m = \\
&&\nonumber
H_{mnp}\Big[
\frac16\Big( (\lambda+ \iota)^3+(\lambda- \iota)^3\Big)
+\frac12\Big( (\lambda + \iota)(\lambda -\iota)^2 
+(\lambda +\iota)^2(\lambda-\iota)\Big)\Big]^{mnp}\sla C_{\mathrm{ev}}= \\
&&\nonumber
H_{mnp}\Big( \frac{\lambda^3}3 + \lambda \iota^2+\lambda^3-\lambda\iota^2\Big)
\sla C_{\mathrm{ev}}=\frac43 H_{mnp}\lambda^{mnp}\,\sla C_{\mathrm{ev}} 
= 8 \!\!\!\!\begin{picture}(10,10)(-10,5)
\put(0,0){\line(4,1){50}}
\end{picture}{ \ H\wedge C_{\mathrm{ev}}}\ . 
\end{eqnarray}
We will now massage the RR part (\ref{Fpart}). First of all, we
notice that one can expand the chirality projector
$1/2(1-\gamma)$ as
\begin{equation}
    \label{eq:1-g}
\frac{1-\gamma}2 = \eta_-\eta_-^\dagger +\frac12\gamma^m \eta_+ \eta_+^\dagger
\gamma_m\ ;    
\end{equation}
as one sees by applying both sides to the complete basis of spinors
 $\{\eta_\pm\ , \gamma^m \eta_\pm\}$\footnote{The $\frac12$ in the right hand 
side comes when multiplying by $\gamma^n \eta_+$ from the right,
which gives {\it twice} a holomorphic projector.}. Let us also  temporarily
pass to spinors of norm one, $\eta^1_+ \to a  \eta^1_+$, 
$\eta^2_+ \to b  \eta^2_+$. Using (\ref{eq:1-g}) and (\ref{eq:moddilB}),
(\ref{Fpart}) becomes then
\begin{eqnarray}
  &&\nonumber
\frac{e^\phi}8\Big[ |a|^2 \Big((1-\gamma)-2 \eta^1_-\eta_-^{1\,\dagger}\Big) \sla F
-|b|^2\sla F \Big( (1-\gamma) -2 \eta^2_-\eta_-^{2\,\dagger}\Big)\Big]=\\
  &&\nonumber
\frac18\Big[|a|^2 e^\phi(1-\gamma)\, \sla F -2\cdot 4\cdot \bar a\eta^1_-\Big(
\bar b \mu\,e^{-A} \eta_+^{2\,\dagger} -b \eta_-^{2\,\dagger}\sla \,\del A \Big) \\
&&\nonumber
-|b|^2 (1+\gamma) \,\sla F 
+2\cdot 4\cdot
\Big(\bar a \,\sla\,\del A\eta^1_- - a\bar \mu\,e^{-A} \eta^1_+\Big)b\eta_-^{2\,\dagger} \Big]=\\
  &&\nonumber
\frac{e^\phi}8
\Big[(|a|^2-|b|^2)\sla F -(|a|^2+|b|^2) \gamma \,\sla F\Big] 
+\bar a b \{ \,\sla \,\del A, \overline{\sla\Phi_+}\}
-2\mathrm{Re}\Big(\bar\mu\,e^{-A}\, a b \ \sla \Phi_-\Big)
\ .
\end{eqnarray}
(Notice the sign change in $(1+\gamma)\to (1-\gamma)$ while passing
through $\sla F$.)
We now get back to non--normalized spinors and reabsorb again 
$\bar a b$ and $ab$ in the pure spinors. 
Finally, we have to be careful in extracting the slash from the $*$
and from the complex conjugation $\overline{(\ )}$. 
For the $*$, (\ref{eq:star}) tells us $\gamma\, \sla F=-i\slashh{*\lambda(F)}$. 
For the $\overline{(\ )}$, we have to remember that gamma matrices 
have been taken to be purely imaginary: complex conjugation and slash
commute on even forms, but {\it anticommute} on odd forms. In particular,
 $\eta^1_-\eta_-^{2\,\dagger}= \overline{\sla\Phi_+}=
\sla {\bar\Phi_+}$, but $\eta^1_-\eta_+^{2\,\dagger}= \overline{\sla\Phi_-}=
- \sla {\bar\Phi_-}$. (Also, $\mathrm{Re}( \mu\ \overline
{\sla \Phi_-})= i \hspace{-.2cm}\begin{picture}(10,10)(-10,5)
\put(0,0){\line(4,1){40}}
\end{picture} \mathrm{Im}(\bar \mu \Phi_-)$\ ).  

Collecting everything, we get
\[
e^{-2A+\phi}(d-H\wedge)(e^{2A-\phi}\Phi_+)= -3i \,e^{-A}
\mathrm{Im} (\bar \mu \Phi_-)+ 
dA\wedge \bar \Phi_+ + 
\frac{e^\phi}{16}\Big[ (|a|^2-|b|^2) F + i (|a|^2 + |b|^2) *\lambda(F) \Big] 
\]
This is the form in which this equation was given in \cite{gmpt2}. We 
can actually
massage a little further by computing (using 
judiciously (\ref{eq:intgravB}) and (\ref{eq:extgravB})) that 
\begin{equation}
    \label{eq:dnorm}
d|\eta^1|^2=|\eta^2|^2\,dA \ , \qquad d|\eta^2|^2=|\eta^1|^2\,dA\ .    
\end{equation}
Recalling that $|a|^2=|\eta^1|^2$
and $|b|^2=|\eta^2|^2$, this gives that
 $|a|^2-|b|^2=c_- e^{-A}$ and $|a|^2+|b|^2=c_+ e^{A}$,
for $c_+>0$ and $c_-\geq0$ two integration constants. Hence we get:
\begin{equation}
    \label{eq:nonintapp}
\fbox{$e^{-2A+\phi}(d-H\wedge)(e^{2A-\phi}\Phi_+)= -3i 
\,e^{-A}\mathrm{Im} (\bar \mu 
\Phi_-)+ 
dA\wedge \bar \Phi_+ + \frac1{16}[
c_- e^{-A+\phi} F + i c_+ e^{A+\phi} *\lambda(F)]$\ }
\end{equation}
with $||\Phi||^2=\frac18||\eta^1||^2||\eta^2||^2=\frac1{32}(c_+^2e^{2A}-c_-^2e^{-2A})$
(see (\ref{eq:eqnorms}) and (\ref{eq:norm})).

\subsection{$d\Phi_-$ (IIB)}
This one is much simpler: indeed
\begin{eqnarray}
 \nonumber
2\slas{\ d \Phi_-}&=&[\gamma^m, D_m (\eta^1_+\eta_-^{2\,\dagger})]= 
\sla D \eta^1_+ \eta_-^{2\,\dagger}
+ \gamma^m \eta^1_+ D_m\eta_-^{2\,\dagger} -D_m\eta^1_+ \eta_-^{2\,\dagger}\gamma^m
-\eta^1_+ \sla D \eta_-^{2\,\dagger} =\\
&&\nonumber
\Big(-2\mu\,e^{-A} \eta^1_- 
-\sla\, \del ( 2A - \phi)\eta^1_+ 
+\frac 14\, \sla H \eta^1_+\Big)\eta_-^{2\,\dagger}
+ \gamma_m \eta^1_+ \Big(\eta_-^{2\,\dagger} \frac {H_m }4 
+\frac18 \eta_-^{1\,\dagger}\gamma^m \,\sla F\Big) \\
&&\nonumber
- \Big(\frac{H_m}4\eta^1_+ -\frac 18 \,\sla F\gamma_m \eta^2_+\Big)
\eta_-^{2\,\dagger} \gamma^m
-\eta^1_+\Big(2\mu\,e^{-A}\, \eta_+^{2\,\dagger}
- \eta_-^{2\,\dagger}\sla\, \del(2A-\phi) +\frac14 
\eta_-^{2\,\dagger}\sla H\Big)=\\
&&\nonumber
-4\mu\,e^{-A}\mathrm{Re}\Big(\sla \Phi_+\Big)
-[\,\sla \del(2A-\phi), \sla \Phi_-] +\frac14 \Big[ [ \sla H, \sla \Phi_-]
+ \gamma_m\, \sla \Phi_- H^m -H_m\, \sla \Phi_- \gamma^m \Big]\ ;
\end{eqnarray}
again the $H$ part reconstructs $H\wedge$, this time because (see footnote 
\ref{foot:H}):
\begin{eqnarray}
  \label{eq:Hwodd}
&&  [\sla H,\sla \,C_{\mathrm{odd}}]+ \gamma^m\, \sla C_{\mathrm{odd}} H_m - 
H_m \,\sla C_{\mathrm{odd}} \gamma^m = \\
&&\nonumber
H_{mnp}\Big[
\frac16\Big( (\lambda+ \iota)^3+(\lambda- \iota)^3\Big)
+\frac12\Big( (\lambda + \iota)(\lambda -\iota)^2 
+(\lambda +\iota)^2(\lambda-\iota)\Big)\Big]^{mnp}\sla C_{\mathrm{odd}}= \\
&&\nonumber
H_{mnp}\Big( \frac{\lambda^3}3 + \lambda \iota^2+\lambda^3-\lambda\iota^2\Big)
\sla C_{\mathrm{odd}}=\frac43 H_{mnp}\lambda^{mnp}\,\sla C_{\mathrm{odd}} 
= 8\!\!\!\!\! \begin{picture}(10,10)(-10,5)
\put(0,0){\line(4,1){50}}
\end{picture}{ \ H\wedge C_{\mathrm{odd}}} \ ;
\end{eqnarray}
crucially, the RR term has
disappeared because $\gamma_m \eta_+^i \eta_-^{i\,\dagger}\gamma^m=0$ -- 
a fact that follows from $\Phi_-$ being a three--form for any SU(3) structure
and from 
\begin{equation}
  \label{eq:gmgm}
\gamma_m \,\sla C_k \gamma^m= (-)^k(6-2k)\,\,\sla C_k  \ .
\end{equation}
Summing up, we get
\begin{equation}
    \label{eq:intapp}
\fbox{$ e^{-2A + \phi} (d- H\wedge) (e^{2A-\phi}\Phi_-) = 
-2\mu \,e^{-A}\mathrm{Re}
(\Phi_+)\ .
$}    
\end{equation}

\subsection{Bianchi and flux equations of motion}

We will now see how the equations imply the EoM for the flux, 
and (if $c_-=0$) their BI. (Remember that we have decided to call
EoM and BI the first and the second equation in (\ref{bianchi})
respectively.)

We start from the EoM. For the Minkowski case, we saw this in 
(\ref{eeom}). The general strategy is the same: we take the
imaginary part of (\ref{eq:nonintapp}) 
\begin{equation}
\label{eq:imB}  c_+ e^{4A}*\lambda(F) = 16 \Big( 
(d- H\wedge) (e^{3A-\phi}\mathrm{Im} \Phi_+)
+3 e^{2A-\phi}\mathrm{Im}(\bar \mu \Phi_+)\Big)\ .
\end{equation}
and act on it with $(d-H\wedge)$; remembering (\ref{eq:intapp}), 
we are left with $(d-H\wedge) (e^{4A}*\lambda(F))=0$ (notice 
that $c_+$ is never zero); pulling out $\lambda$ from $(d-H\wedge)$
we obtain $(d+H\wedge)(e^{4A}*F)=0$ as desired (see the second
equation in (\ref{bianchi})). 

It is not completely surprising that the EoM should follow from 
the conditions for a supersymmetric vacuum: if $(d+H\wedge)(e^{4A}*F)$
were non--zero, one could interpret its right hand side as a source.
These sources would be branes extended along the internal 
directions only; hence they would break Poincar\'e invariance, contrary
to our assumptions. (This argument is not completely rigorous, 
as the branes could be smeared in the external directions.)

We now come to the BI. Taking this time the real part of (\ref{eq:nonintapp}), we get
\begin{equation}
    \label{eq:BIsusy}
c_- F= (d- H\wedge) (e^{A-\phi} \mathrm{Re} \Phi_+) \ .
\end{equation}
If $c_-=0$, the first equation in (\ref{bianchi}) again follows
by acting with $(d-H\wedge)$. This means that no sources extended
along the external directions as well as some of the internal 
ones are present, either. It is only right, then, 
that the orientifold projection makes $c_-=0$: in this case, $(d-H\wedge)F=0$
does not follow any more, and the source required from the presence of an orientifold
is allowed. The orientifold makes room for itself, as it were.

Notice that in this paper we have 
always taken $c_-=0$. In the Minkowski case, we were forced 
 to do so by the 
orientifold projection. It is noticed in \cite{martucci} 
 that this is actually the case whenever the background can admit a 
supersymmetric probe brane. 

In the AdS case, $c_-=0$ is actually necessary: (\ref{eq:intapp})
tells us that $e^{A-\phi}{\rm Re} \Phi_+$ is $(d-H\wedge)$--exact,
and hence must also be $(d-H\wedge)$ closed, which implies
(comparing with  (\ref{eq:BIsusy})) that $c_-=0$. ($F$ is non--zero by 
assumption in this paper.)

As a final remark, notice that, once one takes $c_-=0$, both pure spinors 
have a norm 
$||\Phi||^2=\frac {c^2_+}{32} e^{2A}$. In the main 
text, (\ref{int}) and (\ref{nonint}), we have taken $c_+=2$.

\subsection{Sufficiency ($d\Phi_\pm$ equations
$\Rightarrow $ supersymmetry)}

We will now show explicitly why (\ref{eq:intapp}), (\ref{eq:nonintapp}) and
(\ref{eq:dnorm}) imply the supersymmetry equations (\ref{eq:intgravB}), 
(\ref{eq:extgravB}) and (\ref{eq:moddilB}). This proof was described
in words in \cite{gmpt2}; here we give the details. 

The algebraic part of the proof consists in showing that a pair of compatible
pure spinors $\Phi_\pm$ defines two Weyl spinors $\eta^{1,2}_+$. This
is essentially because $\Phi_\pm$ define an \stt\ structure on \tts, and
by projecting each of the two SU(3)'s one gets an SU(3) structure on the
base; each of these SU(3) structures can then be described by a spinor. 
We saw more details about this in the discussion around equation (\ref{eq:E}). 

The differential part is more complicated, and we will have to introduce
some notation. 
In this subsection indices $i_1, j_1\ldots$ and 
$\bar i_1, \bar j_1\ldots$ are (anti)holomorphic with respect to the almost
complex structure $I_1$ defined by $\eta^1$, and similarly  
$i_2, j_2\ldots$ and 
$\bar i_2, \bar j_2\ldots$ are (anti)holomorphic with respect to $I_2$
defined by $\eta^2$. $m$ is going to be a real index. 
Let us also define
\begin{equation}
\hspace{-.5cm}
    \begin{array}{c}\vspace{.2cm}
    \left(\sla D -\frac14\, \sla H\right) \eta^{1}= 
(T^{1}_m \gamma^m + T^{1}_+ + i T^{1}_-\gamma) \eta^{1}\ ,\\
    \left(\sla D +\frac14\, \sla H\right) \eta^{2}= 
(T^{2}_m \gamma^m + T^{2}_+ + i T^{2}_-\gamma) \eta^{2}\ ,
    \end{array}
\quad 
\begin{array}{c}\vspace{.2cm}
\left(D_m -\frac14 H_m\right) \eta^1=(iQ^1_{mn}\gamma^{n}
+ \del_m \log |a|+ i Q^1_m\gamma ) \eta^1\ ,\\
\left(D_m +\frac14 H_m\right) \eta^2=(iQ^2_{mn}\gamma^{n}
+ \del_m \log |b|+ i Q^2_m\gamma ) \eta^2\ ,
\end{array}
    \label{eq:QT}
\end{equation}
where $\eta^a=\eta^a_++\eta^a_-$, and hence $Q^a_m$, $Q^a_{mn}$ and $T^a_m$
are real. This is the expansion of the left hand sides in the complete
basis of spinors $\gamma^m \eta^a$, $\gamma\eta^a$, $\eta^a$, for $a$ either
1 or 2. 

We can also use the ``pure Hodge diamond'' 
\begin{equation}
  \label{eq:hodge}
  \begin{array}{c}\vspace{.1cm}
\Phi_+ \\ \vspace{.1cm}
\Phi_+\gamma^{i_2}  \hspace{1cm} \gamma^{\bar i_1}\Phi_+ \\ 
\Phi_-\gamma^{\bar i_2} \hspace{1cm} \gamma^{\bar i_1} \Phi_+\gamma^{i_2} 
\hspace{1cm} \gamma^{i_1} \bar \Phi_-\\
\Phi_- \hspace{1.2cm}\gamma^{\bar i_1}\Phi_-\gamma^{\bar j_2} 
\hspace{1cm} \gamma^{i_1}\bar\Phi_-\gamma^{j_2} 
\hspace{1.2cm}\bar\Phi_-\\
 \gamma^{\bar i_1} \Phi_-\hspace{1cm} \gamma^{i_1} \bar \Phi_+ \gamma^{\bar j_2} 
\hspace{1cm}\bar\Phi_-\gamma^{i_2}\\
\gamma^{i_1} \bar\Phi_+ \hspace{1cm}  \Phi_+\gamma^{\bar i_2}\\
 \bar\Phi_+\\
  \end{array}\ 
\end{equation}
to expand all differential forms. For example, for $F$ one defines
\begin{equation}
    \label{eq:expandF}
    \begin{array}{c}
\vspace{.2cm}  
\sla F= R^{10}_{i_2}\,\,\sla \Phi_+\gamma^{i_2} + R^{01}_{\bar i_1}\, \gamma^{\bar i_1} 
\,\sla\Phi_+ + \\\vspace{.2cm}  
 R^{30}\, \sla\Phi_-+ R^{21}_{\bar i_1 \bar j_2} \gamma^{\bar i_1}\,
\sla \Phi_- \,\gamma^{\bar j_2} + 
R^{12}_{i_1 j_2} \gamma^{i_1}\,  \sla\bar\Phi_-\gamma^{j_2}
+R^{03}\, \sla\bar\Phi_-\\
+R^{32}_{i_1} \gamma^{i_1}\, \sla \bar\Phi_+ 
+R^{23}_{\bar i_2}\,  \sla \bar\Phi_+ \gamma^{\bar i_2}\ ;
    \end{array}
\end{equation}
due to $\overline{\sla F}=-\sla F$ (the $\gamma^m$ are purely imaginary) 
one has $\overline{R^{ab}}= R^{3-a,3-b}$. 

The expansion of $d\Phi_\pm$ fortunately 
does not define independent quantities, as they can related to the 
$Q_m$, $Q_{mn}$, $T$, $T_m$ above by reexpressing them as 
$[\gamma_m, D_m(\eta^1_+\eta^{2\,\dagger}_+)]$ and
$\{\gamma_m, D_m(\eta^1_+\eta^{2\,\dagger}_-)\}$ as we did in the
previous subsections. We can now expand both equations (\ref{eq:intapp}), 
(\ref{eq:nonintapp}) to give
\begin{eqnarray}\vspace{.8cm}  
    T^1=-2\mu\,e^{-A}\ , \qquad 
    T^1_{\bar i_1} +i Q^2_{\bar i_1} =-\del_{\bar i_1} (2A-\phi+\log|b|)\ , 
    \  \quad Q^1_{\bar i_2 j_1}=0&&
\label{pure:intB}\\
\begin{array}{c}
    T^1_{\bar i_1} -i Q^2_{\bar i_1} =-\del_{\bar i_1} (2A-\phi+\log|b|)
    +\frac {e^\phi} 4 |a|^2 R^{01}_{\bar i_1} \ , \qquad
    T^1=-3\mu\,e^{-A}-  \frac {e^\phi} 4 |a|^2 R^{03}\vspace{.2cm}   \\
    \frac {e^\phi} 4 |b|^2 R^{01}_{\bar i_1}= \del_{\bar i_1} A \ , \qquad
    i Q^1_{i_2 j_1}=\frac {e^\phi} 4 |b|^2 R^{12}_{j_1 i_2} 
\end{array}\label{pure:nonintB}
\end{eqnarray}
plus another set of equations obtained from these by applying the rule
\fbox{$1\leftrightarrow
2\ , a\leftrightarrow b\ , R^{01}\to - R^{23}$} and by leaving the other
quantities invariant.   The equations (\ref{pure:intB}) and
(\ref{pure:nonintB}) come respectively from expanding (\ref{eq:intapp})
and (\ref{eq:nonintapp}). 
Again, in these expressions for example $\del_{\bar i_1}(\ldots)$ can be
read as $\Pi_{\bar i_1}{}^n \del_n (\ldots)$.

Now by simply 
 taking suitable linear combinations of  (\ref{eq:dnorm}) and 
(\ref{pure:intB}) and (\ref{pure:nonintB}), we can derive the equations
\begin{eqnarray}
    \label{spinor:intgravB}
 \hspace{-2cm}Q^1_{\bar i_2 j_1} =0\ , && 
i Q^1_{ i_2 j_1} =\frac{e^\phi}4|b|^2 R^{12}_{j_1 i_2}\ , \quad
i Q^1_{\bar i_2} + \del_{\bar i_2} \log|a| =0=
i Q^1_{ i_2}+ \del_{i_2} \log|a| 
+\frac{e^\phi}4 |b|^2 R^{10}_{i_2} \\
 &&\hspace{1cm}   \label{spinor:extgravB}
\mu\,e^{-A}+\frac{e^\phi}4 |b|^2 R^{03}=0\ , \qquad
\frac{e^\phi}4 |b|^2 R^{01}_{\bar i_1}=\del_{\bar i_1}A\ , \\
&&  \hspace{1cm} 2\mu\,e^{-A}+T^1=0\ , \qquad     T^1_{\bar i_1}+\del_{\bar i_1}(2A-\phi)=0\ 
\end{eqnarray}
along, again, with another set of equations obtained by the boxed rule above. 
These equations are exactly the content of (\ref{eq:intgravB}), 
(\ref{eq:extgravB}) and (\ref{eq:moddilB}) after expanding in the basis
$\gamma^{\bar i_1} \eta^1_+$, $\eta^1_+$, 
$\gamma^{i_1} \eta^1_-$, $\eta^1_-$ or, for the equation obtained
after the boxed rule, on the basis $\gamma^{\bar i_2} \eta^2_+$, $\eta^2_+$, 
$\gamma^{i_2} \eta^2_-$, $\eta^2_-$. 
This completes the proof. 

\section{Six--dimensional nilmanifolds and compact solvmanifolds} 
\label{app:Nil}

We collect some practical data on nilmanifolds and compact solvmanifolds here. 

As already mentioned, there are 34 nilpotent six--dimensional algebras \cite{MOMA1,MOMA2} all of which  admit cocompact lattice \cite{malcev}. This gives 34 {\it classes}
of nilmanifold; within each class the cocompact subgroup can vary, but for most of the 
paper only left--invariant forms are considered, which do not see the difference.
The algebras are placed in a descending order 
of ``twistedness"  in Table  \ref{ta:nil}  --- the 
bottom of the Table is occupied by the straight $T^6$ while the upper levels are populated by manifolds with lower first Betti numbers.  To make references easier 
we label each manifolds by two numbers --- its first Betti number and the relative position 
in the Table; these labels are in the first column of the Table. In the second 
column the structure constants are given. The way these define the action of the exterior
derivative $d$ on the forms $e^a$ is 
the following. For example, the manifold 2.1 of Table  \ref{ta:nil},  which has 
$b_1=2$ and is in the first line, has  structure constants given by  (0,0,12,13,14,15); 
thus, $de^1=0, de^2=0, de^3=e^1 \wedge e^2, de^4=e^1 \wedge e^3$, and so on. 

\begin{figure}[b]
\begin{center}
\includegraphics[width=3in]{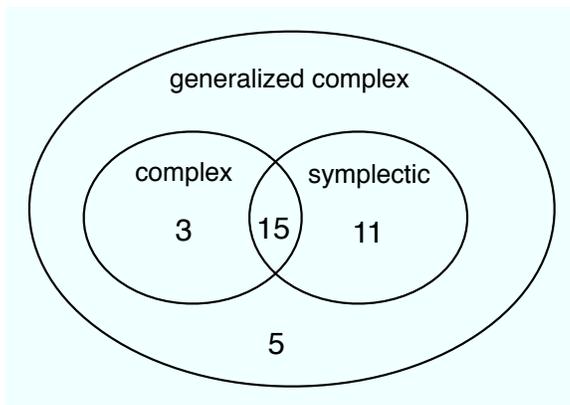}
\caption{Integrable structures on the 34 six--dimensional nilpotent Lie groups, from 
\cite{CG}.}
\label{fig:nil}
\end{center}
\end{figure}

We adopt a similar format for presenting compact (algebraic) solvmanifolds which are collected in  Table \ref{ta:solv}.  A word of caution is due here. 
As already mentioned, for the nilmanifolds,  the de Rham cohomology is 
isomorphic to the Lie algebra cohomology ${\cal G}$ \cite{nomizu}. This is no longer the case for generic solvmanifolds.  Thus the first of the integers labeling each entry in Table \ref{ta:solv} refers to the dimension of the first cohomology on the complex of the  left--invariant forms.

In table \ref{ta:nil}, the next four columns (taken from \cite{CG}) tell about the types of the {\sl closed} pure spinors, i.~e.~some  underlying integrable structure,  admitted by the given nilmanifold. If such a spinor exists,  it is marked by a symbol $\surd$. The construction  of the most general closed pure spinor of the given type is not hard, and examples are given in the text.  In order to compare to Figure \ref{fig:nil}, we can count the number of $\surd$ in each column and confirm the existence of 18 closed type 3 pure spinors (i.s. integrable complex structure) and 26   type 0 (i.e. integrable symplectic structure). Note that we talk here only about the individual pure spinors  --- two closed pure spinors of different types for a given manifold are not compatible and do not define a metric. In particular the 15 nilmanifolds that have both closed type 3 and 0, do not have a metric of special holonomy to go along with these. Finally, we see that there are 5 cases where there are no closed  type 0 or 3 pure spinors. These were the cases shown to admit a generalized complex structure in \cite{CG},  i.e. either type 1 or type 2 closed pure spinor.

 The last five columns (in both Tables) give the possible involutions compatible 
of the algebras. 
We actually only considered involutions that act with $\pm$ on the basis of one--forms 
$e^a$. All other involutions we found were related to these by changes of coordinates. 
It is possible that more involutions exist. 
With this in mind, the action of the involution on the one--forms 
$\sigma(e^i) = \pm e^i$ can be characterized, in each case, by a certain number 
of independent signs. For example, considering the manifold 2.1 
of Table  \ref{ta:nil}, it is not hard to see that choosing the sign 
of $e^1$ and $e^2$ determines all the others. For example, since $f^3_{12}=1$, the sign 
${\rm sign}(e^3)$ is forced to be equal to the product 
${\rm sign}(e^1) \times {\rm sign}(e^2)$. 
The rest of the signs are determined by the sign of these two. 
The manifold 2.3, on the other hand, is more restrictive due to the extra condition 
${\rm sign}(e^1) \times {\rm sign}(e^5)={\rm sign}(e^2) \times {\rm sign}(e^3)$.
 This leaves us with a single one--form whose sign can be chosen independently from 
the others. Of course when we realize all the possibilities, we are going to 
get different assignments of $+$ and $-$ signs which give us all the admissible orientifold 
planes. Clearly the case of six pluses is always allowed, while six minus signs are 
never allowed other than for zero structure constants, i.~e.~$T^6$. Thus it is not hard 
to see that if we sum the number of possible involutions for each manifold, the total 
is always $2^n -1$, where $n$ is the number of independent signs and we have subtracted 
one for the all-plus case which corresponds to an O9. The only exception to this 
is $T^6$, where to the total  on the last line of the Table  \ref{ta:nil}  we should add 
1  for the O3--plane in order to get all of the 63 allowed orientifold planes.

Some additional explanation is due for the second column in Table \ref{ta:solv}.
The criterion in \cite{saito} to check that a Lie group $G$ with given structure 
constants can be made compact by quotienting by a subgroup $\Gamma$ assumes that
$G$ be algebraic (as explained in \ref{sec:compact}). As checking this hypothesis
might be tricky, we have first applied Saito's criterion to all the solvable algebras
in \cite{Tarko} and \cite{Mubara5}. All the ones that passed the test are given in 
Table \ref{ta:solv}. Then, we have tried to check the algebraic hypothesis on each
case of this restricted list.  We have checked this by 
first explicitly finding by hand a faithful (one--to--one) $n$--dimensional 
(for some $n$) 
representation for $G$ in each case (the adjoint representation has usually a kernel and
hence cannot be used; we have found the representations by trial and error). 
Given the representation, we have 
seen if the resulting subgroup of Gl$(n, \Bbb R)$ could
be described by polynomial equations. We denote this by a ``$\yes$" in the second
column of the table. The ones for which this symbol is absent are still listed, because
there might be another representation, which we have not found, in which $G$ is actually
described by polynomial equations.


\begin{landscape}

\begin{table}[h]

\begin{center}
{\small
\begin{tabular}{| l || l || c|c|c|c|c|c|c|c|c|}
\hline
$n$ & Nilmanifold class  & T3 &T2 & T1 & T0 &  O4 & O5 & O6 & O7 & O8\\
\hline

2.1 & $(0,0,12,13,14,15)$ &  -- & -- &
$\surd$  &  $\surd$
& 1 &  35
&  246 & -- & --\\

\hline
2.2 & $(0,0,12,13,14,34+52)$ &  -- & -- & $\surd$ & --&  -- & 35;16;24
&-- & --  & --\\

\hline
2.3 & $(0,0,12,13,14,23+15)$  & -- & -- & $\surd$ &
$\surd$  & -- & 35 & -- & -- & --
\\

\hline
2.4 & $(0,0,12,13,23,14)$ &  -- &  -- & -- & $\surd $ &
  -- & 36;15;24 & -- & -- & --
\\

\hline
2.5 & $(0,0,12,13,23,14-25)$ &-- & -- & -- & $\surd$ &
-- & 36;15;24 &  -- & -- & --
\\

\hline
2.6 & $(0,0,12,13,23,14+25)$ & $\surd$ & $\surd$ &
$\surd$ & $\surd$ & -- & 36;15;24
&-- & --  & --\\

\hline
2.7 & $(0,0,12,13,14+23,34+52)$  & -- & -- & $\surd$  &-- & -- & 24 &-- & --  & -- \\

\hline
2.8 & $(0,0,12,13,14+23,24+15)$ & -- & -- & $\surd$ & $\surd$  & -- &  -- & 246 & -- & --
\\

\hline
\hline
3.1 & $(0,0,0,12,13,14+35)$ & -- & $\surd$ &$\surd$ & --  & -- &  {34};45
& -- & {1246} & --
\\

\hline
3.2 & $(0,0,0,12,13,14+23)$ & $\surd$ & $\surd$ & $\surd$ & $\surd$  &  -- & 34 &
135;236 & -- & -- \\

\hline
3.3 & $(0,0,0,12,13,24)$ &$\surd$ &
$\surd $ &
$\surd$ & $\surd$ & -- & 23;16;25;34;45
  &-- & 1246;1356 & --
\\

\hline
3.4 & $(0,0,0,12,13,14)$  & $\surd$ & $\surd$ & $\surd$ & $\surd$
&1&34;45 &  236;135;{256} &1246 & --
\\

\hline
3.5 & $(0,0,0,12,13,23)$  & $\surd$ &
$\surd$ & $\surd$ &
$\surd$   & -- & 16;25;34
& 124;135;236;456 & -- & --
\\

\hline
3.6 & $(0,0,0,12,14,15+23)$ & -- & -- & $\surd$ & $\surd$  &
-- & 46;13;25 &
-- & -- & --
\\

\hline
3.7 & $(0,0,0,12,14,15 + 23 + 24)$  & -- & -- & $\surd$ &
$\surd$  &
-- & 25 &
-- & -- & --
\\

\hline
3.8 & $(0,0,0,12,14,15+24)$ & -- & -- & $\surd$ &
$\surd$  &
-- & 25 & 235 & --  & $\bot$3 \\

\hline
3.9 & $(0,0,0,12,14,15)$ & -- & -- & $\surd$ &
$\surd$  & 1 &46;13;25
&235;346  & --  & $\bot$3 \\

\hline
3.10 & $(0,0,0,12,14,24)$ & $\surd$ &
$\surd$ & $\surd$ & -- & 4 &16;25;34  &136;235 & --   & $\bot$3\\

\hline
3.11 & $(0,0,0,12,14,13+42)$  & $\surd$ &
-- &  $\surd$ & $\surd$  &
-- & 34  &136;235 & --   & --\\

\hline
3.12 & $(0,0,0,12,14,23+24)$ & $\surd$ & -- &
$\surd$ & $\surd$  &
-- & 34;25;16 & - & --  & -- \\

\hline
3.13 & $(0,0,0,12,23,14+35)$ & -- & $\surd$ & $\surd$ & --
    &
-- & 13;15;26;34;45 &-- &1246;2356 & --
\\

\hline
3.14 & $(0,0,0,12,23,14-35)$ & $\surd$ &
$\surd$ & $\surd$ & -- &
-- & 13;15;26;34;45 &-- &1246;2356 & -- \\

\hline
3.15 & $(0,0,0,12,14-23,15+34)$  & -- &  $\surd$
&
$\surd$  & $\surd$  &
-- & 13  & {235};346 & --  & --\\

\hline
3.16 & $(0,0,0,12,14+23,13+42)$  & $\surd$ &
$\surd$ & $\surd$ &
$\surd$ &
-- & 34 &136;235 & --   & --\\
\hline
\hline

4.1 & $(0,0,0,0,12,15+34)$ & -- & $\surd$ & $\surd$ & --
&-- & 35;45;26;13;14
&-- &1256;2346& --\\

\hline
4.2 & $(0,0,0,0,12,15)$  & -- & $\surd$ &
$\surd$ & $\surd$
& 1;5& 26;13;14;{35};{45}  &{134;236;246;345}
&1256;2346& $\bot$:3;4\\

\hline
4.3 & $(0,0,0,0,12,14+25)$  & $\surd$ &
$\surd$ & $ \surd $& $\surd$ &
-- & 24;45 & 146;234;345 &1346& $\bot$3\\

\hline
4.4 & $(0,0,0,0,12,14+23)$ & $ \surd$ &
$\surd$ & $\surd$ & $\surd$
&-- & 56;13;24 &125;146;{236;345} & --& --\\
\hline

4.5 & $(0,0,0,0,12,34)$ & $\surd$&
$\surd$ & $\surd$ & $\surd$ &-- &
\minicent{2}{56;16;26;14;13\\23;24;35;45} &  -- &\minicent{2.5}{1235;1245;1256\\1346;2346;3456}& -- \\
\hline

4.6 & $(0,0,0,0,12,13)$ & $\surd$ &
$\surd$ & $\surd$ & $\surd$  & 1&
56;14;26;35;{23}  & \minicent{1.5}{234;125;136\\{ 246;345;456}} &1245;1346 & $\bot$4 \\

\hline
4.7 & ($0,0,0,0,13+42,14+23)$  & $\surd$ &
$\surd$ & $\surd$ & $\surd$  &
-- & 56;12;34  & {135;146;236;245} & -- & --\\

\hline \hline
5.1 &  $(0,0,0,0,0,12+34)$  & $\surd$ &
$\surd$ &
$\surd$ & -- & 6 & 56;13;14;23;24  &\minicent{2.3}{126;346;135;145\\235; 245} &1256;3456  & $\bot$5\\
\hline
5.2 & $(0,0,0,0,0,12)$ & $\surd$ &
$\surd$ & $\surd$& $\surd$& 1;2;6 &
\minicent{2}{56;13;14;23;24 \\36;46;15;25}  &
\minicent{2}{126;346;135;145\\235;245;356;456\\134;234} $ \ $ & \minicent{2.3}{1236;1246;1256\\1345;2345;3456}& $\bot$:3;4;5 \\
\hline \hline
6.1 & $(0,0,0,0,0,0)$ &
$\surd$ &
$\surd$ & $\surd$ &  $\surd$  & any (6)  & any (15)  & any (20)  & any (15)  &any (6)
\\
\hline
\end{tabular} }
\end{center}
\caption{\label{ta:nil} Six--dimensional nilmanifolds. 
}
\end{table}
\end{landscape}

\begin{landscape}
\begin{table}[h]
\begin{center}
{\small
\begin{tabular}{| l | c || l || c|c|c|c|c|}
\hline
$s$ & Alg.? & Solvmanifold class  & O4 & O5 & O6 & O7 & O8\\
\hline
1.1 & $\yes \ \forall \alpha\in\Bbb Z$ & $ (23,-36,26,-\alpha 56, \alpha 46,0)$ & 
 - &  16,24,25,34,35
&  -- & 1236,1456 & --
\\
\hline
1.2 &  &$ (24+35,-36, 26, -56, 46, 0)$ & - &  16,25,34
&  124,135,236,456 & -- & --
\\
\hline
2.1 & & $(23,0,26,-56,46,0)$ 
& - &  16,24,25,34,35
&  -- & 1236,1456 & --
\\
\hline
2.2 & $\yes$ &$ (35-26,45+16,-46,36,0,0) $ 
& - &  14,23,56
&  126,135,245,346 & -- & --
\\
\hline
2.3 &$\yes$ & $ (24+35,0,-56,0,36,0)$ 
& - &  16,23,25,34,45
&  -- & 1246,1356 & --
\\
\hline
2.4 & $\yes$&(23,-13,0,56,-46,0)& -- & \minicent{2}{14,15,16,24,25\\26,34,35,36}  & -- &  \minicent{2}{1234,1235,1236\\1456,2456,3456} & --
\\
\hline
2.5& $\yes \ \forall \alpha\in\Bbb Z$& $(25,-15,\alpha 45,- \alpha 35,0,0)$ & 5 & 13,14,23,24,56 & 125,136,146,236,246,345 & 3456, 1256 & $\bot$ 6
\\
\hline
2.6& & (25+35,-15+45,45,-35,0,0) &5 & 14,23,56 & 146,236 & -- & $\bot$ 6
\\
\hline
\hline
3.1 & $\yes$& (23,-13,0,56,0,0) & -- & \minicent{2}{14,15,16,24,25\\26,34,35,36}  & -- &  \minicent{2}{1234,1235,1236\\1456,2456,3456} & --
\\
\hline
3.2 &$\yes$& (23,34,-24,0,0,0) &2, 3 & 14,25,26,35,36 & 146,145,256,356 &  1234, 1456 & $\bot$:5,6
\\
\hline
3.3 & &(25,0,45,-35,0,0) &5 & 13,14,23,24,56 & 125,136,146,236,246,345 & 3456, 1256 & $\bot$6
\\
\hline
3.4& & $(23+ 45,-35,25,0,0,0)$ & -- & 24,34 & 145,246,346 & 1456  & $\bot$ 6
\\
\hline
\hline
4.1 & $\yes$ &$(23,-13,0,0,0,0) $  &1;2;3 &
\minicent{2}{14,15,16,24,25,\\ 26,34,35,36}  &
\minicent{2}{123,145,156,346,256\\ 245,356,345,146,246} $ \ $ & \minicent{2.3}{1236,1234,1235\\1456,2456,3456}& $\bot$:4,5,6 \\
\hline
\end{tabular}}
\end{center}
\caption{\label{ta:solv} Six--dimensional compact solvmanifolds which are possibly
algebraic. }

\end{table}

\noindent
In both Tables we label the manifolds  by the dimension of the first cohomology on the space of left--invariant forms, and the position they occupy in the Table. 
A $\surd$ indicates that the manifold admits a closed pure spinor of the given type (T3=type 3). We indicate the allowed O4,O5, O6 and O7--planes by giving its parallel directions, and O8--planes by giving the transverse ones. The symbol `-'
denotes nonexistence of either a pure spinor or an O--plane. 
In Table \ref{ta:solv}, only the ones with a "$\yes$" in the second column 
are established
as compact solvmanifolds; the other ones might or might not be. Also, we do not 
consider non--algebraic compact solvmanifolds. Notice that {\it s 3.2} is
actually obtained by setting $\alpha=0$ in {\it s 1.1}, and likewise {\it s 4.1}
is obtained by setting $\alpha=0$ in {\it s 2.5}. Given that doing this changes the 
Lie algebra cohomology (the cohomology computed on the left--invariant forms), we have 
decided to list them separately. 

\end{landscape}

\section{Orientifold projections and pure spinor equations}
\label{app:eqs}

In this Appendix we give the form of the supersymmetry 
equations (\ref{int}) and (\ref{nonint}) for the different orientifold 
projections in IIA and IIB in the cases of type0--type3 spinors (\ref{puresu3}) 
and type1--type2 (\ref{puresu2}), subject to the rescaling Ansatz 
supposing the rescalings (\ref{rescale}):
\[
e^\alpha_+ = e^{A} \tilde e^\alpha_+ \, \qquad   e^i_- = e^{-A} \tilde e^i_- \  .
\]
We also give the most 
general form of pure spinors in terms of the 1-forms $e^\alpha, e^i$ transforming under the involution as $\sigma(e^\alpha)=e^\alpha$, $\sigma(e^i)=-e^i$ and the moduli of the 
manifolds.

\subsection{IIB: O5 orientifolds}
\label{sec05A}
For type3--type0 pure spinors, (\ref{puresu3} the O5 projection sets $a=b$.
Inserting the explicit expressions for the spinors in the supersymmetry 
variations (\ref{int}),(\ref{nonint}) with  
$\Phi_1=\Phi_-$, $\Phi_2=\Phi_+$, we obtain 
\bea
&& e^\phi=\gs e^{2A} \nn \\
&& d (e^A \Omega_3) = 0 \nn \\
&& d J^2  = 0 \nn \\
&& d(e^{2A}J) = \gs e^{4A}  *F_3 \nn\\
&& H =0  \ ,
\label{03O5}
\eea
where $\gs$ in the first equation is an  integration constant, that  we take to be  the large volume limit ($A=0$) of $e^\phi$. In this case, these equations
actually follow from (\ref{int}), (\ref{nonint}), without 
assuming (\ref{rescale}).   
 The most general $\Omega_3$ compatible with the O5 projection (which leaves it  invariant) is\footnote{This is the most
 general $\Omega_3$ compatible with an O5 up to an  overall complex scaling which can be absorbed
 by the phase of $b$ and the modulus $|b|^2$ relative to $e^A$. Equations (\ref{int}), (\ref{nonint})
 are invariant under this overall scaling.} 
\beq \label{Om}
\Omega_3 \equiv   z^1  \wedge    z^{2} \wedge    z^{3} =  (  e^{1}_- + 
 i \tau^{1}   e^{3}_ - +  i \tau^2    e^4_-) \wedge (  e^2_- + 
 i \tau^{3}   e^3_-  + i \tau^4    e^4_-) \wedge 
 (  e^1_+ + i \tau^+   e^2_+) 
 \ .
\eeq
where $\tau^+, \tau^1,\ldots, \tau^4$ are complex constants  and by 
definition $  z^i$ are the holomorphic 1-forms. 
The most general fundamental 2-form compatible with $\Omega_3$ (i.~e., 
satisfying (\ref{compsu3})) is 
\beq \label{Je}
J=\frac{i}{2} \left( t_{1} \   z^{1}\wedge   {\bar z}^{1} +  t_2 \   z^{2} \wedge   {\bar z}^{2}+ b \    z^{1} \wedge   {\bar z}^{2} - \bar b \  {\bar z}^{1} \wedge    z^{2} + t_3 \    z^3 \wedge   {\bar z}^{3 } \right) \equiv
J_{--} + J_{++}
\eeq 
where $t_i$ are real and $b$ is complex. The normalization condition (\ref{compsu3}) requires $ t_3 (t_1 t_2 -|b|^2) =1$.

We now turn to the discussion of the hybrid of SU(2) structure.
Inserting the type 1-type 2 pure spinors of Table 2 in 
eqs (\ref{int}), (\ref{nonint}), we get the following system
\bea \label{12O5}
&& e^\phi=\gs e^{2A} \nn \\
&& d(e^{A} \Omega_1) =0 \nn \\
&& \Omega_1 \wedge  dj= -i \ \Omega_1  \wedge  H \nn \\
&& d(e^{2A}{ \R }\Omega_2) =- \gs e^{4A} *F_3 \nn \\
&&  d( {\I} \Omega_2)  = 0 \nn \\
&&  i\, dA \wedge \,  \R \Omega_2 \wedge \Omega_1 \wedge \bar \Omega_1 
-\frac{1}{2}H  \wedge \I \Omega_2  
=  -\frac{i \gs}{4}e^{2 A} *F_3 \wedge  \Omega_1 \wedge \bar \Omega_1  \nn \\
&& \frac{1}{2}H  \wedge {\R} \Omega_2 = \frac{\gs}{2}e^{2A} *F_1 
\eea

The transformation properties of the pure spinors and $H$ under 
the orientifold projection are given in Table 2. These imply
\bea
\Omega_1 &\equiv& 
   z^3 = \tau^-_i    e^{i}_-  \nn \\
\Omega_2&\equiv&   z^1 \wedge   z^2 =   (   e^{1}_+ + i \tau^1_i   e^{i}_-) \wedge (   e^{2}_+ +  i \tau^2_i   e^{i}_-)  \nn \\
j &=&\frac{i}{2} \left( t_1  z^1 \wedge \bar z^1 + t_2 z^2 \wedge \bar z^2 + b ( z^1 \wedge \bar z^2 - \bar z^1 \wedge z^2) \right) \nn \\
H &=& h_i \     e^{i}_-   \wedge e^{1}_+ \wedge   e^{2}_+ + 
h'_i \ \epsilon^{ijkl} \     e^{j}_-   \wedge e^{k}_- \wedge   e^{l}_- \ ,
\eea
where a sum over $i,j,k,l=1,\ldots,4$ is understood. In these equations 
$\tau^-_i$ are complex, while $\tau^1_i$, $\tau^2_i$ are real, which means there are
8 complex structure moduli in all.  On the other hand, $t_1$, $t_2$ 
and $b$ are all real, the same as $h_i$, $h'_i$. 

The volume is given by
\beq
{\rm vol} =\sqrt{\tilde g} \,  e^{1}_- \wedge  e^{2}_- \wedge  e^{3}_- \wedge 
e^{4}_- \wedge  e^{1}_+ \wedge  e^{2}_+ = -8i e^{-2A}   \left< \Phi_\pm, \bar \Phi_\pm \right> =
 \frac{i}{8}  \Omega_2 \wedge \bar \Omega_2 \wedge \Omega_1 \wedge \bar \Omega_1
 \eeq
 where the normalization condition $ \left< \bar \Phi_+, \Phi_+ \right> = \left< \bar \Phi_-, \Phi_- \right>$ requires $\Omega_2\wedge \bar\Omega_2=2 j^2$, or in other
words  $(t_1 t_2 - |b|^2)=1$. 

\subsection{IIB: O7 orientifolds}
\label{secO7}
For the O7 case, the supersymmetry equations (\ref{int}) and (\ref{nonint}) for type0-type3 become
\bea
e^\phi&=& \gs e^{4A} \nn \\
d (e^{-A}  \Omega_3) &=&0 \nn \\
d (e^{-2A} J) &=& 0 \nn \\
d(J^2) &=& - 2 \, \gs\, e^{4A}  *F_1 \nn \\
H &=& e^{4A} \, \gs\, *F_3 \nn \\
H \wedge \Omega_3 &=& H \wedge J = 0 
\label{03O7}
\eea
and the general form of t $\Omega_3$  and $J$ compatible with the orientifold projection is 
\bea \label{OmJ7}
\Omega_3&=&  (e^{1}_+ + 
 i \tau^{1} e^{3}_+  + i \tau^2  e^{4}_+) \wedge (e^{2}_+ + 
 i \tau^{3} e^{3}_+  + i \tau^4  e^{4}_+) \wedge 
 (e^{1}_- + i \tau^- e^{2}_-)  \nn \\
 J&=& \frac{i}{2} \left( t_{1} \ z^{1}\wedge  {\bar z}^{1} +  t_2 \ z^{2} \wedge  {\bar z}^{2}+ b \  z^{1} \wedge  {\bar z}^{2} - \bar b \ {\bar z}^{1} \wedge  z^{2} + t_3 \  z^3 \wedge  {\bar z}^{3 } \right) \ .
\eea
In these expressions, all $\tau$'s as well as $b$ are complex, while $t_i$ are real.


 For type1--type 2  the supersymmetry equations are
\bea \label{12O7}
e^\phi&=& \gs e^{4A} \nn \\
d(e^{-A} \Omega_1)&=&0 \nn \\
\Omega_1 \wedge  dj&=& -i \ \Omega_1  \wedge  H \nn \\
 d(e^{-2A} \R \Omega_2) & =&  0 \nn \\
 d( \I \Omega_2)  & =& -\gs\, e^{4A} *F_3 \nn \\
-2 i\, dA \wedge \,  \R \Omega_2 \wedge \Omega_1 \wedge \bar \Omega_1 - 
H  \wedge \I \Omega_2    &=&  \gs\, e^{4 A} *F_1   \nn \\
2 \, i  H  \wedge \R \Omega_2  &=&  -2 \, i \, \gs\, e^{4A} *F_3 \,  \Omega_1 \wedge \bar \Omega_1 
\eea

The transformations of the spinors imply
\noindent
\bea
\Omega_1 & \equiv & z^3 = \tau^+_\alpha  e^{\alpha}_+  \nn \\
\Omega_2&\equiv & z^1  \wedge z^2= ( e^{1}_- + i \tau^1_i e^{i}_+) \wedge ( e^{2}_- +  i \tau^2_i e^{i}_+)  \nn \\
j &=&\frac{ i}{2} \left( t_1  z^1 \wedge \bar z^1 + t_2 z^2 \wedge \bar z^2 + b \, ( z^1 \wedge \bar z^2 - \bar z^1 \wedge z^2) \right) \nn \\
H &=& h_{1i} \   e^{i}_- \wedge e^{1}_+ \wedge e^{2}_+ + 
 h_{2i} \   e^{i}_- \wedge e^{1}_+ \wedge e^{3}_+ + h_{3i} \   e^{i}_- \wedge e^{1}_+ \wedge e^{4}_+ +  \nn \\
 &&h_{4i} \   e^{i}_- \wedge e^{2}_+ \wedge e^{3}_+ + h_{5i} \   e^{i}_- \wedge e^{2}_+ \wedge e^{4}_+ + h_{6i} \   e^{i}_- \wedge e^{3}_+ \wedge e^{4}_+
 \eea
where $\tau_i^+$ are complex, while all the rest are real.

\subsection{IIA: O6 orientifolds}
\label{secO6}
We start with the SU(3) structure case. 
For type3-type 0 spinors corresponding to O6, eqs. (\ref{int}) and
(\ref{nonint}) become
\bea
e^\phi&=& \gs e^{3A} \nn \\
H &=& 0 \nn \\
d J &=& 0 \nn \\
d (e^{-A}  \R  \Omega_3) &=&0 \nn \\
d(e^{A} \I  \Omega_3) &=&  \, -\gs\, e^{4A}  *F_2\ .
\label{03O6}
\eea

The most general spinors compatible with the orientifold projection are
\bea \label{Om6A}
\Omega_3& \equiv & z^1 \wedge z^2 \wedge 
 z^3  =   (e^{1}_- + 
 i \tau^{1}_\alpha e^{\alpha}_+ ) \wedge (e^{2}_- + 
 i \tau^{2}_\alpha e^{\alpha}_+ ) \wedge 
 (e^{3}_- + i \tau^3_\alpha e^{\alpha}_+ )\nn \\
J&=& \frac{i}{2}  \sum_{i=1}^{3}  t_{i} \ z^{i} \wedge  {\bar z}^{i}  +  \I  \left(  b_1 \  z^{2} \wedge  {\bar z}^{3} + b_2   \  z^{3} \wedge {\bar  z}^{1} + b_3 \  z^1 \wedge {\bar z}^2  \right)
\eea
where all $\tau^i_j$,  $a_i$ and $b_i$ are real.

For type1 -- type 2 spinors, or in other words for SU(2) structure, 
and an O6 projections we get the following equations
\bea \label{12O6}
e^\phi&=& \gs e^{3A} \nn \\
d(\Omega_2)&=&0 \nn \\
d(\Omega_1 \wedge \bar \Omega_1) \wedge \Omega_2&=& 2\,  H \wedge   \Omega_2   \nn \\
 d(e^{-A} j  \wedge \R \Omega_1) & =&  - e^{-A} H \wedge \I \Omega_1 \nn \\
 d( e^{A} \R \Omega_1)  & =& \, \gs\, e^{4A} *F_4 \nn \\
d(e^A j\wedge  \I \Omega_1  )   &=&   e^A H  \wedge \R \Omega_1 
- \, \gs\, e^{4 A} *F_2  
\eea

The general spinor for this case are
\bea
\Omega_1 & \equiv&  z^3 = \tau^-_i   e^{i}_- + i  \tau^+_\alpha     e^{\alpha}_+  \nn \\
\Omega_2& \equiv&  z^1 \wedge  z^2 = (\tau^1_\alpha  e^{\alpha}_+) \wedge (\tau^2_i   e^{i}_-) \nn \\
j &=& \frac{i}{2} \left( t_1 \,  z^1 \wedge \bar z^1 + t_2 \, z^2 \wedge \bar z^2  \right) \nn \\
H &=& h_{-} \    e^{1}_-  \wedge e^{2}_- \wedge  e^{3}_- + 
h_{i \alpha} \ \epsilon^{\alpha \beta \gamma} \   e^{i}_- \wedge  e^{\beta}_+ \wedge  e^{\gamma}_+ \ .   
 \eea
In these expressions, $\tau^{1,2}_{\alpha,i}$ are complex, while $\tau^{\pm}_{\alpha,i}$, $t_i$ and $h$'s are real.

\subsection{IIA: O4 and O8 orientifolds}
\label{secO4}
The O4 and O8 projections do not allow SU(3) structure. Thus we only give the supersymmetry
variations for the type1--type2 pure spinors
\bea
\label{O4eq}
&& e^\phi = \gs e^{A} \, , \nn\\
&& d (e^A v') = 0 \, ,\nn\\
&& d(e^{2 A} \Omega_2) = 0 \, , \nn\\
&& d (e^A v \wedge j) = - e^A  H \wedge v' \, , \\
&& {\rm Re} \Omega_2 \wedge d(v \wedge v') =  -H \wedge {\rm Im} \Omega_2 \, ,\nn\\
&& {\rm Im} \Omega_2 \wedge d(v \wedge v') = H \wedge {\rm Re} \Omega_2 \,  , \nn\\
&& v' \wedge j \wedge dj = -H \wedge v \wedge j \, ,\nn\\
&& d( e^{3 A} v) = \,- \gs\, e^{4 A} * F_4 \, ,\nn\\
&&  d( e^{3 A} v' \wedge j) = e^{3 A} H \wedge v + \, \gs\, e^{4 A} * F_2 \, ,\nn\\
&&  d( e^{3 A} v \wedge j \wedge j ) = - 2 e^{3 A} H \wedge v' \wedge j  +2 \, \gs\, e^{4 A} * F_0 \, .
\eea
The general pure spinors compatible with an O4 and the NS flux are
\bea
\Omega_1&\equiv&z^3 =e_+ + i \tau^3_i e^i_- \nn \\
\Omega_2&\equiv&z^1 \wedge z^2 = (\tau^1_i  e^{i}_- ) \wedge (\tau^2_i   e^{i}_-) \nn \\
j&=&  \frac{i}{2}  \left( t_1 \,  z^1 \wedge \bar z^1 + t_2 \, z^2 \wedge \bar z^2 + b \    z^{1} \wedge   {\bar z}^{2} - \bar b \  {\bar z}^{1} \wedge    z^{2}   \right) \nn \\
H&=& h_{ij} \epsilon^{ijklm} e^k_- \wedge e^l_- \wedge e^m_-  
\eea
where $\tau^3_i$, $t_i$ are real, while $\tau^{1,2}_i$ and $b$ are complex, and we should
bear in mind that $\Omega_1=v+iv'$.
The spinors compatible with an O8 are 
\bea
\Omega_1&\equiv&z^3 =\tau^3_\alpha e^i_\alpha + i e_- \nn \\
\Omega_2&\equiv&z^1 \wedge z^2 = (\tau^1_\alpha  e^{\alpha}_+) \wedge (\tau^2_\alpha   
e^{\alpha}_+) \nn \\
j&=&  \frac{i}{2}  \left( t_1 \,  z^1 \wedge \bar z^1 + t_2 \, z^2 \wedge \bar z^2 + b \    z^{1} \wedge   {\bar z}^{2} - \bar b \  {\bar z}^{1} \wedge    z^{2}   \right) \nn \\
H&=& h_{ij}  e^- \wedge e^i_+ \wedge e^j_+  
\eea
where $\tau^3_i$, $t_i$ are real, while $\tau^{1,2}_i$ and $b$ are complex.

In the presence of O8--planes alone, the only flux allowed is $F_0$, and it comes
purely from a derivative of the warp factor.  In the large volume
limit, there are no RR fluxes and therefore the manifolds are 
generalized K\"ahler. This is similar to the situation for O7 projection. 
However, the pure spinors for O8 have a very nonsymmetric expression, and looking
for two closed pure spinors of this form turns out to be much more involved.
We have checked that there are no solutions among the compact solvable not nilpotent algebras, but 
the possibility of finding a  solution among the nilpotent ones is not ruled out. 




 


\end{document}